\documentclass{article}

\usepackage[main, final]{neurips_2026}
\usepackage[utf8]{inputenc} % allow utf-8 input
\usepackage[T1]{fontenc}    % use 8-bit T1 fonts
\usepackage{hyperref}       % hyperlinks
\usepackage{url}            % simple URL typesetting
\usepackage{booktabs}       % professional-quality tables
\usepackage{amsfonts}       % blackboard math symbols
\usepackage{nicefrac}       % compact symbols for 1/2, etc.
\usepackage{microtype}      % microtypography
\usepackage{xcolor}         % colors

\usepackage{amsmath, amssymb, amsthm}
\usepackage{mathtools}
\usepackage{bm}
\usepackage{tikz} 
\usepackage{subcaption}

\newtheorem{lemma}{Lemma}[section]
\newtheorem{proposition}[lemma]{Proposition}
\newtheorem{assumption}[lemma]{Assumption}
\newtheorem{remark}[lemma]{Remark}
\newtheorem{theorem}[lemma]{Theorem}
\newtheorem{definition}[lemma]{Definition}
\newtheorem{corollary}[lemma]{Corollary}

\DeclareMathOperator{\Var}{Var}
\DeclareMathOperator{\Cov}{Cov}
\DeclareMathOperator{\LLR}{LLR}
\DeclareMathOperator{\expit}{expit}
\DeclareMathOperator{\sgn}{sgn}

\title{AI-Assisted Classification under Correlation Neglect and Trust}

\author{%
Dingwen Kong\\
  MIT\\
  \texttt{dingwenk@mit.edu}
  \And
  Liang Lyu\\
  MIT\\
  \texttt{lianglyu@mit.edu}
  \And
  Saurabh Amin \\
  MIT\\
  \texttt{amins@mit.edu} \\
  \And
  Amine Bennouna \\
  Northwestern University \\ 
\texttt{amine.bennouna@northwestern.edu}
  \And
  Daniel Huttenlocher\\
  MIT\\
  \texttt{huttenlocher@mit.edu}\\
  \And
  Asuman Ozdaglar\\
  MIT\\
  \texttt{asuman@mit.edu}
}

\begin{document}

\maketitle

\begin{abstract}
When does an AI assistant improve human decision-making, and when does it hurt? We argue that the answer depends not just on AI accuracy but also on how much its evidence overlaps with the human's and how the human combines the two. A worker makes a binary decision by observing her own signal and receiving a recommendation from an AI. We develop a human behavioral model with two primitives: her awareness of the overlap and her trust in the AI. Our analysis gives a closed-form phase diagram over AI capability and overlap, with regions of complementarity, impairment, and automation. A worker who neglects the overlap (correlation neglect) can be made less accurate by a more capable AI, double-counting the shared evidence. Trust reshapes the regions - undertrust may compensate for correlation neglect, so that the trust level maximizing accuracy depends on the deployment parameters. Empirically, we identify the two model parameters by pairing the same workers with different AIs, and break the accuracy loss into a correlation-neglect cost and an undertrust cost, one tied to reducing overlap on the AI side, the other to calibrating trust on the human side. On ChatBench, AI assistance raises accuracy by 21 points, yet workers apply only about a quarter of the Bayes-rational weight to AI information; GPT-4o falls in the automation region and Llama-3.1-8b stays at the complementarity boundary; and calibrating trust recovers about 2.6 points for GPT-4o but almost nothing for Llama-3.1-8b.
\end{abstract}

\section{Introduction}

Empirical evaluations of human-AI teams reach very different conclusions. Some studies report large accuracy gains from AI assistance~\citep{tschandl2020human,steyvers2022bayesian}, while many find no improvement over the unaided human, or outright degradation: a more capable AI can hurt team performance~\citep{bansal2019updates}, AI explanations need not help~\citep{bansal2021does}, AI predictions may not improve expert radiologists~\citep{agarwal2023combining}, and a meta-analysis across tasks and systems finds that human-AI teams on average perform worse~\citep{vaccaro2024when}. We argue that whether assistance helps turns not on the AI's accuracy alone, but also on how much its evidence overlaps the human's and on how the human combines it with her own. For example, when a worker treats overlapping evidence as if it were independent, she counts the shared part twice, a bias documented as correlation neglect in belief formation~\citep{enke2019correlation,eyster2010correlation}, and the team can end up less accurate than the worker alone. This paper constructs a model built on these factors, uses it to characterize when AI assistance helps and when it hurts, and develops an empirical estimation strategy that exploits variation of outputs obtained by pairing the same workers with different AIs.

We study a binary classification problem in which the human and the AI each hold evidence about the state, summarized by their log-likelihood ratios $\ell_H$ and $\ell_A$, and the worker decides by the sign of a weighted sum $w_H \ell_H + w_A \ell_A$. The model has two parts. In the \emph{observation model}, each agent's evidence is a collection of \emph{cues}, and the key primitive is the \emph{overlap coefficient} $\lambda$, the share of the AI's cues the human already holds. In the \emph{behavioral model}, the worker combines the two signals through correlation awareness $\hat\alpha \in [0, 1]$, representing how well she accounts for that overlap, and trust $\hat\beta > 0$, how much weight she puts on the AI information. Two benchmarks sit inside this family. The \emph{Bayes-rational} worker ($\hat\alpha = 1, \hat\beta = 1$) de-correlates and weights optimally, while the \emph{correlation-neglect} worker ($\hat\alpha = 0$) treats the two signals as independent. Overlap is a property of the information, awareness and trust are properties of the worker, and together they determine the team's accuracy.

% The model yields three sets of results. First, closed-form thresholds in AI capability and signal overlap partition the deployment space into impairment, complementarity, and automation regions; the impairment region exists when overlap exceeds $1 - \hat\beta/2$, and higher trust amplifies double-counting and expands impairment but also extends complementarity to higher AI capability. Second, the behavioral primitives $(\hat\alpha, \hat\beta)$ are jointly identified from variation in signal overlap across AI systems paired with a common human population. Third, for correlation-neglect workers, the welfare loss against the Bayes-rational benchmark decomposes additively into a correlation-neglect cost and a trust-miscalibration cost, each linked to a distinct intervention point on the human side (trust calibration) or the AI design side (overlap reduction).

Our first results compare team performance against human and AI alone. Under the Bayes-rational benchmark the team is at least as accurate as either source alone, and strictly more accurate unless the AI's evidence already contains the human's. The impairment and automation outcomes seen in the literature, outside that case, must then come from the worker's behavior rather than from the information structure. Away from that benchmark, we obtain a closed-form phase diagram over AI capability $\tau_A$ and overlap $\lambda$ with three regions, complementarity where the team is most accurate, impairment where the unaided human is, and automation where the AI alone is. Two of its features run against intuition. The first is that a more capable AI can make the team less accurate. When overlap is high ($\lambda > 1 - \hat\beta/2$), a stronger AI only gives a correlation-neglecting worker more evidence to double-count, and that can lower accuracy. The second is that undertrust is not simply an error to be corrected. Reducing trust shrinks the impairment region but brings automation on at lower capability, so the trust level that maximizes team accuracy is not always $\hat\beta = 1$ but depends on the deployment parameters.

% Third, the behavioral primitives $(\hat\alpha, \hat\beta)$ are jointly identified from variation in signal overlap across AI systems paired with a common human population. Fourth, for correlation-neglect workers, the welfare loss against the Bayes-rational benchmark decomposes additively into a correlation-neglect cost and a trust-miscalibration cost, each linked to a distinct intervention point on the human side (trust calibration) or the AI design side (overlap reduction).

% We instantiate the framework empirically on ChatBench~\citep{chang2025chatbench}, in which 643 workers answer multiple-choice questions paired with either GPT-4o or Llama-3.1-8b. AI assistance raises worker accuracy by 21 percentage points, but workers undertrust the AI by roughly a factor of four. GPT-4o lies in the automation regime while Llama-3.1-8b sits near the complementarity-automation boundary, and the residual welfare loss is primarily attributable to undertrust.

Our second set of results turns the model into a measurement tool. A Bayes-rational worker's weight on the AI changes with overlap, while a correlation-neglecting worker's does not, and this difference lets us estimate the parameters by pairing the same workers with two AIs of different overlap. Trust $\hat\beta$ is sharply identified, while awareness $\hat\alpha$ is only weakly identified because the overlap coefficient in our experiment is relatively small. We then split the worker's accuracy loss against the Bayes-rational benchmark into two parts that add up exactly at any level of awareness, a correlation-neglect cost from misperceiving the overlap and an undertrust cost from underweighting the AI. Each is tied to a distinct lever on opposite sides of the information-behavior divide: reducing overlap on the AI side by training it to supply what the human lacks, and calibrating trust on the human side.

We examine the framework in the context of ChatBench~\citep{chang2025chatbench}, where 643 workers answer multiple-choice questions paired with either GPT-4o or Llama-3.1-8b. AI assistance raises accuracy by 21 percentage points, yet workers put only about a quarter of the Bayes-rational weight on the AI ($\hat\beta = 0.285$). GPT-4o is more capable and overlaps the human less, so it lies in the automation regime, where the AI alone beats the team; Llama-3.1-8b is weaker and overlaps more, so it sits at the boundary between complementarity and automation. The worker's behavior is nearly the same in both, so it is the difference in AI capability between the two deployments that shifts the deployment from one regime to the other, as the framework predicts. The cost of undertrust is therefore uneven. We simulate a trust-calibrated decision rule, which recovers about $2.6$ points accuracy for the strong AI and moves the team toward the boundary, but almost nothing for the weak AI.

The two sides of the information-behavior divide provide two design levers. On the human side, trust should be set to the deployment's accuracy-maximizing level, not reflexively to $\hat\beta = 1$, since some undertrust is protective in the impairment region. On the AI side, reducing overlap by training the AI to supply what the human lacks rather than what she already knows can raise team accuracy even without raising the AI's own accuracy. The phase diagram also says when to show the AI at all: withhold its signal when the AI is too weak, provide it in the complementarity band, and automate once the AI is strong enough.

Our work sits among several literatures but differs in what it treats as endogenous. Studies of optimal human-AI collaboration through sufficient statistics~\citep{agarwal2025designing} leave out the conditionally dependent signals that are central here. Work on reducing overreliance~\citep{bucinca2021trust} and on testing whether AI adds to human expertise~\citep{alur2023auditing, alur2024human} takes the intervention or the test as given, whereas we characterize the best deployment as a function of capability, overlap, and behavior. Closest in spirit, \citet{Guo2024Reliance} define over- and under-reliance from the human's adoption or switch rates against a Bayesian benchmark; our two parameters give a micro-foundation for their two sources of loss, trust for their reliance loss and correlation awareness for their discrimination loss, with a model of a worker who combines a continuous AI signal with her own rather than choosing per instance between the two recommendations. Other work on combining expert information~\citep{donahue2022human, peng2025no} and learning-to-defer~\citep{madras2018predict, mozannar2020consistent, verma2023learning, wilder2020learning} takes the human's decision rule as exogenous, whereas we derive it from behavioral primitives. Section~\ref{sec:model} introduces the model, Section~\ref{sec:analysis} develops the phase diagram and welfare decomposition, and Section~\ref{sec:experiments} presents the empirical framework and results.

\section{Model}
\label{sec:model}

A worker forms a tentative answer to a question, sees an AI assistant's recommendation, and decides whether to revise. The evidence available about a question is a collection of \emph{cues}, atomic units of evidence about the underlying answer that an agent might or might not access. For a question on classical mechanics, the cues are the relevant physical principles, the diagnostic features of the answer choices, and the patterns from prior similar problems. If a human and an AI both lean on the standard textbook derivation, there will be substantial overlap in cues whereas if they each approach the question through different reasoning paths there will be little overlap.

Formally, a binary state $Y \in \{0,1\}$ is drawn from the symmetric prior $P(Y=1) = P(Y=0) = 1/2$.\footnote{We assume symmetric prior in the main text for simplicity. Appendix~\ref{sec:appx_asymmetric_priors} describes an asymmetric prior extension with the main results and insights.} There is a finite collection of $N$ cues $\{X_1, \ldots, X_N\}$, conditionally independent given $Y$:
%There is a finite collection of $N$ i.i.d. primitive cues $\{X_1, \ldots, X_N\}$:
\[
\textstyle
X_k \mid Y = 1 \sim \mathcal{N}(\mu_1, \sigma^2), \qquad X_k \mid Y = 0 \sim \mathcal{N}(\mu_0, \sigma^2), \qquad \mu_1 \neq \mu_0,
\]
where $\mu_1, \mu_0 \in \mathbb{R}$ are the state-conditional means and $\sigma^2 > 0$ is the common conditional variance; the assumption $\mu_1 \neq \mu_0$ ensures each cue is informative about $Y$. We denote the \emph{per-cue capability} by $\theta := (\mu_1 - \mu_0)^2 / (2\sigma^2)$, the signal-to-noise ratio of a single cue. \footnote{Our results generalize to heterogeneous cues; see Appendix~\ref{app:heterogeneous-cues} for details.} 

The human observes the cues in $\mathcal{X}_H \subseteq \{1,\ldots,N\}$ with cardinality $n_H$; the AI observes the cues in $\mathcal{X}_A$ with cardinality $n_A$; their shared cues have cardinality $n_\cap := |\mathcal{X}_H \cap \mathcal{X}_A|$. Each agent forms a log-likelihood ratio summarizing her evidence,
\[
\textstyle
\ell_H := \sum_{k \in \mathcal{X}_H} \LLR(X_k), \qquad \ell_A := \sum_{k \in \mathcal{X}_A} \LLR(X_k),
\]
where $\LLR(X_k) := \log[\phi(X_k; \mu_1, \sigma^2) / \phi(X_k; \mu_0, \sigma^2)]$ is the standard Gaussian log-likelihood ratio, evaluating to $(\mu_1 - \mu_0)(X_k - (\mu_1+\mu_0)/2)/\sigma^2$. Here, $\phi(\cdot; \mu, \sigma^2)$ denotes the Gaussian density with mean $\mu$ and variance $\sigma^2$. Since posterior log-odds equal prior log-odds plus the log-likelihood ratio, and the prior is symmetric (log-prior-odds equal $0$), for either agent $B \in \{H, A\}$, $\ell_B$ is exactly $B$'s posterior log-odds for $Y = 1$ given her cues.

The \emph{overlap coefficient}
\[
\lambda := \frac{n_\cap}{n_A} \in [0, 1]
\]
measures redundancy of the AI's evidence from the human's standpoint: if $\lambda = 0.5$, half of the AI's cues are also possessed by the human, so only half of the AI's evidence is new to the human. Equivalently, $\lambda$ is the population regression coefficient of $\ell_H$ on $\ell_A$ given $Y$.

%The \emph{overlap coefficient} $\lambda$ measures redundancy of the AI's evidence from the human's standpoint: if $\lambda = 0.5$, half of the AI's cues are also possessed by the human, so only half of the AI's evidence is new to the human,
%\[
%\textstyle
%\lambda := \frac{n_\cap}{n_A} \in [0, 1]
%\]
%equal to the population regression coefficient of $\ell_H$ on $\ell_A$ given $Y$.

\begin{lemma}[Joint distribution of log-likelihood ratios]
\label{lem:joint}
Conditional on $Y$,
\[
\textstyle
\begin{pmatrix}\ell_H \\ \ell_A\end{pmatrix} \,\Big|\, Y \sim \mathcal{N}\!\left((2Y-1)\begin{pmatrix}\tau_H \\ \tau_A\end{pmatrix},\ \Sigma\right), \qquad \Sigma = \begin{pmatrix}2\tau_H & 2\tau_A\lambda \\ 2\tau_A\lambda & 2\tau_A\end{pmatrix},
\]
where $\tau_H := \theta n_H$ and $\tau_A := \theta n_A$ are the \emph{agent capabilities}.
\end{lemma}

The \emph{capability} $\tau_B$ measures how informative agent $B$'s evidence is. By Lemma~\ref{lem:joint}, $\ell_B$ has conditional mean of magnitude $\tau_B$ and conditional standard deviation $\sqrt{2\tau_B}$, so its \emph{signal-to-noise ratio} (SNR) given $Y$ is $\tau_B/\sqrt{2\tau_B} = \sqrt{\tau_B/2}$. An agent who classifies via $\mathbf{1}\{\ell_B > 0\}$ is correct iff $\ell_B$ shares the sign of $(2Y-1)$; by symmetry, this happens with probability $\Phi(\sqrt{\tau_B/2})$, where $\Phi$ is the standard normal CDF. Standalone classification accuracy is thus $\Phi$ evaluated at the SNR, and is strictly increasing in $\tau_B$.

%Asymptotic statements such as $\tau_A \to \infty$ at fixed $\lambda$ describe trends across deployments (increasingly capable AIs in domains with given overlap), not within a fixed cue pool. An additional constraint comes from Lemma~\ref{lem:joint} itself: positive semidefiniteness of the conditional covariance matrix requires $\tau_A \le \tau_H/\lambda^2$, which holds independently of the cue-bank story. At fixed $\lambda > 0$, $\tau_A$ is therefore bounded above; asymptotic statements like $\tau_A \to \infty$ at fixed $\lambda$ are read as deployment paths along which $\lambda$ shrinks as $\tau_A$ grows.

%A natural question is whether $\lambda$ is a property of the human, the AI, or the deployment. The following gives a precise answer.
We refer to the cue pool $\{1,\ldots,N\}$ as the \emph{domain} (a property of the question type), and to $\mathcal{X}_H$ as the human's \emph{access} within that domain. The following shows that, under uniform sampling, overlap depends on the domain and the human's access but not on the AI's capability.

\begin{proposition}[Overlap as a domain property]
\label{prop:overlap-domain}
Fix $\mathcal{X}_H$ with $|\mathcal{X}_H| = n_H$. Suppose the AI's cue access $\mathcal{X}_A$ is a uniformly random subset of size $n_A$ drawn without replacement from $\{1,\ldots,N\}$. Then $\mathbb{E}[\lambda] = n_H/N$, regardless of $n_A$.
\end{proposition}

%\begin{proposition}[Overlap as a domain property]
%\label{prop:overlap-domain}
%Fix $\mathcal{X}_H$ with $|\mathcal{X}_H| = n_H$. Suppose the AI's cue access $\mathcal{X}_A^{(n)}$ is a uniformly random subset of size $n$ drawn without replacement from $\{1,\ldots,N\}$, so that the AI accesses $n$ distinct cues. Then $\mathbb{E}[\lambda^{(n)}] = n_H/N$ for every $n$, and $\lambda^{(n)} \to n_H/N$ almost surely as $n \to \infty$.
%\end{proposition}

If an AI scales by uniformly sampling more cues from a fixed pool, its overlap with a given human equals the human's coverage fraction $n_H/N$ in expectation (the share of all cues the human accesses), regardless of the AI's capability. Overlap is therefore not tunable through AI scale alone. The invariance is a statement in expectation under uniform random sampling of $\mathcal{X}_A$, not a claim that all AI systems in a domain share the same realized $\lambda$: two AI systems whose cue access is drawn from different sampling distributions (different training corpora, training objectives, or architectures) can yield different realized $\lambda$ even within the same domain.
%two AI systems whose cue access is drawn from different sampling distributions (different training data corpus, non-uniform coverage of the cue pool) can yield different realized $\lambda$ even paired with the same human on the same task. 
%This cross-AI variation in cue access also enables identification: pairing a single worker population with two AI systems generates $\lambda^d \neq \lambda^{d'}$, the variation needed to jointly identify $\hat\alpha$ and $\hat\beta$ (Appendix~\ref{app:identification}).

In the analysis that follows we consider $(\tau_H, \tau_A, \lambda)$ as separate deployment parameters, because different humans paired with different AIs in different domains generate different parameter combinations. 
Two feasibility constraints on $(\tau_H, \tau_A, \lambda)$ deserve note. The model-agnostic constraint, from positive semidefiniteness of the covariance matrix in Lemma~\ref{lem:joint}, is $\tau_A \le \tau_H/\lambda^2$. The cue-pool generative model imposes the tighter bound $\tau_A \le \tau_H/\lambda$ (since $n_\cap \le n_H$); at this boundary, the AI's evidence is Blackwell-equivalent to the human-AI team's. All quantitative results in our analysis live within the cue-pool feasible region (which is contained in the broader PSD region). 
% and the phase thresholds $\tau_{\mathrm{aug}}, \tau_{\mathrm{auto}}$ are reached at finite $\tau_A$ values inside it for any $\lambda \in (0, 1)$.
%If an AI scales by uniformly sampling more cues from a fixed pool, its overlap with a given human stabilizes at the human's coverage fraction $n_H/N$ (the share of all cues the human accesses), regardless of the AI's capability. Overlap is therefore a property of the human-domain pair, not tunable through AI scale alone. A consequence is that when two AIs differ in their overlap with the same human, the difference reflects \emph{which} cues each AI accesses, not the number of cues each processes.

\vspace{-1em}
\subsection{The Worker's Decision Rule}
\label{sec:beliefs}

After processing its cues, the AI communicates a sufficient statistic for $\ell_A$ to the human, such as a recommended answer, a probability report, or a textual explanation. We assume this communication is informationally equivalent to $\ell_A$ (the below analysis is agnostic to the format of the information). We refer to the human decision-maker who has communicated with AI and has  access to $(\ell_H, \ell_A)$ as the \emph{worker}. The worker forms a posterior probability for $Y=1$ and provides a binary decision. We model the worker via two parameters that capture distinct primitives: one a belief about the joint distribution of signals, the other a behavioral discount on the AI's contribution.
\begin{itemize}
\item \textbf{Correlation awareness $\hat\alpha \in [0, 1]$ (a belief).} The worker's perceived overlap is $\hat\alpha \cdot \lambda$. At $\hat\alpha = 1$ she perceives overlap correctly; at $\hat\alpha = 0$ she perceives the signals as conditionally independent (correlation neglect); intermediate values represent partial awareness.
\item \textbf{AI trust $\hat\beta > 0$ (a behavioral discount).} The worker applies a multiplicative weight $\hat\beta$ to the AI signal $\ell_A$ in her decision statistic. At $\hat\beta = 1$ she applies no behavioral discount; $\hat\beta < 1$ corresponds to AI undertrust; $\hat\beta > 1$ corresponds to overtrust.
\end{itemize}
These parameters are jointly identified from cross-AI variation in $\lambda$ (Appendix~\ref{app:identification}). 

\begin{assumption}[Worker's behavior]
\label{ass:worker}
The worker has parameters $(\hat\alpha, \hat\beta) \in [0, 1] \times \mathbb{R}_{>0}$ and behaves as follows:
\begin{enumerate}
\item[(i)] \emph{Bayesian updating under perceived overlap.} The worker performs Bayesian updating on $(\ell_H, \ell_A, Y)$ under a belief that perceives overlap as $\hat\alpha\lambda$ and capabilities $(\tau_H, \tau_A)$ at their true values. This produces a Bayesian posterior log-odds.
\item[(ii)] \emph{Behavioral trust discount.} After Bayesian updating, the worker scales the AI signal $l_A$ by a multiplicative factor $\hat\beta$ in the posterior log-odds. This discount is applied to the AI signal \emph{after} Bayesian updating, with no underlying belief interpretation.
\end{enumerate}
\end{assumption}

%The worker's behavior under these conditions is given by the following rule.
Under this assumption and the Gaussian signal structure, the worker's decision rule satisfies the following linear aggregation of the agent log-likelihood ratios. \footnote{Our linear behavioral rule, characterized by two behavioral parameters, is a substantial simplification adopted for conceptual clarity. Human and LLM reasoning in real-world settings may be considerably more complex, and we leave modeling this complexity to future work.}
\begin{lemma}[Worker's decision weights]
\label{lem:behavioral}
Under Assumption~\ref{ass:worker}, a worker with parameters $(\hat\alpha, \hat\beta) \in [0, 1] \times \mathbb{R}_{>0}$ uses the linear rule $\hat\ell = w_H \ell_H + w_A \ell_A$ with weights
\[
\textstyle
w_H(\hat\alpha) = \frac{\tau_H - \tau_A \hat\alpha \lambda}{\tau_H - \tau_A \hat\alpha^2 \lambda^2}, \qquad w_A(\hat\alpha, \hat\beta) = \hat\beta \cdot \frac{\tau_H(1 - \hat\alpha \lambda)}{\tau_H - \tau_A \hat\alpha^2 \lambda^2}.
\]
The human evidence weight depends only on $\hat\alpha$. The AI evidence weight equals $\hat\beta$ times the Bayesian weight $w_A^{\mathrm{Bayes}}(\hat\alpha) := \tau_H(1 - \hat\alpha \lambda)/(\tau_H - \tau_A \hat\alpha^2 \lambda^2)$.
\end{lemma}
The worker treats $\hat\ell$ as her log-odds for $Y = 1$. Under the symmetric prior, this gives the binary decision $\hat Y := \mathbf{1}\{\hat\ell > 0\}$ (for 0-1 loss) and the probability report $\hat p := \expit(\hat\ell)$ (for log loss).
\iffalse
\begin{remark}[Why $\hat\beta$ is a behavioral discount, not a belief]
\label{rem:bayesian-alternative}
A natural alternative is to assume the agent has a perceived AI capability $\hat \beta \tau_A$  and Bayesian updates the joint distribution using this AI capability (similar to how correlation awareness is modeled in Assumption \ref{ass:worker}). This would yield a linear decision rule that combines $l_H$ and $l_A$ using the weights:

\[
w_H^{\mathrm{alt}} = \frac{\tau_H - \hat\beta\hat\alpha\tau_A\lambda}{\tau_H - \hat\beta\hat\alpha^2\tau_A\lambda^2}, \qquad w_A^{\mathrm{alt}} = \frac{\tau_H(1 - \hat\alpha\lambda)}{\tau_H - \hat\beta\hat\alpha^2\tau_A\lambda^2}.
\]

At $\hat\alpha = 0$ both weights collapse to $1$ regardless of $\hat\beta$, so the decision statistic reduces to $\ell_H + \ell_A$ and trust is unidentifiable from the worker's weights. Intuitively, a worker who thinks the AI is more capable expects larger AI signals but also more variance; the two scale together and the Bayesian weight on $\ell_A$ is unchanged. At interior $\hat\alpha$, $\hat\beta$ enters nonlinearly through the denominator, and the simple ``AI weight equals trust'' reading is lost.

The behavioral-discount framing of Lemma~\ref{lem:behavioral} sidesteps both: $\hat\beta$ multiplies $\ell_A$ directly, identification holds at every $\hat\alpha$, and trust shapes welfare. This seems analogous to the algorithm-aversion literature, which documents context-dependent under- and over-weighting as behavioral rather than belief-based [TO BE CHECKED or REMOVED].
\end{remark}
\fi
\begin{remark}[Why $\hat\beta$ is a behavioral discount, not a belief]
A natural alternative interprets trust discount as a subjective belief about AI capability, replacing $\tau_A$ by $\hat\beta\tau_A$ in the worker's perceived signal distribution. Because any such subjective-capability model leaves the Bayesian weight on $\ell_A$ unchanged, rendering trust unidentifiable, we model $\hat\beta$ as a behavioral discount applied after Bayesian updating; see Appendix~\ref{app:behavioral}.
\end{remark}

\subsection{Two Benchmarks}
\label{sec:benchmarks}
The two-parameter family $(\hat\alpha, \hat\beta) \in [0, 1] \times \mathbb{R}_{>0}$ contains a continuum of worker behaviors. We consider two benchmarks at the extremes of correlation awareness.

\textbf{Bayes-rational (BR).} $\hat\alpha = 1$, $\hat\beta = 1$. The worker perceives overlap correctly and applies no trust discount. Substituting into Lemma~\ref{lem:behavioral} gives weights $w_H^{\mathrm{BR}} = \frac{\tau_H - \tau_A \lambda}{\tau_H - \tau_A \lambda^2}$ and $w_A^{\mathrm{BR}} = \frac{\tau_H(1 - \lambda)}{\tau_H - \tau_A \lambda^2}$.
In the symmetric case $\tau_H = \tau_A$, both weights equal $1/(1+\lambda)$, with the Bayesian shrinkage correcting for redundant evidence on both sides.

\textbf{Correlation neglect (CN).} $\hat\alpha = 0$, $\hat\beta$ arbitrary. 
The worker treats $(\ell_H, \ell_A)$ as conditionally independent given $Y$, yielding $\hat\ell = \ell_H + \hat\beta\,\ell_A$. The case $\hat\beta = 1$ is the classical CN agent; $\hat\beta < 1$ also discounts the AI signal.

The benchmarks differ in how the AI weight responds to true overlap: $w_A^{\mathrm{BR}}$ depends on $\lambda$, whereas $w_A^{\mathrm{CN}} = \hat\beta$ is invariant to $\lambda$ --- the CN worker treats AI evidence the same regardless of true redundancy. The interior of $(\hat\alpha, \hat\beta)$ space combines partial correlation awareness and trust miscalibration.
%The worker perceives the two log-likelihood ratios $(\ell_H, \ell_A)$ as conditionally independent given $Y$, and uses the rule
%\[
%\hat\ell = \ell_H + \hat\beta \ell_A.
%\]
%The CN family is one-dimensional, parameterized by the trust factor $\hat\beta$. The case $\hat\beta = 1$ is the classical correlation-neglect agent; $\hat\beta < 1$ describes a worker who simultaneously neglects correlation and discounts the AI signal.

%%\paragraph{First-order distinguishing property.} The two benchmarks differ in how the AI weight responds to true overlap $\lambda$. Under BR,
%\[
%\frac{\partial w_A^{\mathrm{BR}}}{\partial \lambda} = -\frac{\tau_H\bigl[\tau_H - \tau_A\lambda(2 - \lambda)\bigr]}{(\tau_H - \tau_A\lambda^2)^2},
%\]
%which is generically nonzero, with sign determined jointly by $(\tau_H, \tau_A, \lambda)$. Under CN, $w_A^{\mathrm{CN}} = \hat\beta$, so $\partial w_A^{\mathrm{CN}}/\partial\lambda = 0$. The CN worker treats the AI's evidence the same regardless of true redundancy. We develop the implications of this structural difference later.

%The interior of the parameter space admits combinations of partial correlation awareness and trust miscalibration. 

%In next Section, we develop the accuracy formula and identification result for general $(\hat\alpha, \hat\beta)$, while the phase diagram and welfare decomposition are derived under correlation neglect ($\hat\alpha = 0$). 
\vspace{-0.5em}
\section{Analysis}
\label{sec:analysis}

This section develops three results that guide the empirical study presented later. The first is a closed-form expression for the worker's classification accuracy that, under correlation neglect ($\hat \alpha=0$), partitions the deployment space $(\tau_A, \lambda)$ into three regimes: \emph{complementarity} (human-AI team is most accurate), \emph{impairment} (human alone is most accurate), and \emph{automation} (AI alone is most accurate). 
%The second is an identification result: variation across two AI systems with different overlap coefficients on a common worker population pins down $\hat\alpha$ and $\hat\beta$ separately. 
The second is a welfare decomposition: the worker's loss against the Bayes-rational benchmark splits additively into a correlation-neglect cost and an undertrust cost, which could then be calculated by the identified behavioral parameters $\hat\alpha$ and $\hat\beta$. Our results depend only on the joint-Gaussian structure of Lemma~\ref{lem:joint}; the cue-pool model is one interpretable construction yielding it, not a requirement.

\subsection{Performance Objective}
\label{sec:accuracy}
The worker's decision rule under Assumption~\ref{ass:worker} can be evaluated using different performance objectives. We focus on binary-decision accuracy in the main text, which is the natural welfare statistic when the worker's output is a binary label. Formally, the \emph{accuracy} is the probability that the worker's binary decision is correct: $\mathcal{A} := P(\hat Y = Y)$, where $\hat Y = \mathbf{1}\{\hat\ell > 0\}$ is her hardmax decision. Other objectives like log loss and switching-rate are derived and discussed in Appendix~\ref{sec:more_objective} and are useful for probability reports and revision data.

%For the thereotical analysis, we work primarily with the error rate, the welfare-relevant objective for the binary decision setting. We defer the closed-form derivations and the discussions of the other two in the appendix, since they are relevant mainly for our empirical analysis.

\textbf{Accuracy via the signal-to-noise (SNR) ratio.} Under Assumption~\ref{ass:worker}, the worker's decision statistic $\hat\ell = w_H \ell_H + w_A \ell_A$ is a linear combination of the jointly Gaussian pair $(\ell_H, \ell_A)$ from Lemma~\ref{lem:joint}; hence $\hat\ell \mid Y$ is itself Gaussian. Accuracy under the rule $\mathbf{1}\{\hat\ell > 0\}$ is a function of the SNR ratio of $\hat\ell$, namely the ratio of the conditional mean's magnitude to the conditional standard deviation. 
%Computing this ratio yields the following.

\begin{lemma}[Accuracy of the worker's decision rule]
\label{lem:accuracy}
Under Assumption~\ref{ass:worker} and the symmetric prior, the worker classifying via $\hat Y = \mathbf{1}\{\hat\ell > 0\}$ attains accuracy $\mathcal{A}_{HA} = \Phi(r_{HA})$,
where $r_{HA}$ is the signal-to-noise ratio of $\hat\ell$:
\[
\textstyle
r_{HA}(\hat\alpha, \hat\beta; \tau_H, \tau_A, \lambda) = \frac{w_H \tau_H + w_A \tau_A}{\sqrt{2\bigl(w_H^2 \tau_H + w_A^2 \tau_A + 2 w_H w_A \tau_A \lambda\bigr)}},
\]
and the weights $(w_H, w_A)$ are given by Lemma~\ref{lem:behavioral}.
\end{lemma}

The signal-to-noise ratio $r_{HA}$ has a clear reading. Conditional on $Y$, the worker's decision statistic $\hat\ell$ has mean $(2Y-1)(w_H\tau_H + w_A\tau_A)$ and variance $2(w_H^2\tau_H + w_A^2\tau_A + 2w_Hw_A\tau_A\lambda)$. The variance decomposes into three terms: the human's contribution $w_H^2 \tau_H$, the AI's contribution $w_A^2 \tau_A$, and a cross-term $2 w_H w_A \tau_A \lambda$ capturing the double-counting penalty from overlapped evidence.
%from the conditional covariance between $\ell_H$ and $\ell_A$. Under correlation neglect ($\hat\alpha = 0$), the weights collapse to $w_H = 1$ and $w_A = \hat\beta$; in this case, as overlap $\lambda$ rises, the cross-term rises, the variance rises, and $r_{HA}$ falls. Therefore, the accuracy of a correlation-neglect worker is always strictly decreasing in $\lambda$.

\textbf{Bayes-rational rule is welfare-optimal.} By Lemma~\ref{lem:joint}, $(\ell_H, \ell_A) \mid Y$ is Gaussian with shared covariance across $Y$; the linear direction maximizing $r_{HA}$ from Lemma~\ref{lem:accuracy} therefore coincides with the Bayes-rational rule from \S\ref{sec:benchmarks}, giving maximum SNR $r^* := r_{HA}(\hat\alpha = 1, \hat\beta = 1)$ and optimal accuracy $\mathcal{A}^{\mathrm{BR}} = \Phi(r^*)$.
%\textbf{Bayes-rational rule is welfare-optimal.} By the classical Gaussian-classification result, the linear weights maximizing $r_{HA}$ in Lemma~\ref{lem:accuracy} coincide (up to scaling) with the Bayes-rational rule from \S\ref{sec:benchmarks}, giving maximum SNR $r^* := r_{HA}(\hat\alpha = 1, \hat\beta = 1)$ and optimal accuracy $\mathcal{A}^{\mathrm{BR}} = \Phi(r^*)$. 
%For any $(\hat\alpha, \hat\beta)$, $r^* - r_{HA}(\hat\alpha, \hat\beta) \ge 0$, with equality iff $(w_H, w_A) \propto (w_H^{\mathrm{BR}}, w_A^{\mathrm{BR}})$ — a one-dimensional set including $(1,1)$, since SNR depends only on the ratio $w_A/w_H$.
%At $\lambda = 0$, $r^{*2} = r_H^2 + r_A^2$: independent signals contribute additively. At $\lambda = 1$ with $\tau_A \neq \tau_H$, $r^* = r_H$: the AI is fully redundant.
The accuracy gap $\mathcal{A}^{\mathrm{BR}} - \mathcal{A}_{HA}(\hat\alpha, \hat\beta) \ge 0$ is the worker's loss relative to the BR benchmark.

%When the worker reports probabilities rather than a binary decision (e.g., medical diagnosis with uncertainty estimates), log loss is the appropriate welfare objective: it captures the calibration of probability reports, not just point-estimate correctness; we defer the closed form and its discussion to Appendix~\ref{app:log-loss}.
%The \emph{switching rate} $\mathcal{S} = P(\hat Y_{HA} \ne \hat Y_H)$ measures how often AI assistance changes the worker's binary decision and serves as a behavioral model-fit check whenever data record whether tentative answers were revised (Appendix~\ref{app:switching}).
%The \emph{exploitable-to-dangerous ratio} -- exploitable cases (human wrong, AI right) over dangerous cases (human right, AI wrong) -- exceeds $1$ iff $\tau_A > \tau_H$, giving a quick deployment-level handle on whether AI assistance has the capability to help (Appendix~\ref{app:exploitable}).

\subsection{Phase Diagram}
\label{sec:phase}

Comparing the three accuracies $\mathcal{A}_H = \Phi(\sqrt{\tau_H/2})$, $\mathcal{A}_{HA}$ from Lemma~\ref{lem:accuracy}, and $\mathcal{A}_{AI} = \Phi(\sqrt{\tau_A/2})$ partitions the deployment space $(\tau_A, \lambda)$, for each fixed $\hat\beta > 0$, into three regions: \emph{complementarity} (HA optimal, i.e., the human-AI team is most accurate), \emph{impairment} (H optimal), and \emph{automation} (AI optimal). For clarity, we state the closed-form phase diagram at strict correlation neglect ($\hat\alpha = 0$); the same qualitative forces extend continuously to partial correlation awareness, with larger $\hat\alpha$ attenuating double-counting.

%Comparing the three accuracies $\mathcal{A}_H = \Phi(\sqrt{\tau_H/2})$, $\mathcal{A}_{HA}$ from Lemma~\ref{lem:accuracy}, and $\mathcal{A}_{AI} = \Phi(\sqrt{\tau_A/2})$ partitions the deployment space $(\tau_A, \lambda)$, for each fixed $\hat\beta > 0$, into three regions: \emph{complementarity} (HA optimal), \emph{impairment} (H optimal), and \emph{automation} (AI optimal). Under correlation neglect ($\hat\alpha = 0$) with the symmetric prior, the partitioning admits closed-form thresholds, given in the following theorem.

\begin{theorem}[Phase diagram under correlation neglect]
\label{thm:phase}
Fix $\hat\alpha = 0$, $\hat\beta > 0$, and the symmetric prior.\footnote{An extension of this theorem to asymmetric priors, under sufficient conditions, is shown in Appendix~\ref{sec:appx_asymmetric_priors}.} Define
\[
\textstyle
\tau_{\mathrm{aug}}(\lambda; \hat\beta) := \frac{\hat\beta + 2\lambda - 2}{\hat\beta}\,\tau_H, \qquad \tau_{\mathrm{auto}}(\lambda; \hat\beta) := \frac{2\hat\beta - 1 + \sqrt{(2\hat\beta-1)^2 + 8\hat\beta\lambda}}{4\hat\beta\lambda}\,\tau_H \ \ \text{for } \lambda > 0,
\]
with $\tau_{\mathrm{auto}}(0; \hat\beta) = +\infty$ if $\hat\beta \ge 1/2$ and $\tau_H/(1-2\hat\beta)$ otherwise. Then for $\tau_A > 0$:
\begin{itemize}
\item \emph{Low overlap} ($\lambda \le 1 - \hat\beta/2$, when $\hat\beta < 2$): complementarity if $\tau_A < \tau_{\mathrm{auto}}(\lambda; \hat\beta)$; automation if $\tau_A \ge \tau_{\mathrm{auto}}(\lambda; \hat\beta)$.
\item \emph{Moderate overlap} ($1 - \hat\beta/2 < \lambda < 1$): impairment if $\tau_A < \tau_{\mathrm{aug}}(\lambda; \hat\beta)$; complementarity if $\tau_{\mathrm{aug}}(\lambda; \hat\beta) \le \tau_A \le \tau_{\mathrm{auto}}(\lambda; \hat\beta)$; automation if $\tau_A > \tau_{\mathrm{auto}}(\lambda; \hat\beta)$.
\item \emph{High overlap} ($\lambda = 1$): the two thresholds coincide, $\tau_{\mathrm{aug}}(1; \hat\beta) = \tau_{\mathrm{auto}}(1; \hat\beta) = \tau_H$, so the complementarity region collapses to the single point $\tau_A = \tau_H$. 
%At that point, $\mathcal{A}_H = \mathcal{A}_{HA} = \mathcal{A}_{AI}$ (a triple tie). For $\tau_A < \tau_H$, $\mathcal{A}_H > \mathcal{A}_{HA}$ and $\mathcal{A}_H > \mathcal{A}_{AI}$, so $H$ alone is best; for $\tau_A > \tau_H$, $\mathcal{A}_{AI} > \mathcal{A}_{HA}$ and $\mathcal{A}_{AI} > \mathcal{A}_H$, so $AI$ alone is best.
\end{itemize}
Here, $\tau_{\mathrm{aug}}$ is strictly increasing in $\lambda$ on $(1 - \hat\beta/2, 1]$ and $\tau_{\mathrm{auto}}$ is strictly decreasing in $\lambda$ on $(0, 1]$.
\end{theorem}

\begin{figure}[h]
\centering

\begin{minipage}{\textwidth}
\centering

% --- Panel 1: Undertrust (hatbeta = 0.5) ---
\begin{subfigure}{0.31\textwidth}
\centering
\begin{tikzpicture}[scale=0.65]

% Parameters: tauH = 1, hatbeta = 0.5
% Augmentation positive for lambda > 0.75 (y = 3)
% Automation: tau_auto(lambda) = 1/sqrt(lambda); at lambda = 0.04, tau_auto = 5
% Reference: tau_auto(0.5) = 1.41, tau_aug(0.9) = 0.60
% Cue-pool feasibility: tau_A <= 1/lambda, i.e., lambda <= 1/tau_A; curve y = 4/x

\fill[red!15] (0, 0) rectangle (5, 4);

\fill[blue!15]
  plot[domain=0.04:1.0, samples=50, smooth, variable=\l] ({1/sqrt(\l)}, {\l*4})
  -- (1, 4) -- (5, 4) -- (5, 0.16) -- cycle;

\fill[gray!25]
  (0, 3) -- (0, 4) -- (1, 4)
  -- plot[domain=1.0:0.75, samples=20, smooth, variable=\l] ({4*\l - 3}, {\l*4})
  -- cycle;

% Mask out the cue-pool-infeasible region
\fill[white, opacity=0.6]
  plot[domain=1.0:5.0, samples=30, smooth, variable=\x] ({\x}, {4/\x})
  -- (5, 4) -- (1, 4) -- cycle;

\draw[thick, red!80!black, domain=0.75:1.0, samples=20, smooth, variable=\l]
  plot ({4*\l - 3}, {\l*4});
\draw[thick, blue!80!black, domain=0.04:1.0, samples=50, smooth, variable=\l]
  plot ({1/sqrt(\l)}, {\l*4});

% Cue-pool feasibility curve: tau_A = 1/lambda
\draw[dashed, thick, black]
  plot[domain=1.0:5.0, samples=30, smooth, variable=\x] ({\x}, {4/\x});

% Reference line at lambda = 0.5 (y = 2), showing tau_auto(0.5) = 1.41
\draw[dotted, blue!80!black, thick] (0, 2) -- (1.41, 2);
\draw[dotted, blue!80!black, thick] (1.41, 2) -- (1.41, 0);
\filldraw[blue!80!black] (1.41, 2) circle (1.2pt);
\node[font=\tiny, blue!80!black, anchor=west] at (1.5, 2) {$1.41$};

% Reference line at lambda = 0.9 (y = 3.6), showing tau_aug(0.9) = 0.60
\draw[dotted, red!80!black, thick] (0, 3.6) -- (0.6, 3.6);
\draw[dotted, red!80!black, thick] (0.6, 3.6) -- (0.6, 0);
\filldraw[red!80!black] (0.6, 3.6) circle (1.2pt);
\node[font=\tiny, red!80!black, anchor=west] at (0.7, 3.6) {$0.60$};

\draw[->, thick] (0, 0) -- (5.3, 0) node[right, font=\small] {$\tau_A$};
\draw[->, thick] (0, 0) -- (0, 4.4) node[above, font=\small] {$\lambda$};
\draw[font=\footnotesize] (0, 0) node[below left] {$0$};
\draw (-0.08, 4) -- (0.08, 4);
\draw[font=\footnotesize] (0, 4) node[left] {$1$};
\draw (-0.08, 3.6) -- (0.08, 3.6);
\draw[font=\footnotesize] (0, 3.6) node[left] {$0.9$};
\draw (-0.08, 3) -- (0.08, 3);
\draw[font=\footnotesize] (0, 3) node[left] {$0.75$};
\draw (-0.08, 2) -- (0.08, 2);
\draw[font=\footnotesize] (0, 2) node[left] {$0.5$};
\draw[dashed, gray!60] (1, 0) -- (1, 4);
\draw[font=\footnotesize] (1, 0) node[below] {$\tau_H$};

\node[font=\scriptsize] at (0.7, 1.3) {Comp.};
\node[font=\scriptsize] at (0.5, 3.95) {Imp.};
\node[font=\scriptsize] at (3.7, 1.0) {Auto.};

\filldraw[black] (1, 4) circle (1.5pt);

\end{tikzpicture}
\caption{Undertrust ($\hat\beta = 0.5$)}
\end{subfigure}
\hfill
% --- Panel 2: Calibrated (hatbeta = 1) ---
\begin{subfigure}{0.31\textwidth}
\centering
\begin{tikzpicture}[scale=0.65]

% Parameters: tauH = 1, hatbeta = 1
% Reference: tau_auto(0.5) = 1.62, tau_aug(0.9) = 0.80

\fill[red!15] (0, 0) rectangle (5, 4);

\fill[blue!15]
  plot[domain=0.13:1.0, samples=50, smooth, variable=\l] ({(1 + sqrt(1 + 8*\l))/(4*\l)}, {\l*4})
  -- (1, 4) -- (5, 4) -- (5, 0.52) -- cycle;

\fill[gray!25]
  (0, 2) -- (0, 4) -- (1, 4)
  -- plot[domain=1.0:0.5, samples=20, smooth, variable=\l] ({2*\l - 1}, {\l*4})
  -- cycle;

% Mask out cue-pool-infeasible region
\fill[white, opacity=0.6]
  plot[domain=1.0:5.0, samples=30, smooth, variable=\x] ({\x}, {4/\x})
  -- (5, 4) -- (1, 4) -- cycle;

\draw[thick, red!80!black, domain=0.5:1.0, samples=20, smooth, variable=\l]
  plot ({2*\l - 1}, {\l*4});
\draw[thick, blue!80!black, domain=0.13:1.0, samples=50, smooth, variable=\l]
  plot ({(1 + sqrt(1 + 8*\l))/(4*\l)}, {\l*4});

% Cue-pool feasibility curve
\draw[dashed, thick, black]
  plot[domain=1.0:5.0, samples=30, smooth, variable=\x] ({\x}, {4/\x});

% Reference line at lambda = 0.5, tau_auto = 1.62
\draw[dotted, blue!80!black, thick] (0, 2) -- (1.62, 2);
\draw[dotted, blue!80!black, thick] (1.62, 2) -- (1.62, 0);
\filldraw[blue!80!black] (1.62, 2) circle (1.2pt);
\node[font=\tiny, blue!80!black, anchor=west] at (1.7, 2) {$1.62$};

% Reference line at lambda = 0.9, tau_aug = 0.80
\draw[dotted, red!80!black, thick] (0, 3.6) -- (0.8, 3.6);
\draw[dotted, red!80!black, thick] (0.8, 3.6) -- (0.8, 0);
\filldraw[red!80!black] (0.8, 3.6) circle (1.2pt);
\node[font=\tiny, red!80!black, anchor=west] at (0.9, 3.6) {$0.80$};

\draw[->, thick] (0, 0) -- (5.3, 0) node[right, font=\small] {$\tau_A$};
\draw[->, thick] (0, 0) -- (0, 4.4) node[above, font=\small] {$\lambda$};
\draw[font=\footnotesize] (0, 0) node[below left] {$0$};
\draw (-0.08, 4) -- (0.08, 4);
\draw[font=\footnotesize] (0, 4) node[left] {$1$};
\draw (-0.08, 3.6) -- (0.08, 3.6);
\draw[font=\footnotesize] (0, 3.6) node[left] {$0.9$};
\draw (-0.08, 2) -- (0.08, 2);
\draw[font=\footnotesize] (0, 2) node[left] {$0.5$};
\draw[dashed, gray!60] (1, 0) -- (1, 4);
\draw[font=\footnotesize] (1, 0) node[below] {$\tau_H$};

\node[font=\scriptsize] at (0.7, 1) {Comp.};
\node[font=\scriptsize] at (0.5, 3.95) {Imp.};
\node[font=\scriptsize] at (3.7, 1.0) {Auto.};

\filldraw[black] (1, 4) circle (1.5pt);

\end{tikzpicture}
\caption{Calibrated ($\hat\beta = 1$)}
\end{subfigure}
\hfill
% --- Panel 3: Overtrust (hatbeta = 1.5) ---
\begin{subfigure}{0.31\textwidth}
\centering
\begin{tikzpicture}[scale=0.65]

% Parameters: tauH = 1, hatbeta = 1.5
% Reference: tau_auto(0.5) = 1.72, tau_aug(0.9) = 0.87

\fill[red!15] (0, 0) rectangle (5, 4);

\fill[blue!15]
  plot[domain=0.18:1.0, samples=50, smooth, variable=\l] ({(2 + sqrt(4 + 12*\l))/(6*\l)}, {\l*4})
  -- (1, 4) -- (5, 4) -- (5, 0.72) -- cycle;

\fill[gray!25]
  (0, 1) -- (0, 4) -- (1, 4)
  -- plot[domain=1.0:0.25, samples=30, smooth, variable=\l] ({(2*\l - 0.5)/1.5}, {\l*4})
  -- cycle;

% Mask out cue-pool-infeasible region
\fill[white, opacity=0.6]
  plot[domain=1.0:5.0, samples=30, smooth, variable=\x] ({\x}, {4/\x})
  -- (5, 4) -- (1, 4) -- cycle;

\draw[thick, red!80!black, domain=0.25:1.0, samples=30, smooth, variable=\l]
  plot ({(2*\l - 0.5)/1.5}, {\l*4});
\draw[thick, blue!80!black, domain=0.18:1.0, samples=50, smooth, variable=\l]
  plot ({(2 + sqrt(4 + 12*\l))/(6*\l)}, {\l*4});

% Cue-pool feasibility curve
\draw[dashed, thick, black]
  plot[domain=1.0:5.0, samples=30, smooth, variable=\x] ({\x}, {4/\x});

% Reference line at lambda = 0.5, tau_auto = 1.72
\draw[dotted, blue!80!black, thick] (0, 2) -- (1.72, 2);
\draw[dotted, blue!80!black, thick] (1.72, 2) -- (1.72, 0);
\filldraw[blue!80!black] (1.72, 2) circle (1.2pt);
\node[font=\tiny, blue!80!black, anchor=west] at (1.8, 2) {$1.72$};

% Reference line at lambda = 0.9, tau_aug = 0.87
\draw[dotted, red!80!black, thick] (0, 3.6) -- (0.87, 3.6);
\draw[dotted, red!80!black, thick] (0.87, 3.6) -- (0.87, 0);
\filldraw[red!80!black] (0.87, 3.6) circle (1.2pt);
\node[font=\tiny, red!80!black, anchor=west] at (0.97, 3.6) {$0.87$};

\draw[->, thick] (0, 0) -- (5.3, 0) node[right, font=\small] {$\tau_A$};
\draw[->, thick] (0, 0) -- (0, 4.4) node[above, font=\small] {$\lambda$};
\draw[font=\footnotesize] (0, 0) node[below left] {$0$};
\draw (-0.08, 4) -- (0.08, 4);
\draw[font=\footnotesize] (0, 4) node[left] {$1$};
\draw (-0.08, 3.6) -- (0.08, 3.6);
\draw[font=\footnotesize] (0, 3.6) node[left] {$0.9$};
\draw (-0.08, 1) -- (0.08, 1);
\draw[font=\footnotesize] (0, 1) node[left] {$0.25$};
\draw (-0.08, 2) -- (0.08, 2);
\draw[font=\footnotesize] (0, 2) node[left] {$0.5$};
\draw[dashed, gray!60] (1, 0) -- (1, 4);
\draw[font=\footnotesize] (1, 0) node[below] {$\tau_H$};

\node[font=\scriptsize] at (0.5, 0.5) {Comp.};
\node[font=\scriptsize] at (0.5, 3.95) {Imp.};
\node[font=\scriptsize] at (3.7, 1.0) {Auto.};

\filldraw[black] (1, 4) circle (1.5pt);

\end{tikzpicture}
\caption{Overtrust ($\hat\beta = 1.5$)}
\end{subfigure}

\end{minipage}

\caption{Phase diagrams under correlation neglect ($\hat\alpha = 0$) for three trust regimes. The deployment space $(\tau_A, \lambda)$ is partitioned into \emph{complementarity} (HA optimal, red), \emph{impairment} (H optimal, gray), and \emph{automation} (AI optimal, blue). The dashed black curve marks the cue-pool feasibility boundary $\tau_A = \tau_H/\lambda$. %beyond which the cue-pool model is infeasible (the looser PSD constraint $\tau_A \le \tau_H/\lambda^2$ is automatically satisfied within this region). 
The augmentation threshold becomes positive at $\lambda > 1 - \hat\beta/2$ (vertical-axis tick): $0.75, 0.50, 0.25$ across panels (a)--(c), so the impairment region becomes active at lower overlap as trust rises. Dotted reference lines mark threshold values at fixed $\lambda$: at $\lambda = 0.9$ (red), the augmentation threshold $\tau_{\mathrm{aug}}$ shifts from $0.60$ to $0.80$ to $0.87$, and at $\lambda = 0.5$ (blue), the automation threshold $\tau_{\mathrm{auto}}$ shifts from $1.41$ to $1.62$ to $1.72$.}
%All thresholds lie strictly within the cue-pool feasible region. The triple-tie point $(\tau_A, \lambda) = (\tau_H, 1)$ is fixed across panels.}
\label{fig:phase-diagrams}
\end{figure}
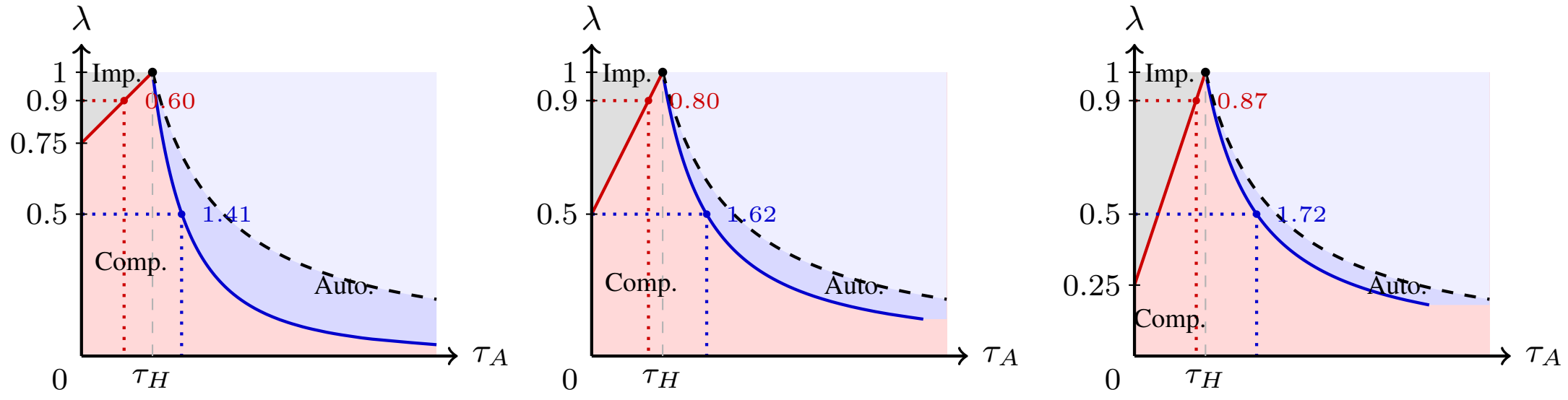

The theorem yields several substantive insights that we elaborate below. 
%Each can be understood as the outcome of how two effects play against each other as the overlap $\lambda$ and the worker's trust $\hat\beta$ vary. As $\lambda$ rises, the AI's evidence becomes more redundant with the human's (reducing the AI's marginal contribution to the team's accuracy), but the correlation-neglect worker's mishandling of overlap also becomes more severe (raising the welfare cost of using the AI). The trust parameter $\hat\beta$ scales the magnitude of both effects: a worker who underweights the AI signal benefits less from its marginal information but also incurs a smaller correlation-neglect penalty. 
%The three regimes (impairment, complementarity, automation) reflect different balances between these effects, and the closed-form thresholds in Theorem~\ref{thm:phase} make the balance explicit.

\textbf{Insight 1: (Impairment Region) -- A weak AI can hurt a worker who neglects correlation.} The phase diagram has an \emph{impairment} region where the worker assisted by the AI does worse than the worker alone ($\mathcal{A}_{HA} < \mathcal{A}_H$). From Theorem~\ref{thm:phase}, this region exists whenever overlap satisfies $\lambda > 1 - \hat\beta/2$.
The mechanism is direct. When overlap is high, much of the AI's evidence is content the worker already has. The correlation-neglect worker, using the rule $\hat\ell = \ell_H + \hat\beta\ell_A$, treats $\ell_A$ as if it were a fresh, independent signal. By giving $\ell_A$ full weight $\hat\beta$, she effectively double-counts the overlapping content. When the AI is weak (small $\tau_A$), the genuinely new information in $\ell_A$ is too small to offset the double-counting penalty, and her accuracy falls below the human-alone benchmark. 
%The cue-bank model imposes a feasibility constraint $\tau_A \le \tau_H/\lambda$ at fixed $\lambda > 0$, since the AI's overlapping cues are a subset of the human's $\tau_H$ cues; we describe the impairment region within this feasible range.

\textbf{Insight 2: (Automation Region) -- Automation is inevitable as AI capability grows.} From Theorem~\ref{thm:phase}, the automation threshold $\tau_{\mathrm{auto}}(\lambda; \hat\beta)$ is finite for any $\lambda > 0$, so AI alone outperforms the team whenever $\tau_A > \tau_{\mathrm{auto}}$. The cue-pool interpretation imposes an upper bound $\tau_A \le \tau_H/\lambda$; at this boundary the AI's evidence is Blackwell-equivalent to the team's, so automation is welfare-optimal regardless of trust calibration. Since $\tau_{\mathrm{auto}}(\lambda; \hat\beta) < \tau_H/\lambda$ for all $\lambda \in (0, 1)$, automation is reached at finite AI capability strictly inside the feasible region. This contrasts with the Bayes-rational case, where the team weakly outperforms either modality and strict automation is never optimal; therefore, under correlation neglect, inevitable automation is a signature of behavioral misspecification, not of the underlying information structure.

\textbf{Insight 3: (Complementarity Region) -- The complementarity region narrows as overlap grows.} For any fixed $\hat\beta > 0$, the complementarity region in the $(\tau_A, \lambda)$ plane has the form $\tau_A \in [\tau_{\mathrm{aug}}(\lambda; \hat\beta), \tau_{\mathrm{auto}}(\lambda; \hat\beta)]$ when both thresholds are positive. As $\lambda$ rises: the augmentation threshold $\tau_{\mathrm{aug}}$ is strictly increasing: more overlap requires a more capable AI before the team beats the human alone; meanwhile, the automation threshold $\tau_{\mathrm{auto}}$ is strictly decreasing: more overlap means a less capable AI suffices for AI alone to beat the team. Therefore, the complementarity region shrinks from both ends as $\lambda$ grows, and at $\lambda = 1$ collapses to the single tie point $\tau_A = \tau_H$.

%, since reducing $\lambda$ (the redundancy of AI evidence with human evidence) expands the parameter space where complementarity is achievable.

\textbf{Insight 4: Trust is a double-edged lever under correlation neglect.}
The worker's trust $\hat\beta$ scales how strongly she weights the AI signal, with opposing effects across the two phase boundaries.
\begin{corollary}[Comparative statics of phase boundaries in trust]
\label{cor:trust-boundaries}
The phase boundaries are strictly increasing in $\hat\beta$ for $\lambda \in (0, 1)$: $\frac{\partial \tau_{\mathrm{aug}}}{\partial \hat\beta} > 0$ and $\frac{\partial \tau_{\mathrm{auto}}}{\partial \hat\beta} > 0$.
\end{corollary}
A worker who trusts the AI signal more benefits more from its marginal information but also incurs a larger double-counting penalty. This has opposite effects at the two boundaries. At the lower boundary, higher trust amplifies double-counting when the AI is weak and redundant, so a more capable AI is needed before assistance beats the human-alone benchmark. Similarly, higher trust will also make the lowest level of overlap needed for impairment, $1-\hat\beta/2$, lower. At the upper boundary, higher trust also lets the human-AI team make better use of a strong AI signal, so AI-alone dominance is delayed. Thus undertrust is protective against harmful redundant AI but makes automation arrive sooner, while overtrust expands impairment but can preserve complementarity up to higher AI capability.

\begin{figure}[h]
\centering

% --- Panel 1: Low overlap (lambda = 0.3) ---
\begin{subfigure}[t]{0.32\textwidth}
\centering
\begin{tikzpicture}[scale=0.6]

% Parameters: tauH = 1, hatbeta = 1, lambda = 0.3
\fill[red!10] (0, 0) rectangle (2.37, 4);
\fill[blue!10] (2.37, 0) rectangle (4, 4);

\draw[dashed, gray!50] (1, 0) -- (1, 4);
\draw[dotted, blue!60] (2.37, 0) -- (2.37, 4);

\draw[thick, gray!80!black] (0, 1.83) -- (4, 1.83);
\node[gray!80!black, font=\tiny, anchor=south west] at (3.4, 1.85) {$\mathcal{A}_H$};

\draw[thick, red!80!black] plot[smooth] coordinates {
  (0, 1.83) (0.5, 2.11) (1, 2.40) (1.5, 2.64) (2, 2.86)
  (2.5, 3.04) (3, 3.20) (3.5, 3.34) (4, 3.46)
};
\node[red!80!black, font=\tiny, anchor=south west] at (3.4, 3.5) {$\mathcal{A}_{HA}$};

\draw[thick, blue!80!black] plot[smooth] coordinates {
  (0.13, 0) (0.5, 1.04) (1, 1.83) (1.5, 2.36) (2, 2.75)
  (2.5, 3.06) (3, 3.32) (3.5, 3.51) (4, 3.67)
};
\node[blue!80!black, font=\tiny, anchor=south west] at (3.4, 3.7) {$\mathcal{A}_{AI}$};

\draw[->, thick] (0, 0) -- (4.3, 0) node[right, font=\small] {$\tau_A$};
\draw[->, thick] (0, 0) -- (0, 4.4) node[above, font=\small] {accuracy};
\draw[font=\footnotesize] (1, 0) node[below] {$\tau_H$};
\draw[font=\footnotesize] (2.37, 0) node[below, blue!60] {$\tau_{\mathrm{auto}}$};

\draw (-0.08, 0) -- (0.08, 0);
\draw[font=\footnotesize] (0, 0) node[left] {$0.6$};
\draw (-0.08, 4) -- (0.08, 4);
\draw[font=\footnotesize] (0, 4) node[left] {$0.95$};

\end{tikzpicture}
\caption{Low overlap ($\lambda = 0.3$)}
\end{subfigure}%
\hfill
% --- Panel 2: Moderate overlap (lambda = 0.9) ---
\begin{subfigure}[t]{0.32\textwidth}
\centering
\begin{tikzpicture}[scale=0.6]

% Parameters: tauH = 1, hatbeta = 1, lambda = 0.9
\fill[gray!15] (0, 0) rectangle (0.8, 4);
\fill[red!10] (0.8, 0) rectangle (1.073, 4);
\fill[blue!10] (1.073, 0) rectangle (4, 4);

\draw[dashed, gray!50] (1, 0) -- (1, 4);
\draw[dotted, red!60] (0.8, 0) -- (0.8, 4);
\draw[dotted, blue!60] (1.073, 0) -- (1.073, 4);

\draw[thick, gray!80!black] (0, 1.83) -- (4, 1.83);

\draw[thick, red!80!black] plot[smooth] coordinates {
  (0, 1.83) (0.1, 1.78) (0.2, 1.73) (0.3, 1.73) (0.4, 1.74)
  (0.5, 1.76) (0.6, 1.78) (0.7, 1.81) (0.8, 1.83) (1, 1.89)
  (1.5, 2.07) (2, 2.23) (2.5, 2.39) (3, 2.54) (3.5, 2.67) (4, 2.79)
};

\draw[thick, blue!80!black] plot[smooth] coordinates {
  (0.13, 0) (0.5, 1.04) (1, 1.83) (1.5, 2.36) (2, 2.75)
  (2.5, 3.06) (3, 3.32) (3.5, 3.51) (4, 3.67)
};

\draw[->, thick] (0, 0) -- (4.3, 0) node[right, font=\small] {$\tau_A$};
\draw[->, thick] (0, 0) -- (0, 4.4) node[above, font=\small] {accuracy};
\draw[font=\footnotesize] (0.8, 0) node[below, red!60, xshift=-3pt] {$\tau_{\mathrm{aug}}$};
\draw[font=\footnotesize] (1.073, 0) node[below, blue!60, xshift=15pt] {$\tau_{\mathrm{auto}}$};

\draw (-0.08, 0) -- (0.08, 0);
\draw[font=\footnotesize] (0, 0) node[left] {$0.6$};
\draw (-0.08, 4) -- (0.08, 4);
\draw[font=\footnotesize] (0, 4) node[left] {$0.95$};

\end{tikzpicture}
\caption{Moderate overlap ($\lambda = 0.9$)}
\end{subfigure}%
\hfill
% --- Panel 3: High overlap (lambda = 1) ---
\begin{subfigure}[t]{0.32\textwidth}
\centering
\begin{tikzpicture}[scale=0.6]

% Parameters: tauH = 1, hatbeta = 1, lambda = 1
\fill[gray!15] (0, 0) rectangle (1, 4);
\fill[blue!10] (1, 0) rectangle (4, 4);

\draw[dashed, gray!50] (1, 0) -- (1, 4);

\draw[thick, gray!80!black] (0, 1.83) -- (4, 1.83);

\draw[thick, red!80!black] plot[smooth] coordinates {
  (0, 1.83) (0.2, 1.74) (0.4, 1.71) (0.6, 1.71) (0.8, 1.76) (1, 1.83)
  (1.5, 2.00) (2, 2.16) (2.5, 2.31) (3, 2.45) (3.5, 2.58) (4, 2.70)
};

\draw[thick, blue!80!black] plot[smooth] coordinates {
  (0.13, 0) (0.5, 1.04) (1, 1.83) (1.5, 2.36) (2, 2.75)
  (2.5, 3.06) (3, 3.32) (3.5, 3.51) (4, 3.67)
};

\filldraw[black] (1, 1.83) circle (1.5pt);

\draw[->, thick] (0, 0) -- (4.3, 0) node[right, font=\small] {$\tau_A$};
\draw[->, thick] (0, 0) -- (0, 4.4) node[above, font=\small] {accuracy};
\draw[font=\footnotesize] (1, 0) node[below] {$\tau_H$};

\draw (-0.08, 0) -- (0.08, 0);
\draw[font=\footnotesize] (0, 0) node[left] {$0.6$};
\draw (-0.08, 4) -- (0.08, 4);
\draw[font=\footnotesize] (0, 4) node[left] {$0.95$};

\end{tikzpicture}
\caption{High overlap ($\lambda = 1$)}
\end{subfigure}

\caption{Accuracy vs.\ AI capability $\tau_A$ at three overlap levels $\lambda \in \{0.3, 0.9, 1.0\}$ under correlation neglect with calibrated trust ($\hat\alpha = 0, \hat\beta = 1, \tau_H = 1$). Curves: human alone $\mathcal{A}_H$ (gray), team $\mathcal{A}_{HA}$ (red), AI alone $\mathcal{A}_{AI}$ (blue). Shaded background marks the welfare-optimal regime (gray: impairment, red: complementarity, blue: automation). As $\lambda$ rises, the team curve flattens, the impairment region appears (panel b), and widens until the AI is fully redundant at $\tau_A = \tau_H$ (panel c).}
\label{fig:accuracy-vs-tauA}
\end{figure}
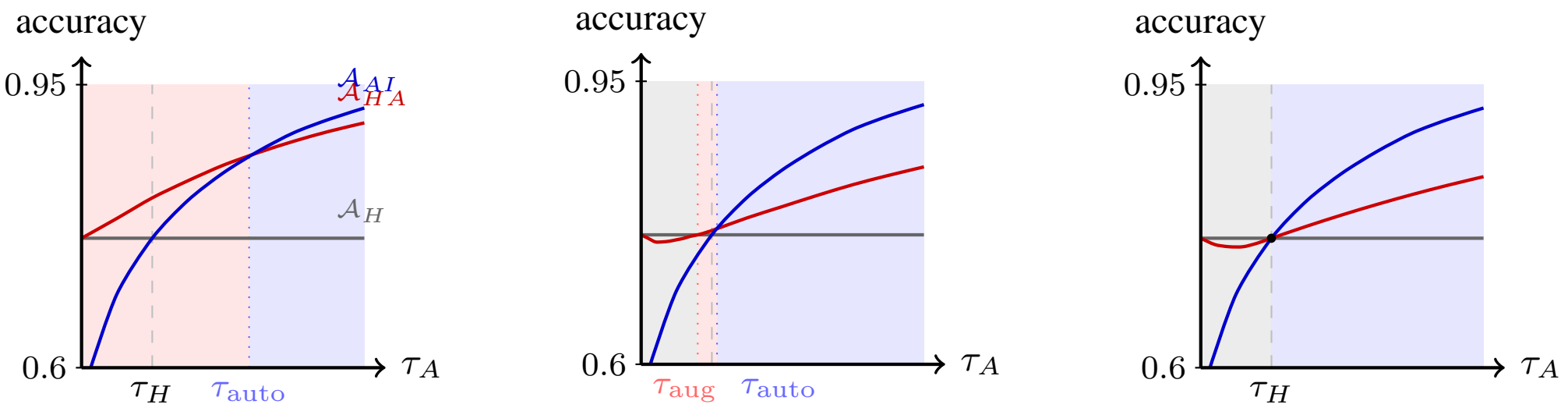

\textbf{Accuracy as a function of AI capability.} Figure~\ref{fig:accuracy-vs-tauA} traces $\mathcal{A}_H$, $\mathcal{A}_{HA}$, $\mathcal{A}_{AI}$ vs.\ $\tau_A$ at three overlap levels while fixing $\hat\beta = 1$, $\tau_H = 1$. At low overlap, $\mathcal{A}_{HA}$ rises monotonically and dominates $\mathcal{A}_H$ until $\mathcal{A}_{AI}$ overtakes it at $\tau_{\mathrm{auto}}$. At moderate overlap, $\mathcal{A}_{HA}$ is non-monotone: it dips below $\mathcal{A}_H$ for small $\tau_A$ (impairment), crosses at $\tau_{\mathrm{aug}}$, and is overtaken by $\mathcal{A}_{AI}$ at $\tau_{\mathrm{auto}}$. At $\lambda = 1$, the complementarity interval collapses to the triple tie at $\tau_A = \tau_H$. The non-monotonicity is the signature of correlation neglect: weak AI hurts via double-counting, moderate AI helps once its marginal information dominates, strong AI eventually dominates as the worker becomes a noise channel.

The \textbf{design implications} are threefold. First, AI systems serving correlation-neglect workers benefit from \emph{reducing} effective overlap with human evidence, not just from increasing $\tau_A$: at fixed $\tau_A$ the team SNR is strictly decreasing in $\lambda$ ($\partial r_{HA}^2/\partial \lambda |_{\hat\alpha=0} < 0$ for any $\hat\beta > 0$), giving the principle of training AI to predict what humans miss rather than what they already know. Second, whether to provide $\ell_A$ at all is a complementary lever: from Theorem~\ref{thm:phase}, the team is welfare-optimal only on $\tau_A \in [\tau_{\mathrm{aug}}(\lambda;\hat\beta),\, \tau_{\mathrm{auto}}(\lambda;\hat\beta)]$, so the optimal disclosure rule is to withhold $\ell_A$ when $\tau_A < \tau_{\mathrm{aug}}$, provide it when $\tau_A \in [\tau_{\mathrm{aug}}, \tau_{\mathrm{auto}}]$, and automate when $\tau_A > \tau_{\mathrm{auto}}$. Third, in the impairment region, behavioral interventions encouraging \emph{undertrust} ($\hat\beta < 1$) can improve welfare for correlation-neglect workers: $\hat\beta < 1$ does not correct the neglect but dampens its double-counting cost (Corollary~\ref{cor:trust-boundaries}), so the conventional calibration target $\hat\beta = 1$ is not welfare-optimal there.

%\vspace{-1.5em}
\subsection{Welfare Decomposition}
\label{sec:welfare}

The phase diagram characterizes which deployment mode yields the highest accuracy. We now consider a complementary question: how much accuracy does the worker lose relative to the Bayes-rational benchmark, and how much of this loss is attributable to correlation neglect versus undertrust? %Appendix~\ref{app:identification} describes identification of the worker's behavioral parameters from cross-AI variation.

Write $\mathcal{A}(\hat\alpha, \hat\beta)$ for the worker's accuracy under behavioral primitives $\hat\alpha$ and $\beta$, and let $\mathcal{A}^{\mathrm{BR}} = \mathcal{A}(1, 1)$ be the Bayesian rational benchmark. We define
\begin{align*}
\mathrm{TWL}(\hat\alpha, \hat\beta) &:= \mathcal{A}^{\mathrm{BR}} - \mathcal{A}(\hat\alpha, \hat\beta), \\
\mathrm{CNC}(\hat\alpha) &:= \mathcal{A}^{\mathrm{BR}} - \mathcal{A}(\hat\alpha, 1), \\
\mathrm{UTC}(\hat\alpha, \hat\beta) &:= \mathcal{A}(\hat\alpha, 1) - \mathcal{A}(\hat\alpha, \hat\beta).
\end{align*}
Here, $\mathrm{TWL}$ is the worker's total welfare loss compared to the BR benchmark, which subsequently decomposes into two types of loss, $\mathrm{CNC}$ and $\mathrm{UTC}$. The correlation-neglect cost $\mathrm{CNC}$ measures the loss from imperfect correlation awareness when trust is calibrated. The undertrust cost $\mathrm{UTC}$ measures the gain from calibrating trust while holding the worker's actual awareness level fixed. We have the following exact decomposition.

\begin{proposition}[Welfare decomposition]
\label{prop:welfare}
Within the cue-pool feasible region, for every worker $(\hat\alpha, \hat\beta) \in [0, 1] \times \mathbb{R}_{>0}$, we have $\mathrm{TWL}(\hat\alpha, \hat\beta) = \mathrm{CNC}(\hat\alpha) + \mathrm{UTC}(\hat\alpha, \hat\beta)$.
The correlation-neglect component is nonnegative: $\mathrm{CNC}(\hat\alpha) \ge 0$. If $\hat\beta \in (0, 1]$ and $\tau_A \ge \tau_H$, the undertrust component is also nonnegative; it vanishes when $\hat\beta = 1$.
\end{proposition}

As mentioned before, the two types of loss correspond to two design levers. Calibrating trust first recovers $\mathrm{UTC}$; correcting the remaining correlation misperception then recovers $\mathrm{CNC}$. We will later link this welfare decomposition to our empirical study.
Finally, we define the welfare gain relative to the unaided human,
\[
\mathrm{GAP}(\hat\alpha, \hat\beta) := \mathcal{A}(\hat\alpha, \hat\beta) - \mathcal{A}_H.
\]
It is used to measure the benefit of AI assistance relative to human-alone performance. It is also used in our empirical study.

\section{Empirical Evaluation}
\label{sec:experiments}

We instantiate the framework on ChatBench~\citep{chang2025chatbench}, a controlled MTurk study in which $643$ workers answer multiple-choice questions from MMLU~\citep{hendrycks2020measuring} across math, physics, and moral reasoning. Each worker is randomly assigned to one of two AI assistants (GPT-4o or Llama-3.1-8b) under an answer-first protocol: she records an unassisted answer with stated confidence, then sees the AI's response and records a final answer. After matching the answer-first user-AI rows to their corresponding user-alone rows and restricting to pairs where the AI committed to a definite letter recommendation, we obtain $2{,}199$ paired observations covering $315$ workers and $389$ questions. We construct $\ell_H = \mathrm{logit}(p_H(c))$ from worker confidence and $\ell_A^{m,q} = \mathrm{logit}(p_A^{m,q})$ from AI model $m$'s empirical accuracy on question $q$ across up to $50$ generations. The binary outcome $Y$ is the ground-truth letter. Appendix~\ref{app:sample} details sample construction.

\textbf{AI assistant produces significant welfare gains; different AI models lead to different optimal deployment modes}
Workers go from $\mathcal{A}_H = 0.612$ alone to $\mathcal{A}_{HA} = 0.824$ with AI assistance, a pooled welfare gain of $\mathrm{GAP} = +0.213$ (95\% CI $[0.191, 0.235]$, never negative in $1{,}000$ bootstrap samples). The two deployments fall in different empirical regimes (Table~\ref{tab:emp-summary}). For GPT-4o, the AI alone outperforms the team ($\mathcal{A}_{AI} = 0.884 > \mathcal{A}_{HA} = 0.851$, $p = 0.002$): GPT is firmly in the framework's \emph{automation} regime. For Llama-3.1-8b, $\mathcal{A}_{HA} = 0.795$ and $\mathcal{A}_{AI} = 0.787$ are statistically indistinguishable ($p = 0.55$), placing Llama at the \emph{boundary} between complementarity and automation. The cross-deployment GAP gap of $+0.081$ is approximately $93\%$ explained by capability composition (95\% CI $[68\%, 148\%]$); a likelihood-ratio test of cell-by-AI interactions gives $\chi^2(4) = 0.96$, $p = 0.92$. This validates the framework's central qualitative claim that AI capability $\tau_A^m$ governs deployment-level welfare while behavior is approximately invariant across deployments.

\begin{table}[h]
\centering
\caption{Per-deployment summary on the Has-Rec sample. Regimes are classified by paired comparison of $\mathcal{A}_{HA}$ and $\mathcal{A}_{AI}$ on the same observations; see Appendix~\ref{app:identification} for test details.}
\label{tab:emp-summary}
\small
\begin{tabular}{lcccccl}
\toprule
Deployment & $n$ & $\mathcal{A}_H$ & $\mathcal{A}_{HA}$ & $\mathcal{A}_{AI}$ & GAP & Regime ($p$) \\
\midrule
GPT-4o          & $1{,}160$ & $0.600$ & $0.851$ & $0.884$ & $+0.251$ & Automation ($p = 0.002$) \\
Llama-3.1-8b    & $1{,}039$ & $0.625$ & $0.795$ & $0.787$ & $+0.170$ & Boundary ($p = 0.55$) \\
Pooled          & $2{,}199$ & $0.612$ & $0.824$ & ---     & $+0.213$ & --- \\
\bottomrule
\end{tabular}
\end{table}

\textbf{Workers substantially undertrust the AI; strict correlation neglect is rejected.}
Inverting Lemma~\ref{lem:behavioral} from per-deployment logistic regression weights identifies $\hat\beta = 0.285$ (95\% CI $[0.246, 0.322]$): workers apply roughly a quarter of the Bayes-rational weight to the AI signal. A second approach using cross-question variation to estimate $\lambda$ directly gives $\hat\beta = 0.254$, in close agreement; see Appendix~\ref{app:identification}. Both decisively reject $\hat\beta = 1$. Strict correlation neglect ($\hat\alpha = 0$) is also rejected: the inversion's residual sum of squares jumps $141$-fold when $\hat\alpha = 0$ is forced, and direct $\hat\lambda$ is positive in $100\%$ of bootstrap samples for both deployments. The data does not pin down $\hat\alpha$ to a narrow range but decisively rejects this corner case.

\textbf{Overlap $\lambda$ varies meaningfully across AIs and across subject domains.}
Direct $\lambda$ estimation from cross-question variation gives two empirical patterns that connect back to the framework's cue-pool picture of overlap (Proposition~\ref{prop:overlap-domain}). First, $\lambda$ differs significantly across the two AIs: $\hat\lambda^{\text{Llama}} = 0.127$ exceeds $\hat\lambda^{\text{GPT}} = 0.071$ by $+0.056$ (95\% CI $[0.015, 0.097]$, two-sided $p \approx 0.009$). The smaller model's evidence overlaps more with what humans access; GPT brings more genuinely novel information. Second, $\lambda$ varies dramatically across subjects, pooled across both AIs: math $\hat\lambda \approx 0.20$, physics $\hat\lambda \approx 0.04$, moral reasoning $\hat\lambda \approx 0$, with the same pattern holding within each AI separately. Domains where humans and AI rely on shared informational cues exhibit substantial overlap; domains where reasoning paths are more idiosyncratic exhibit essentially none. Proposition~\ref{prop:overlap-domain} characterizes overlap as a property of the human-domain pair under uniform cue sampling; the cross-AI heterogeneity indicates that real AIs' training data is more complicated.

\textbf{Trust calibration and design implications.}
Section~\ref{sec:welfare} splits the worker's accuracy loss against the Bayes-rational benchmark into a correlation-neglect cost $\mathrm{CNC}$ and an undertrust cost $\mathrm{UTC}$, with $\mathrm{UTC}$ the accuracy recovered by calibrating trust (Proposition~\ref{prop:welfare}). Here, we measure the empirical counterpart of $\mathrm{UTC}$. On items where the worker's unaided answer and the AI's recommendation disagree, we replace her final choice with a simulated trust-calibrated rule that selects whichever answer is more likely correct given the same two signals used to identify the behavioral parameters, her stated confidence $\ell_H$ and the AI's question-level accuracy $\ell_A$. Here, the correctness probabilities are estimated by cross-validated logistic regression (Appendix~\ref{app:welfare}). Like $\mathrm{UTC}$, the resulting gain isolates the trust channel: it changes how she relies on the AI. For GPT-4o, the simulated rule follows the AI on about $92\%$ of disagreements, versus the real workers' $70\%$, and raises team accuracy by $2.6$ percentage points, from $0.851$ to $0.877$, close to the AI-alone accuracy of $0.884$. So for GPT-4o, which lies in the automation regime, calibrating trust moves the team nearly to the AI-alone benchmark. For Llama-3.1-8b, which sits at the complementarity-automation boundary, the simulated rule instead does worse than the actual workers ($-1.4$ percentage points). The reason is informational rather than behavioral: the simulated rule sees only two coarse signals, whereas the worker also reads the AI's specific answer and can judge its reasoning, which for the weaker AI evidently outweighs calibrating trust. Two design implications follow from the result. First, trust calibration pays off more where the AI is strong, so interventions that raise trust should be targeted to AI-strong cases rather than applied uniformly. Meanwhile, besides behavior calibration, the granularity of information the worker receives is equally important.  

\textbf{Extensions to other applications.} The contribution of our empirical framework extends beyond ChatBench's setup with MMLU questions, and can be applied to other domains such as medicine, law, or programming, where human and AI expertise are combined. Estimating the model's primitives
%deployment parameters (capabilities $\tau_H, \tau_A$ and overlap $\lambda$), and workers' behavioral parameters (correlation awareness $\hat\alpha$ and trust $\hat\beta$) 
requires only the observables present in any binary classification setting: the human-alone report $\ell_H$, the AI-alone report $\ell_A$, the AI-assisted human's report $\hat\ell$, and the ground-truth label $Y$. An example is professional radiologists diagnosing from X-ray imagery with and without AI assistance as in \citet{agarwal2023combining}: ground truth comes from an expert diagnostic panel, the human and AI signals are the prior-corrected logits of the human-alone and AI-alone probability reports, and the same class-conditional moments identify $\tau_H, \tau_A, \lambda, \hat\alpha$, and $\hat\beta$.

\textbf{Broader impacts.} Our framework characterizes when sharing an AI recommendation helps or harms a human decision-maker, and identifies two levers for improving deployed human-AI teams: calibrating trust and reducing information overlap. Positively, the phase diagram offers deployment guidance (withhold, assist, or automate) that can prevent the accuracy losses documented when capable AI is naively added to human workflows, and the overlap lever suggests training assistants to supply complementary rather than redundant evidence. Risks include over-reliance on a stylized two-signal model: the behavioral parameters are estimated from a specific crowdsourced setting (ChatBench/MMLU) and need not transfer to high-stakes domains such as medicine or hiring, where miscalibrated `automation' recommendations could cause harm. Our prescriptions should be treated as a modeling-informed starting point for extensive domain-specific validation.
\vspace{-1em}
\section{Conclusion}
\label{sec:conclusion}

We build a model of AI-assisted classification in which a worker combines her own evidence with an AI's through two behavioral primitives, correlation awareness and trust. We derive a closed-form phase diagram over capability and overlap that separates complementarity, impairment, and automation. We examine the framework on ChatBench, finding substantial undertrust, regimes decided by AI capability and overlap, and heterogeneous gains from trust calibration. The framework potentially applies to any setup of AI-assisted decision making where human-alone, AI-alone, AI-assisted assessment, and ground truth are all observed.

\bibliographystyle{plainnat}
\bibliography{references_new}

@article{eyster2010correlation,
  title={Correlation neglect in financial decision-making},
  author={Eyster, Erik and Weizsacker, Georg},
  year={2010},
  publisher={DIW Berlin Discussion Paper}
}

@article{steyvers2022bayesian,
  title={Bayesian modeling of human--AI complementarity},
  author={Steyvers, Mark and Tejeda, Heliodoro and Kerrigan, Gavin and Smyth, Padhraic},
  journal={Proceedings of the National Academy of Sciences},
  volume={119},
  number={11},
  pages={e2111547119},
  year={2022},
  publisher={National Academy of Sciences}
}

@techreport{agarwal2025designing,
  title={Designing human-ai collaboration: A sufficient-statistic approach},
  author={Agarwal, Nikhil and Moehring, Alex and Wolitzky, Alexander},
  year={2025},
  institution={National Bureau of Economic Research}
}

@article{enke2019correlation,
  title={Correlation neglect in belief formation},
  author={Enke, Benjamin and Zimmermann, Florian},
  journal={The review of economic studies},
  volume={86},
  number={1},
  pages={313--332},
  year={2019},
  publisher={Oxford University Press}
}

@article{hendrycks2020measuring,
  title={Measuring massive multitask language understanding},
  author={Hendrycks, Dan and Burns, Collin and Basart, Steven and Zou, Andy and Mazeika, Mantas and Song, Dawn and Steinhardt, Jacob},
  journal={arXiv preprint arXiv:2009.03300},
  year={2020}
}

@inproceedings{chang2025chatbench,
  title={Chatbench: From static benchmarks to human-ai evaluation},
  author={Chang, Serina and Anderson, Ashton and Hofman, Jake M},
  booktitle={Proceedings of the 63rd Annual Meeting of the Association for Computational Linguistics (Volume 1: Long Papers)},
  pages={26009--26038},
  year={2025}
}

@article{vaccaro2024when,
  title     = {When combinations of humans and {AI} are useful: A systematic review and meta-analysis},
  author    = {Vaccaro, Michelle and Almaatouq, Abdullah and Malone, Thomas},
  journal   = {Nature Human Behaviour},
  volume    = {8},
  number    = {12},
  pages     = {2293--2303},
  year      = {2024}
}

@inproceedings{bansal2019updates,
  title     = {Updates in Human-{AI} Teams: Understanding and Addressing the Performance/Compatibility Tradeoff},
  author    = {Bansal, Gagan and Nushi, Besmira and Kamar, Ece and Weld, Daniel S. and Lasecki, Walter S. and Horvitz, Eric},
  booktitle = {Proceedings of the AAAI Conference on Artificial Intelligence},
  volume    = {33},
  number    = {01},
  pages     = {2429--2437},
  year      = {2019}
}

@inproceedings{bansal2021does,
  title     = {Does the Whole Exceed Its Parts? The Effect of {AI} Explanations on Complementary Team Performance},
  author    = {Bansal, Gagan and Wu, Tongshuang and Zhou, Joyce and Fok, Raymond and Nushi, Besmira and Kamar, Ece and Ribeiro, Marco Tulio and Weld, Daniel S.},
  booktitle = {Proceedings of the 2021 CHI Conference on Human Factors in Computing Systems (CHI)},
  year      = {2021},
  doi       = {10.1145/3411764.3445717}
}

@article{bucinca2021trust,
  title     = {To Trust or to Think: Cognitive Forcing Functions Can Reduce Overreliance on {AI} in {AI}-Assisted Decision-Making},
  author    = {Bu{\c{c}}inca, Zana and Malaya, Maja Barbara and Gajos, Krzysztof Z.},
  journal   = {Proceedings of the ACM on Human-Computer Interaction},
  volume    = {5},
  number    = {CSCW1},
  pages     = {1--21},
  year      = {2021},
  doi       = {10.1145/3449287}
}

@article{tschandl2020human,
  title     = {Human--computer collaboration for skin cancer recognition},
  author    = {Tschandl, Philipp and Rinner, Christoph and Apalla, Zoe and Argenziano, Giuseppe and Codella, Noel and Halpern, Allan and Janda, Monika and Lallas, Aimilios and Longo, Caterina and Malvehy, Josep and Paoli, John and Puig, Susana and Rosendahl, Cliff and Soyer, H. Peter and Zalaudek, Iris and Kittler, Harald},
  journal   = {Nature Medicine},
  volume    = {26},
  number    = {8},
  pages     = {1229--1234},
  year      = {2020},
  doi       = {10.1038/s41591-020-0942-0}
}

@techreport{agarwal2023combining,
  title       = {Combining Human Expertise with Artificial Intelligence: Experimental Evidence from Radiology},
  author      = {Agarwal, Nikhil and Moehring, Alex and Rajpurkar, Pranav and Salz, Tobias},
  institution = {National Bureau of Economic Research},
  type        = {NBER Working Paper},
  number      = {31422},
  year        = {2023}
}

@inproceedings{donahue2022human,
  title     = {Human-Algorithm Collaboration: Achieving Complementarity and Avoiding Unfairness},
  author    = {Donahue, Kate and Chouldechova, Alexandra and Kenthapadi, Krishnaram},
  booktitle = {Proceedings of the 2022 ACM Conference on Fairness, Accountability, and Transparency (FAccT)},
  pages     = {1639--1656},
  year      = {2022},
  doi       = {10.1145/3531146.3533221}
}

@inproceedings{peng2025no,
  title={A no free lunch theorem for human-ai collaboration},
  author={Peng, Kenny and Garg, Nikhil and Kleinberg, Jon},
  booktitle={Proceedings of the AAAI Conference on Artificial Intelligence},
  volume={39},
  number={13},
  pages={14369--14376},
  year={2025}
}

@inproceedings{madras2018predict,
  title     = {Predict Responsibly: Improving Fairness and Accuracy by Learning to Defer},
  author    = {Madras, David and Pitassi, Toniann and Zemel, Richard},
  booktitle = {Advances in Neural Information Processing Systems 32 (NeurIPS)},
  year      = {2018}
}

@inproceedings{mozannar2020consistent,
  title     = {Consistent Estimators for Learning to Defer to an Expert},
  author    = {Mozannar, Hussein and Sontag, David},
  booktitle = {Proceedings of the 37th International Conference on Machine Learning (ICML)},
  series    = {Proceedings of Machine Learning Research},
  volume    = {119},
  pages     = {7076--7087},
  year      = {2020},
  publisher = {PMLR}
}

@inproceedings{verma2023learning,
  title     = {Learning to Defer to Multiple Experts: Consistent Surrogate Losses, Confidence Calibration, and Conformal Ensembles},
  author    = {Verma, Rajeev and Barrej{\'o}n, Daniel and Nalisnick, Eric},
  booktitle = {Proceedings of the 26th International Conference on Artificial Intelligence and Statistics (AISTATS)},
  series    = {Proceedings of Machine Learning Research},
  volume    = {206},
  pages     = {11415--11434},
  year      = {2023},
  publisher = {PMLR}
}

@inproceedings{wilder2020learning,
  title     = {Learning to Complement Humans},
  author    = {Wilder, Bryan and Horvitz, Eric and Kamar, Ece},
  booktitle = {Proceedings of the 29th International Joint Conference on Artificial Intelligence (IJCAI)},
  pages     = {1526--1533},
  year      = {2020}
}

@inproceedings{alur2023auditing,
  title     = {Auditing for Human Expertise},
  author    = {Alur, Rohan and Laine, Loren and Li, Darrick K. and Raghavan, Manish and Shah, Devavrat and Shung, Dennis},
  booktitle = {Advances in Neural Information Processing Systems 37 (NeurIPS)},
  year      = {2023}
}

@inproceedings{alur2024human,
  title     = {Human Expertise in Algorithmic Prediction},
  author    = {Alur, Rohan and Raghavan, Manish and Shah, Devavrat},
  booktitle = {Advances in Neural Information Processing Systems 38 (NeurIPS)},
  year      = {2024}
}

@inproceedings{Guo2024Reliance,
	author = {Guo, Ziyang and Wu, Yifan and Hartline, Jason D. and Hullman, Jessica},
	title = {A Decision Theoretic Framework for Measuring AI Reliance},
	year = {2024},
	isbn = {9798400704505},
	publisher = {Association for Computing Machinery},
	address = {New York, NY, USA},
	url = {https://doi.org/10.1145/3630106.3658901},
	doi = {10.1145/3630106.3658901},
	booktitle = {Proceedings of the 2024 ACM Conference on Fairness, Accountability, and Transparency},
	pages = {221–236},
	numpages = {16},
	location = {Rio de Janeiro, Brazil},
	series = {FAccT '24}
}

%%%%%%%%%%%%%%%%%%%%%%%%%%%%%%%%%%%%%%%%%%%%%%%%%%%%%%%%%%%%
\newpage
\appendix
\section{Behavioral modeling of trust miscalibration}
\label{app:behavioral}
A natural alternative to model trust miscalibration is to interpret trust as a subjective belief about the AI's capability, replacing $\tau_A$ by $\hat\beta \tau_A$ in the worker's perceived signal distribution. This interpretation is not appropriate for the present classification model. The AI's message is assumed to be informationally equivalent to its log-likelihood ratio $\ell_A$, or equivalently to a calibrated posterior probability. Once $\ell_A$ is observed, its Bayesian contribution to posterior log-odds is already encoded in its realized value.

To verify this, suppose first that the worker perceives the AI signal as conditionally independent of her own signal and believes that
\[
\ell_A \mid Y=1 \sim N(\hat\tau_A,2\hat\tau_A),
\qquad
\ell_A \mid Y=0 \sim N(-\hat\tau_A,2\hat\tau_A).
\]
Then the log-likelihood ratio implied by this subjective model is
\[
\log
\frac{\phi(\ell_A;\hat\tau_A,2\hat\tau_A)}
     {\phi(\ell_A;-\hat\tau_A,2\hat\tau_A)}
=
\ell_A,
\]
which is independent of the perceived capability $\hat\tau_A$. Thus, changing the worker's perceived AI precision changes the ex ante distribution she assigns to $\ell_A$, but it does not change the Bayesian weight placed on $\ell_A$ as long as the worker still believes $\expit(\ell_A)$ is a calibrated forecast. In particular, under perceived independence, any subjective-capability model yields the decision statistic $\hat \ell = \ell_H+\ell_A$, so trust would be behaviorally unidentifiable.

For this reason, we model $\hat\beta$ as a behavioral weight on the AI's
communicated likelihood-ratio contribution, rather than as a subjective precision.
$\hat\beta\neq 1$ should be interpreted as a departure from treating the AI’s probability report as a fully trusted calibrated posterior. If $\hat\beta<1$, the worker behaves as though the AI’s log-odds are too extreme and shrinks them toward zero; if $\hat\beta>1$, she behaves as though the AI’s log-odds are too conservative and amplifies them.
\section{More performance objectives}
\label{sec:more_objective}
\subsection{Log Loss}
\label{app:log-loss}
The \emph{expected log loss} is $\mathcal{L}_{\log} := \mathbb{E}[-\log P(Y \mid \hat\ell)]$, where $P(Y \mid \hat\ell) = \expit((2Y-1)\hat\ell)$ is the worker's reported probability for the realized state. The log loss is the natural statistic when the worker's output is a probability rather than a binary decision. Unlike the error rate, which only depends on whether the decision is correct, log loss also penalizes the confidence with which the worker reports a wrong answer.
\begin{lemma}[Expected log loss]
\label{lem:log-loss}
Define
\[
\Psi(m, v) := \mathbb{E}_{Z \sim \mathcal{N}(0, 1)}\bigl[\log(1 + e^{-(m + \sqrt{v}\,Z)})\bigr].
\]
Under Assumption~\ref{ass:worker} and the symmetric prior, the worker reporting posterior probability $\hat p = \expit(\hat\ell)$ incurs expected log loss
\[
\mathcal{L}_{\log}(\hat\alpha, \hat\beta; \tau_H, \tau_A, \lambda) = \Psi\!\left(m_{HA},\, \sigma_{HA}^2\right),
\]
where $m_{HA} = w_H \tau_H + w_A \tau_A$ and $\sigma_{HA}^2 = 2\bigl(w_H^2 \tau_H + w_A^2 \tau_A + 2 w_H w_A \tau_A \lambda\bigr)$ are the conditional mean and variance of $\hat\ell$ given $Y = 1$, and the weights $(w_H, w_A)$ are given by Lemma~\ref{lem:behavioral}. The function $\Psi$ admits no elementary closed form but is computable by numerical integration. The standalone log-losses are recovered as $\mathcal{L}_H = \Psi(\tau_H, 2\tau_H)$ and $\mathcal{L}_A = \Psi(\tau_A, 2\tau_A)$.
\end{lemma}

The log loss differs from the error rate in a substantive way. The error rate scores only whether the binary decision $\hat Y = \mathbf{1}\{\hat\ell > 0\}$ is correct; it ignores the magnitude $|\hat\ell|$. The log loss scores both the direction of the worker's belief and her confidence in it. Two workers with the same correctness rate can have different log losses if their probability reports differ in how confidently wrong they are: a worker who reports $\hat p = 0.99$ on an answer that turns out wrong pays more under log loss than a worker who reports $\hat p = 0.6$ on the same wrong answer. The log loss therefore captures \emph{calibration} of probability reports, while the error rate captures only point-estimate correctness.

Log loss is the relevant welfare objective when the worker's task involves communicating probabilities rather than binary decisions: medical diagnosis with uncertainty estimates or any setting where downstream users act on the worker's confidence and not just her hard recommendation. In binary-decision settings, where the worker's output is a hard label, error rate is the appropriate welfare objective.

\begin{proof}[Proof of Lemma~\ref{lem:log-loss}]
The worker's posterior report is $\hat p = \expit(\hat\ell)$. The log loss for outcome $Y$ is
\[
L_{\log}(Y, \hat p) = -Y \log \hat p - (1 - Y) \log(1 - \hat p) = \log(1 + e^{-(2Y-1)\hat\ell}),
\]
using $\log \expit(x) = -\log(1 + e^{-x})$ and $\log(1 - \expit(x)) = -\log(1 + e^{x})$. Taking expectations over the joint distribution of $(\ell_H, \ell_A, Y)$ from Lemma~\ref{lem:joint},
\[
\mathcal{L}_{\log}(\hat\alpha, \hat\beta; \tau_H, \tau_A, \lambda) = \mathbb{E}\bigl[\log(1 + e^{-(2Y-1)\hat\ell})\bigr].
\]
From the proof of Lemma~\ref{lem:accuracy}, $\hat\ell \mid Y \sim \mathcal{N}((2Y-1)\, m_{HA},\, \sigma_{HA}^2)$, where $m_{HA} = w_H \tau_H + w_A \tau_A$ and $\sigma_{HA}^2 = 2(w_H^2 \tau_H + w_A^2 \tau_A + 2 w_H w_A \tau_A \lambda)$. By symmetry of the conditional distributions under $Y = 1$ versus $Y = 0$ and the symmetric prior, the two conditional log-loss expectations are equal, so the unconditional expectation reduces to a one-dimensional Gaussian integral:
\[
\mathcal{L}_{\log} = \mathbb{E}_{Z \sim \mathcal{N}(0, 1)}\!\left[\log\bigl(1 + e^{-(m_{HA} + \sigma_{HA} Z)}\bigr)\right] = \Psi\!\left(m_{HA},\, \sigma_{HA}^2\right). \qedhere
\]
\end{proof}

\subsection{Switching Rate}
The \emph{switching rate} is $\mathcal{S} := P(\hat Y_{\mathrm{HA}} \ne \hat Y_H)$, the probability that AI assistance changes the worker's decision relative to her unaided choice $\hat Y_H = \mathbf{1}\{\ell_H > 0\}$. The switching rate is not a welfare objective but a behavioral one: it measures how often AI assistance changes the worker's decision. When the data record whether workers revised their tentative answers, the switching rate is what the data directly measure. 
\label{app:switching}
\begin{lemma}[Switching rate]
\label{lem:switching}
Under Assumption~\ref{ass:worker} and the symmetric prior, the probability that AI assistance changes the worker's decision is
\[
\mathcal{S}(\hat\alpha, \hat\beta; \tau_H, \tau_A, \lambda) = P\!\left(\sgn(\hat\ell) \ne \sgn(\ell_H)\right) = \Phi(\sqrt{\tau_H/2}) + \Phi(r_{HA}) - 2\,\Phi_2\!\left(\sqrt{\tau_H/2},\, r_{HA};\, \rho\right),
\]
where $\Phi_2(a, b; r) := \Pr(Z_1 \le a, Z_2 \le b)$ for $(Z_1, Z_2) \sim \mathcal{N}(0, \Sigma_r)$ with unit variances and correlation $r$, $r_{HA}$ is the team SNR from Lemma~\ref{lem:accuracy}, and the conditional correlation between $\ell_H$ and $\hat\ell$ given $Y$ is
\[
\rho = \frac{w_H \tau_H + w_A \tau_A \lambda}{\sqrt{\tau_H\bigl(w_H^2 \tau_H + w_A^2 \tau_A + 2 w_H w_A \tau_A \lambda\bigr)}}.
\]
\end{lemma}

%\begin{lemma}[Switching rate]
%\label{lem:switching}
%Under Assumption~\ref{ass:worker} and the symmetric prior, the probability that AI assistance changes the worker's decision is
%\[
%\mathcal{S}(\hat\alpha, \hat\beta; \tau_H, \tau_A, \lambda) = P\!\left(\sgn(\hat\ell) \ne \sgn(\ell_H)\right) = \Phi(\sqrt{\tau_H/2}) + \Phi(r_{HA}) - 2\,\Phi_2\!\left(\sqrt{\tau_H/2},\, r_{HA};\, \rho\right),
%\]
%where $\Phi_2(\cdot, \cdot; \rho)$ is the bivariate standard normal CDF, $r_{HA}$ is the team SNR from Lemma~\ref{lem:accuracy}, and the conditional correlation between $\ell_H$ and $\hat\ell$ given $Y$ is
%\[
%\rho = \frac{w_H \tau_H + w_A \tau_A \lambda}{\sqrt{\tau_H\bigl(w_H^2 \tau_H + w_A^2 \tau_A + 2 w_H w_A \tau_A \lambda\bigr)}}.
%\]
%\end{lemma}

The switching rate is a behavioral objective: it measures how often AI assistance changes the worker's binary decision, holding the worker's beliefs $(\hat\alpha, \hat\beta)$ and deployment parameters fixed. The closed form has a structural reading. The first two terms, $\Phi(\sqrt{\tau_H/2})$ and $\Phi(r_{HA})$, are the marginal correctness probabilities of the human alone and the team. The third, $\Phi_2(\sqrt{\tau_H/2}, r_{HA}; \rho)$, is the joint probability that both the human alone and the team are correct, with the conditional correlation $\rho$ measuring how aligned their decisions are. The switching rate equals the sum of marginals minus twice the joint, isolating exactly the disagreement events — the cases where AI assistance changed the worker's binary decision. When the data record whether the worker's tentative answer was revised after seeing the AI, the empirical switching rate is what the data measure directly, providing a model-fit check beyond the accuracy comparison.

\subsection{Exploitable-to-dangerous ratio}
\label{app:exploitable}
Another useful deployment-level diagnostic is the ratio of probabilities of two events: \emph{exploitable} cases where the human is wrong but the AI is right, and \emph{dangerous} cases where the human is right but the AI is wrong. Define the standardized signals $W_H := (2Y-1)\ell_H/\sqrt{2\tau_H}$ and $W_A := (2Y-1)\ell_A/\sqrt{2\tau_A}$, so that $W_B > 0$ means agent $B$ is correct. Under Lemma~\ref{lem:joint},
\[
(W_H, W_A) \sim \mathcal{N}\!\left(\begin{pmatrix}\sqrt{\tau_H/2} \\ \sqrt{\tau_A/2}\end{pmatrix},\ \begin{pmatrix}1 & \rho_{HA} \\ \rho_{HA} & 1\end{pmatrix}\right) \text{ unconditionally}, \qquad \rho_{HA} := \lambda\sqrt{\tau_A/\tau_H}.
\]
The two probabilities are
\[
\Pr(\text{exploitable}) = \Pr(W_H \le 0, W_A > 0) = \Phi_2\!\left(-\sqrt{\tau_H/2}, \sqrt{\tau_A/2}; -\rho_{HA}\right),
\]
\[
\Pr(\text{dangerous}) = \Pr(W_H > 0, W_A \le 0) = \Phi_2\!\left(\sqrt{\tau_H/2}, -\sqrt{\tau_A/2}; -\rho_{HA}\right),
\]
where the negative correlation in $\Phi_2$ reflects the standard form (each event is rewritten as a $\le$ inequality, flipping one of the two variables and hence the sign of their correlation). The ratio $\Pr(\text{exploitable})/\Pr(\text{dangerous})$ exceeds $1$ if and only if $\tau_A > \tau_H$: more capable AIs generate more exploitable than dangerous cases, providing a deployment-level handle on whether AI assistance has the potential to help.

\section{Proofs for Section~\ref{sec:model}}
\label{app:section1-proofs}

\begin{proof}[Proof of Lemma~\ref{lem:joint}]
\emph{Marginal moments.} Conditional on $Y = 1$, $X_k \sim \mathcal{N}(\mu_1, \sigma^2)$, so
\[
\mathbb{E}[\LLR(X_k) \mid Y = 1] = \frac{\mu_1 - \mu_0}{\sigma^2}\!\left(\mu_1 - \frac{\mu_1 + \mu_0}{2}\right) = \theta,
\]
and similarly $\mathbb{E}[\LLR(X_k) \mid Y = 0] = -\theta$. Since $\LLR(X_k)$ is a linear transform of $X_k$ with slope $(\mu_1 - \mu_0)/\sigma^2$,
\[
\Var(\LLR(X_k) \mid Y) = \left(\frac{\mu_1 - \mu_0}{\sigma^2}\right)^{\!2}\!\sigma^2 = 2\theta.
\]
Thus $\LLR(X_k) \mid Y \sim \mathcal{N}((2Y-1)\theta,\, 2\theta)$. By conditional independence of cues given $Y$ and the additivity $\ell_B = \sum_{k \in \mathcal{X}_B} \LLR(X_k)$,
\[
\mathbb{E}[\ell_B \mid Y] = (2Y-1)\tau_B, \qquad \Var(\ell_B \mid Y) = 2\tau_B.
\]

\emph{Conditional covariance.} By conditional independence, $\Cov(\LLR(X_j), \LLR(X_k) \mid Y) = 2\theta \cdot \mathbf{1}\{j = k\}$. Hence
\[
\Cov(\ell_H, \ell_A \mid Y) = \sum_{j \in \mathcal{X}_H}\sum_{k \in \mathcal{X}_A} \Cov(\LLR(X_j), \LLR(X_k) \mid Y) = 2\theta \, n_\cap = 2\tau_A \lambda.
\]

\emph{Joint Gaussianity.} Every linear combination $a\ell_H + b\ell_A = \sum_k c_k \LLR(X_k)$ with $c_k \in \{a, b, a+b, 0\}$ is a sum of independent Gaussians, so $(\ell_H, \ell_A)$ is jointly Gaussian given $Y$.

\emph{Regression interpretation of $\lambda$.} The population regression coefficient of $\ell_H$ on $\ell_A$ given $Y$ is $\Cov(\ell_H, \ell_A \mid Y) / \Var(\ell_A \mid Y) = 2\tau_A\lambda / (2\tau_A) = \lambda$.
\end{proof}

\begin{proof}[Proof of Proposition~\ref{prop:overlap-domain}]
Fix $\mathcal{X}_H$ with $|\mathcal{X}_H| = n_H$, and let $\mathcal{X}_A$ be a uniformly random subset of size $n_A$ drawn without replacement from $\{1,\ldots,N\}$. Write the overlap count as $n_\cap = \sum_{i \in \mathcal{X}_H} \mathbf{1}\{i \in \mathcal{X}_A\}$. By symmetry of the uniform draw, $P(i \in \mathcal{X}_A) = n_A/N$ for each $i$. Hence
\[
\mathbb{E}[n_\cap] = n_H \cdot \frac{n_A}{N}, \qquad \mathbb{E}[\lambda] = \frac{\mathbb{E}[n_\cap]}{n_A} = \frac{n_H}{N},
\]
independent of $n_A$.
\end{proof}

%\begin{proof}[Proof of Proposition~\ref{prop:overlap-domain}]
%Fix $\mathcal{X}_H$ with $|\mathcal{X}_H| = n_H$. Let $\mathcal{X}_A^{(n)}$ be a uniformly random subset of size $n$ drawn without replacement from $\{1,\ldots,N\}$, and let $n_\cap^{(n)} := |\mathcal{X}_H \cap \mathcal{X}_A^{(n)}|$.
%
%\emph{Expectation.} Write $n_\cap^{(n)} = \sum_{i \in \mathcal{X}_H} \mathbf{1}\{i \in \mathcal{X}_A^{(n)}\}$. By symmetry of the uniform draw, $P(i \in \mathcal{X}_A^{(n)}) = n/N$ for each $i$. Therefore
%\[
%\mathbb{E}[n_\cap^{(n)}] = n_H \cdot \frac{n}{N}, \qquad \mathbb{E}[\lambda^{(n)}] = \frac{\mathbb{E}[n_\cap^{(n)}]}{n} = \frac{n_H}{N},
%\]
%independent of $n$.
%
%\emph{Concentration and almost-sure convergence.} The variable $n_\cap^{(n)}$ follows the hypergeometric distribution with parameters $(N, n_H, n)$, with variance
%\[
%\mathrm{Var}(n_\cap^{(n)}) = n \cdot \frac{n_H}{N} \cdot \left(1 - \frac{n_H}{N}\right) \cdot \frac{N - n}{N - 1} \le n.
%\]
%Hence $\mathrm{Var}(\lambda^{(n)}) = \mathrm{Var}(n_\cap^{(n)})/n^2 \le 1/n$. By Hoeffding's inequality for sampling without replacement (Serfling, 1974),
%\[
%P\!\left(|\lambda^{(n)} - n_H/N| > \varepsilon\right) \le 2\exp(-2\varepsilon^2 n).
%\]
%The right-hand side is summable in $n$, so by Borel-Cantelli, $\lambda^{(n)} \to n_H/N$ almost surely as $n \to \infty$.
%\end{proof}

\begin{proof}[Proof of Lemma~\ref{lem:behavioral}]
We construct the rule in two steps, corresponding to parts~(i) and~(ii) of Assumption~\ref{ass:worker}.

\emph{Step 1: Bayesian rule under perceived overlap.} The worker's belief about the joint distribution of $(\ell_H, \ell_A, Y)$ replaces the true overlap $\lambda$ with the perceived overlap $\hat\alpha\lambda$ in Lemma~\ref{lem:joint}, retaining the true capabilities $(\tau_H, \tau_A)$. Her believed conditional joint distribution is therefore Gaussian with mean $(2Y-1)(\tau_H, \tau_A)^\top$ and covariance
\[
\widehat\Sigma = \begin{pmatrix} 2\tau_H & 2\tau_A\hat\alpha\lambda \\ 2\tau_A\hat\alpha\lambda & 2\tau_A \end{pmatrix}.
\]
Under the symmetric prior, the worker's believed posterior log-odds equals the believed log-likelihood ratio, which by the standard Gaussian formula is
\[
\hat\ell^{\mathrm{Bayes}} = 2\widehat\mu_1^\top \widehat\Sigma^{-1} (\ell_H, \ell_A)^\top, \qquad \widehat\mu_1 := (\tau_H, \tau_A)^\top.
\]
The determinant is $\det\widehat\Sigma = 4\tau_A(\tau_H - \tau_A\hat\alpha^2\lambda^2) =: 4\tau_A D$, where $D := \tau_H - \tau_A\hat\alpha^2\lambda^2$. The inverse is
\[
\widehat\Sigma^{-1} = \frac{1}{2D}\begin{pmatrix} 1 & -\hat\alpha\lambda \\ -\hat\alpha\lambda & \tau_H/\tau_A \end{pmatrix}.
\]
The row vector $2\widehat\mu_1^\top \widehat\Sigma^{-1}$ has components
\[
\frac{\tau_H - \tau_A\hat\alpha\lambda}{D} \quad \text{and} \quad \frac{\tau_H(1-\hat\alpha\lambda)}{D},
\]
giving the Bayesian-rule weights
\[
w_H^{\mathrm{Bayes}}(\hat\alpha) = \frac{\tau_H - \tau_A\hat\alpha\lambda}{\tau_H - \tau_A\hat\alpha^2\lambda^2}, \qquad w_A^{\mathrm{Bayes}}(\hat\alpha) = \frac{\tau_H(1-\hat\alpha\lambda)}{\tau_H - \tau_A\hat\alpha^2\lambda^2}.
\]

\emph{Step 2: Trust discount on the AI signal.} By part~(ii) of Assumption~\ref{ass:worker}, the worker applies a multiplicative trust factor $\hat\beta$ to the AI signal contribution, leaving the human-side weight unchanged. Her decision rule is
\[
\hat\ell = w_H^{\mathrm{Bayes}}(\hat\alpha) \cdot \ell_H + \hat\beta \cdot w_A^{\mathrm{Bayes}}(\hat\alpha) \cdot \ell_A,
\]
which is linear in $(\ell_H, \ell_A)$ and yields the weights stated in the lemma.
\end{proof}

\section{Proofs for Section~\ref{sec:analysis}}
\label{app:section2-proofs}

\begin{proof}[Proof of Lemma~\ref{lem:accuracy}]
The worker's decision statistic is $\hat\ell = w_H \ell_H + w_A \ell_A$, a linear combination of the jointly Gaussian pair $(\ell_H, \ell_A)$ from Lemma~\ref{lem:joint}. Hence $\hat\ell \mid Y$ is Gaussian. We compute its conditional mean and standard deviation, then derive the accuracy.

\emph{Conditional mean.} Using $\mathbb{E}[\ell_H \mid Y] = (2Y-1)\tau_H$ and $\mathbb{E}[\ell_A \mid Y] = (2Y-1)\tau_A$ from Lemma~\ref{lem:joint},
\[
\mathbb{E}[\hat\ell \mid Y] = (2Y-1)(w_H \tau_H + w_A \tau_A).
\]

\emph{Conditional variance.} Using $\Var(\ell_H \mid Y) = 2\tau_H$, $\Var(\ell_A \mid Y) = 2\tau_A$, and $\Cov(\ell_H, \ell_A \mid Y) = 2\tau_A \lambda$ from Lemma~\ref{lem:joint},
\[
\Var(\hat\ell \mid Y) = w_H^2 \cdot 2\tau_H + w_A^2 \cdot 2\tau_A + 2 w_H w_A \cdot 2\tau_A\lambda = 2\bigl(w_H^2 \tau_H + w_A^2 \tau_A + 2 w_H w_A \tau_A \lambda\bigr).
\]

\emph{Accuracy.} Classification is correct when $\hat\ell$ has the same sign as $(2Y-1)$. Standardizing $\hat\ell$ given $Y = 1$:
\[
P(\hat\ell > 0 \mid Y = 1) = P\!\left(Z > -\frac{w_H \tau_H + w_A \tau_A}{\sqrt{\Var(\hat\ell \mid Y)}}\right) = \Phi\!\left(\frac{w_H \tau_H + w_A \tau_A}{\sqrt{\Var(\hat\ell \mid Y)}}\right) = \Phi(r_{HA}),
\]
where $Z \sim \mathcal{N}(0, 1)$ and $r_{HA} = (w_H \tau_H + w_A \tau_A)/\sqrt{\Var(\hat\ell \mid Y)}$. By symmetry of the conditional distributions, $P(\hat\ell < 0 \mid Y = 0) = \Phi(r_{HA})$ as well. Under the symmetric prior, the unconditional accuracy is
\[
\mathcal{A}_{HA} = \tfrac{1}{2}P(\hat\ell > 0 \mid Y = 1) + \tfrac{1}{2}P(\hat\ell < 0 \mid Y = 0) = \Phi(r_{HA}). \qedhere
\]
\end{proof}

\begin{proof}[Proof of Lemma~\ref{lem:switching}]
AI assistance changes the worker's decision when $\sgn(\hat\ell) \neq \sgn(\ell_H)$:
\[
\mathcal{S} = P(\ell_H > 0, \hat\ell < 0) + P(\ell_H < 0, \hat\ell > 0).
\]
The pair $(\ell_H, \hat\ell)$ is jointly Gaussian conditional on $Y$, with conditional means $((2Y-1)\tau_H, (2Y-1) m_{HA})$, conditional variances $(2\tau_H, \sigma_{HA}^2)$ where $m_{HA}, \sigma_{HA}^2$ are as in the proof of Lemma~\ref{lem:accuracy}, and conditional covariance
\[
\Cov(\ell_H, \hat\ell \mid Y) = w_H \cdot 2\tau_H + w_A \cdot 2\tau_A \lambda = 2(w_H \tau_H + w_A \tau_A \lambda).
\]
The conditional correlation between $\ell_H$ and $\hat\ell$ given $Y$ is therefore
\[
\rho = \frac{\Cov(\ell_H, \hat\ell \mid Y)}{\sqrt{\Var(\ell_H \mid Y) \Var(\hat\ell \mid Y)}} = \frac{w_H \tau_H + w_A \tau_A \lambda}{\sqrt{\tau_H (w_H^2 \tau_H + w_A^2 \tau_A + 2 w_H w_A \tau_A \lambda)}}.
\]
By symmetry of the conditional distributions under $Y = 1$ versus $Y = 0$ and the symmetric prior, $\mathcal{S}$ equals twice the conditional probability of the disagreement event under $Y = 1$. Standardizing $\ell_H$ and $\hat\ell$ to mean zero and unit variance under $Y = 1$:
\begin{align*}
\ell_H > 0 &\iff Z_H > -\sqrt{\tau_H/2}, \\
\hat\ell > 0 &\iff Z_{HA} > -r_{HA},
\end{align*}
where $(Z_H, Z_{HA})$ is jointly standard normal with correlation $\rho$ and $r_{HA} = m_{HA}/\sigma_{HA}$ from Lemma~\ref{lem:accuracy}. Using the identity $P(Z_1 > a, Z_2 < b) = \Phi(b) - \Phi_2(a, b; \rho)$ for jointly standard normal $(Z_1, Z_2)$ with correlation $\rho$, and similarly for the symmetric event,
\begin{align*}
P(\ell_H > 0, \hat\ell < 0 \mid Y=1) &= \Phi(-r_{HA}) - \Phi_2(-\sqrt{\tau_H/2}, -r_{HA}; \rho), \\
P(\ell_H < 0, \hat\ell > 0 \mid Y=1) &= \Phi(-\sqrt{\tau_H/2}) - \Phi_2(-r_{HA}, -\sqrt{\tau_H/2}; \rho).
\end{align*}
Adding and using $\Phi(-x) = 1 - \Phi(x)$ and the symmetry $\Phi_2(-a, -b; \rho) = \Phi_2(a, b; \rho) - \Phi(a) - \Phi(b) + 1$ (from rotation-invariance of the standard bivariate normal),
\[
\mathcal{S} = \Phi(\sqrt{\tau_H/2}) + \Phi(r_{HA}) - 2\Phi_2(\sqrt{\tau_H/2}, r_{HA}; \rho). \qedhere
\]
\end{proof}

\begin{proof}[Proof of Theorem~\ref{thm:phase}]
Under $\hat\alpha = 0$ and the symmetric prior, the worker's decision statistic is $\hat\ell = \ell_H + \hat\beta \ell_A$. The three modality accuracies are determined by their respective signal-to-noise ratios via Lemma~\ref{lem:accuracy}. Since $\Phi$ is strictly increasing, comparing accuracies reduces to comparing SNRs. Denote the three modality SNRs by $r_H, r_A, r_{HA}$. Direct substitution into Lemma~\ref{lem:accuracy} gives
\[
r_H = \sqrt{\tau_H/2}, \qquad r_A = \sqrt{\tau_A/2}, \qquad r_{HA} = \frac{\tau_H + \hat\beta\tau_A}{\sqrt{2\bigl(\tau_H + \hat\beta(\hat\beta + 2\lambda)\tau_A\bigr)}},
\]
where $r_{HA}$ uses $w_H = 1, w_A = \hat\beta$ from Lemma~\ref{lem:behavioral} at $\hat\alpha = 0$.

The proof proceeds in five steps: derive the augmentation threshold (Step~1), the automation threshold (Step~2), the human-vs-AI comparison (Step~3), partition the deployment space by overlap regime (Step~4), and verify monotonicity of the thresholds (Step~5).

\emph{Step 1: Augmentation threshold.} We derive the condition $r_{HA} > r_H$. Since both are positive, this is equivalent to $r_{HA}^2 > r_H^2$, i.e.,
\[
\frac{(\tau_H + \hat\beta\tau_A)^2}{2(\tau_H + \hat\beta(\hat\beta + 2\lambda)\tau_A)} > \frac{\tau_H}{2}.
\]
The denominator $\tau_H + \hat\beta(\hat\beta + 2\lambda)\tau_A > 0$ for all $(\tau_H, \tau_A) > 0$ and $\lambda \ge 0$. Multiplying through and expanding:
\begin{align*}
(\tau_H + \hat\beta\tau_A)^2 &> \tau_H(\tau_H + \hat\beta(\hat\beta + 2\lambda)\tau_A) \\
\Leftrightarrow \quad \tau_H^2 + 2\hat\beta\tau_H\tau_A + \hat\beta^2\tau_A^2 &> \tau_H^2 + \hat\beta^2\tau_H\tau_A + 2\hat\beta\lambda\tau_H\tau_A \\
\Leftrightarrow \quad \hat\beta\tau_A\bigl[\hat\beta\tau_A + (2 - \hat\beta - 2\lambda)\tau_H\bigr] &> 0.
\end{align*}
Since $\hat\beta\tau_A > 0$, this reduces to $\hat\beta\tau_A > (\hat\beta + 2\lambda - 2)\tau_H$, equivalently $\tau_A > \tau_{\mathrm{aug}}(\lambda; \hat\beta) := ((\hat\beta + 2\lambda - 2)/\hat\beta)\tau_H$. The threshold is positive iff $\hat\beta + 2\lambda - 2 > 0$, i.e., $\lambda > 1 - \hat\beta/2$. When $\lambda \le 1 - \hat\beta/2$, the threshold is non-positive and $r_{HA} > r_H$ holds for every $\tau_A > 0$.

\emph{Step 2: Automation threshold.} We derive the condition $r_{HA} > r_A$. By the same squaring,
\[
\frac{(\tau_H + \hat\beta\tau_A)^2}{2(\tau_H + \hat\beta(\hat\beta + 2\lambda)\tau_A)} > \frac{\tau_A}{2},
\]
and multiplying through gives
\begin{align*}
(\tau_H + \hat\beta\tau_A)^2 &> \tau_A(\tau_H + \hat\beta(\hat\beta + 2\lambda)\tau_A) \\
\tau_H^2 + 2\hat\beta\tau_H\tau_A + \hat\beta^2\tau_A^2 &> \tau_H\tau_A + \hat\beta^2\tau_A^2 + 2\hat\beta\lambda\tau_A^2 \\
\tau_H^2 + (2\hat\beta - 1)\tau_H\tau_A - 2\hat\beta\lambda\tau_A^2 &> 0. \quad (\star)
\end{align*}
For $\lambda > 0$, $(\star)$ is a quadratic in $\tau_A$ with leading coefficient $-2\hat\beta\lambda < 0$ (downward parabola). At $\tau_A = 0$, the LHS equals $\tau_H^2 > 0$; as $\tau_A \to \infty$, the LHS tends to $-\infty$. By continuity, the quadratic has a unique positive root $\tau_{\mathrm{auto}}(\lambda; \hat\beta)$, with $(\star)$ holding iff $\tau_A < \tau_{\mathrm{auto}}$. The quadratic formula gives
\[
\tau_{\mathrm{auto}}(\lambda; \hat\beta) = \frac{(2\hat\beta - 1) + \sqrt{(2\hat\beta - 1)^2 + 8\hat\beta\lambda}}{4\hat\beta\lambda}\,\tau_H,
\]
where the positive sign is selected because the product of the two roots is $-\tau_H^2/(2\hat\beta\lambda) < 0$, forcing one positive and one negative root, and we take the positive one.

For $\lambda = 0$, $(\star)$ reduces to $\tau_H^2 + (2\hat\beta - 1)\tau_H\tau_A > 0$, equivalently $\tau_H + (2\hat\beta - 1)\tau_A > 0$. If $\hat\beta \ge 1/2$, both terms are non-negative and the condition holds for every $\tau_A > 0$, so $\tau_{\mathrm{auto}}(0; \hat\beta) = +\infty$. If $\hat\beta < 1/2$, the condition becomes $\tau_A < \tau_H/(1 - 2\hat\beta)$, giving $\tau_{\mathrm{auto}}(0; \hat\beta) = \tau_H/(1-2\hat\beta)$.

\emph{Step 3: Comparing $r_H$ and $r_A$.} Since $r_H = \sqrt{\tau_H/2}$ and $r_A = \sqrt{\tau_A/2}$, $r_H > r_A \iff \tau_H > \tau_A$.

\emph{Step 4: Region partitioning.} For each $\lambda$, the three score relationships partition $\tau_A > 0$ into the three regimes.

\emph{Low overlap}, $0 \le \lambda \le 1 - \hat\beta/2$ (requires $\hat\beta \le 2$): Step~1 gives $r_{HA} > r_H$ for every $\tau_A > 0$, so $H$ is never the top modality. We claim $\tau_{\mathrm{auto}} > \tau_H$ in this regime. Evaluate $(\star)$ at $\tau_A = \tau_H$: the LHS becomes $2\hat\beta(1 - \lambda)\tau_H^2 > 0$ for $\lambda < 1$, so $r_{HA} > r_A$ at $\tau_A = \tau_H$, giving $\tau_H < \tau_{\mathrm{auto}}$. The choice between $HA$ and $AI$ then follows from Step~2:
\begin{itemize}
\item $\tau_A < \tau_{\mathrm{auto}}$: $r_{HA} > r_A$ and $r_{HA} > r_H$. Complementarity ($HA$ wins).
\item $\tau_A \ge \tau_{\mathrm{auto}}$: $r_A \ge r_{HA}$, and $\tau_A \ge \tau_{\mathrm{auto}} > \tau_H$ gives $r_A > r_H$ (Step~3). Automation ($AI$ wins).
\end{itemize}

\emph{Moderate overlap}, $1 - \hat\beta/2 < \lambda < 1$: Step~1 gives a positive augmentation threshold $\tau_{\mathrm{aug}} > 0$. We show $\tau_{\mathrm{aug}} < \tau_H < \tau_{\mathrm{auto}}$. The first inequality follows from $\hat\beta + 2\lambda - 2 < \hat\beta$ (since $\lambda < 1$), so $\tau_{\mathrm{aug}} = (\hat\beta + 2\lambda - 2)/\hat\beta \cdot \tau_H < \tau_H$. For the second, evaluate $(\star)$ at $\tau_A = \tau_H$: the LHS becomes
\[
\tau_H^2 + (2\hat\beta-1)\tau_H^2 - 2\hat\beta\lambda\tau_H^2 = 2\hat\beta(1 - \lambda)\tau_H^2 > 0
\]
for $\lambda < 1$. So $r_{HA} > r_A$ at $\tau_A = \tau_H$, meaning $\tau_H < \tau_{\mathrm{auto}}$. The three regimes follow:
\begin{itemize}
\item $\tau_A < \tau_{\mathrm{aug}}$: $r_H > r_{HA}$ (Step~1) and $\tau_A < \tau_H$ implies $r_H > r_A$ (Step~3). Impairment ($H$ wins).
\item $\tau_{\mathrm{aug}} \le \tau_A \le \tau_{\mathrm{auto}}$: $r_{HA} \ge r_H$ and $r_{HA} \ge r_A$. Complementarity ($HA$ wins).
\item $\tau_A > \tau_{\mathrm{auto}}$: $r_A > r_{HA}$ and $\tau_A > \tau_H$ (since $\tau_{\mathrm{auto}} > \tau_H$) gives $r_A > r_H$. Automation ($AI$ wins).
\end{itemize}

\emph{High overlap}, $\lambda = 1$: substituting into $\tau_{\mathrm{aug}}$, $\tau_{\mathrm{aug}}(1; \hat\beta) = ((\hat\beta + 2 - 2)/\hat\beta) \tau_H = \tau_H$. Substituting into $\tau_{\mathrm{auto}}$, the radical equals $\sqrt{(2\hat\beta - 1)^2 + 8\hat\beta} = \sqrt{(2\hat\beta + 1)^2} = 2\hat\beta + 1$ (since $\hat\beta > 0$ gives $2\hat\beta + 1 > 0$), so $\tau_{\mathrm{auto}}(1; \hat\beta) = ((2\hat\beta - 1) + (2\hat\beta + 1))/(4\hat\beta) \cdot \tau_H = \tau_H$. Both thresholds equal $\tau_H$ at $\lambda = 1$, and the complementarity region collapses to the single point $\tau_A = \tau_H$. At this point, direct substitution gives $r_H = r_A = r_{HA} = \sqrt{\tau_H/2}$ (a triple tie). For $\tau_A < \tau_H$: $r_H > r_{HA}$ (Step~1, since $\tau_{\mathrm{aug}} = \tau_H$ and $\tau_A < \tau_{\mathrm{aug}}$) and $r_H > r_A$ (Step~3), so $H$ wins. For $\tau_A > \tau_H$: $r_A > r_{HA}$ (Step~2, since $\tau_{\mathrm{auto}} = \tau_H$ and $\tau_A > \tau_{\mathrm{auto}}$) and $r_A > r_H$ (Step~3), so $AI$ wins.

\emph{Step 5: Monotonicity of thresholds.} For the augmentation threshold,
\[
\frac{\partial \tau_{\mathrm{aug}}}{\partial \lambda} = \frac{2\tau_H}{\hat\beta} > 0,
\]
so $\tau_{\mathrm{aug}}$ is strictly increasing in $\lambda$ on its positive domain $(1 - \hat\beta/2, 1]$.

For the automation threshold, write $\tau_{\mathrm{auto}}(\lambda; \hat\beta)/\tau_H = f(\lambda)/g(\lambda)$ with $f(\lambda) = (2\hat\beta - 1) + S(\lambda)$, $g(\lambda) = 4\hat\beta\lambda$, and $S(\lambda) := \sqrt{(2\hat\beta-1)^2 + 8\hat\beta\lambda}$. Direct differentiation:
\[
f'(\lambda) = \frac{4\hat\beta}{S(\lambda)}, \qquad g'(\lambda) = 4\hat\beta,
\]
and by the quotient rule,
\[
\frac{d}{d\lambda}\frac{f}{g} = \frac{f'g - fg'}{g^2} = \frac{(4\hat\beta/S)(4\hat\beta\lambda) - (4\hat\beta) f}{16\hat\beta^2\lambda^2} = \frac{4\hat\beta\lambda - f \cdot S}{4\hat\beta\lambda^2 \cdot S}.
\]
The denominator is positive, so the sign of the derivative equals the sign of $4\hat\beta\lambda - fS$. Computing $fS$ using $S^2 = (2\hat\beta-1)^2 + 8\hat\beta\lambda$:
\[
fS = ((2\hat\beta-1) + S)S = (2\hat\beta-1)S + S^2 = (2\hat\beta-1)S + (2\hat\beta-1)^2 + 8\hat\beta\lambda.
\]
Hence
\[
4\hat\beta\lambda - fS = -(2\hat\beta-1)S - (2\hat\beta-1)^2 - 4\hat\beta\lambda.
\]
We show this is strictly negative for $\lambda > 0$ in two cases.

\emph{Case $\hat\beta \ge 1/2$:} all three terms on the RHS are non-positive (since $S > 0$, $(2\hat\beta-1)^2 \ge 0$, $4\hat\beta\lambda > 0$), and at least the last is strictly negative, so the sum is strictly negative.

\emph{Case $\hat\beta < 1/2$:} $-(2\hat\beta-1) > 0$, so the first term is positive. We must verify that $|2\hat\beta-1| \cdot S < (2\hat\beta-1)^2 + 4\hat\beta\lambda$. Both sides are positive; squaring and using $S^2 = (2\hat\beta-1)^2 + 8\hat\beta\lambda$, the inequality reduces (with $u := (2\hat\beta-1)^2$ and $v := 4\hat\beta\lambda$) to
\[
u(u + 2v) < (u + v)^2 = u^2 + 2uv + v^2 \quad \Longleftrightarrow \quad 0 < v^2,
\]
which holds for $\lambda > 0$.

In both cases, $4\hat\beta\lambda - fS < 0$ for $\lambda > 0$, so $\partial \tau_{\mathrm{auto}}/\partial \lambda < 0$ on $(0, 1]$. Hence $\tau_{\mathrm{auto}}$ is strictly decreasing in $\lambda$ on $(0, 1]$.

The coincidence $\tau_{\mathrm{aug}}(1; \hat\beta) = \tau_{\mathrm{auto}}(1; \hat\beta) = \tau_H$ is verified in Step~4 (high-overlap case).

\end{proof}
\begin{proof}[Proof of Proposition~\ref{prop:welfare}]

\emph{Step 1: General split.} Adding and subtracting $\mathcal{A}(\hat\alpha, 1)$ inside $\mathrm{TWL}$ gives
\[
\mathrm{TWL}(\hat\alpha, \hat\beta) = \mathcal{A}^{\mathrm{BR}} - \mathcal{A}(\hat\alpha, \hat\beta) = \bigl[\mathcal{A}^{\mathrm{BR}} - \mathcal{A}(\hat\alpha, 1)\bigr] + \bigl[\mathcal{A}(\hat\alpha, 1) - \mathcal{A}(\hat\alpha, \hat\beta)\bigr].
\]
The first bracket is $\mathrm{CNC}(\hat\alpha)$ by definition; the second is $\mathrm{UTC}(\hat\alpha, \hat\beta)$.

\emph{Step 2: Sign of $\mathrm{CNC}$.} The Bayes-rational SNR $r^*$ is defined as the maximum of $r_{HA}(\hat\alpha, \hat\beta)$ over all weight choices, so $r^* \ge r_{HA}(\hat\alpha, 1)$ for any deployment, giving $\mathrm{CNC}(\hat\alpha) \ge 0$.

\emph{Step 3: Sign of $\mathrm{UTC}$ for $\hat\beta \in (0, 1]$ and $\tau_A \ge \tau_H$.} Fix $\hat\alpha \in [0, 1]$ and write $u := w_H(\hat\alpha)$ and $v := w_A(\hat\alpha, 1)$. Within the cue-pool feasible region, $u, v \ge 0$ and $\tau_H - \tau_A\lambda \ge 0$. The worker's SNR is
\[
r_{HA}(\hat\alpha, \hat\beta) = \frac{u\tau_H + \hat\beta v\tau_A}{\sqrt{2(u^2\tau_H + \hat\beta^2 v^2\tau_A + 2u\hat\beta v\tau_A\lambda)}}.
\]
Differentiating $r_{HA}^2$ with respect to $\hat\beta$ gives a nonnegative multiple of the linear function $g(\hat\beta) = u\tau_H(1-\lambda) - \hat\beta v(\tau_H - \tau_A\lambda)$. Substituting the weights from Lemma~\ref{lem:behavioral} gives
\[
g(1) = \frac{\tau_H\lambda(1-\hat\alpha)(\tau_A-\tau_H)}{\tau_H-\tau_A\hat\alpha^2\lambda^2} \ge 0.
\]
For $\hat\beta \in (0, 1]$, we therefore have $g(\hat\beta) = g(1) + (1-\hat\beta)v(\tau_H-\tau_A\lambda) \ge 0$. Thus $r_{HA}(\hat\alpha, \cdot)$ is nondecreasing on $(0, 1]$, so $r_{HA}(\hat\alpha, \hat\beta) \le r_{HA}(\hat\alpha, 1)$, hence $\mathrm{UTC}(\hat\alpha, \hat\beta) \ge 0$; it vanishes when $\hat\beta = 1$. The degenerate case $\tau_H = \tau_A$ and $\lambda = 1$ follows by continuity, since the two signals coincide.
\end{proof}

\section{Generalization of the cue-pool construction}
\label{app:heterogeneous-cues}

First, the assumption of equal per-cue capability can be relaxed while retaining conditional independence and Gaussian signals. Suppose
\[
X_k \mid Y = y \sim \mathcal{N}(\mu_{yk}, \sigma_k^2),
\qquad
\theta_k := \frac{(\mu_{1k}-\mu_{0k})^2}{2\sigma_k^2},
\]
where $\sigma_k^2 > 0$ is common across states but may differ across cues. This permits some cues to be highly informative and others nearly uninformative. Define
\[
\tau_B := \sum_{k \in \mathcal{X}_B}\theta_k,
\qquad
\tau_\cap := \sum_{k \in \mathcal{X}_H \cap \mathcal{X}_A}\theta_k,
\qquad
\lambda := \frac{\tau_\cap}{\tau_A},
\]
for $B \in \{H,A\}$, assuming positive agent capabilities. The overlap coefficient now measures the fraction of AI capability contributed by shared cues.

\begin{proposition}[Heterogeneous cues]
\label{prop:heterogeneous-cues}
For fixed access sets, the joint conditional distribution retains the form of Lemma~\ref{lem:joint}:
\[
\begin{pmatrix}\ell_H \\ \ell_A\end{pmatrix}\Bigm|Y
\sim
\mathcal{N}\!\left(
(2Y-1)\begin{pmatrix}\tau_H \\ \tau_A\end{pmatrix},
\begin{pmatrix}
2\tau_H & 2\tau_A\lambda \\
2\tau_A\lambda & 2\tau_A
\end{pmatrix}
\right).
\]
\end{proposition}

\begin{proof}
Cue $k$'s log-likelihood ratio is Gaussian with conditional mean $(2Y-1)\theta_k$ and variance $2\theta_k$. Conditional independence implies that their sums are jointly Gaussian, with agent variances $2\tau_B$ and covariance $2\tau_\cap = 2\tau_A\lambda$.
\end{proof}

Consequently, the decision rule, phase diagram, identification results, and welfare decomposition retain their forms under their existing assumptions, with $\lambda$ interpreted as capability-weighted overlap. Heterogeneity concerns informativeness; individual cue realizations can still favor the wrong state.

The uniform-sampling argument in Proposition~\ref{prop:overlap-domain} also has a weighted analogue. Fix $\mathcal{X}_H$, let $\tau_N := \sum_{k=1}^{N}\theta_k$, and sample $\mathcal{X}_A$ uniformly among subsets of size $n_A$, independently of cue realizations. Each cue has inclusion probability $n_A/N$, so
\[
\frac{\mathbb{E}[\tau_\cap]}{\mathbb{E}[\tau_A]}
=
\frac{(n_A/N)\tau_H}{(n_A/N)\tau_N}
=
\frac{\tau_H}{\tau_N}.
\]
This ratio equals the human's capability-weighted coverage of the domain and is independent of $n_A$. With heterogeneous capabilities, this is a ratio-of-expectations statement; it does not generally imply $\mathbb{E}[\lambda] = \tau_H/\tau_N$.

Second, we note that the accuracy and phase-diagram results depend only on the conditional joint-Gaussian law of $(\ell_H, \ell_A)$ in Lemma~\ref{lem:joint}, not on the cue-pool construction that produces it. That law arises as a Lyapunov-CLT type limit of aggregating many conditionally independent cues; for finite cue pools or heavy-tailed cues it holds only approximately.
\section{Asymmetric Priors}  \label{sec:appx_asymmetric_priors}
This section extends our model to arbitrary prior probability $\pi := \Pr \left(Y=1\right) \in [0,1]$. We show that the qualitative structure that AI can impair the human or make automation optimal survives at every prior, with one sufficient condition on prior asymmetry (which we can define exactly) for establishing the complementarity region analogously to Theorem~\ref{thm:phase}.

Under prior $\pi$, we retain the definition of primitive cues $X_k$, each agent's log-likelihood ratios $\ell_H$ and $\ell_A$, and the definition of overlap $\lambda$, correlation awareness $\hat{\alpha}$ and trust $\hat{\beta}$. Note that each agent $B\in \{H,A\}$'s posterior log-odds equals $\ell_{\mathrm{prior}} + \ell_B$, where $\ell_{\mathrm{prior}} := \log \left( \pi / (1-\pi) \right)$ is the prior log-odds. We can show that $\left( \ell_H, \ell_A \right) \mid Y$ follows the same joint distribution as Lemma~\ref{lem:joint}, and that a worker with behavior given by Assumption~\ref{ass:worker} uses the linear rule $\hat{\ell} = \ell_{\mathrm{prior}} + w_H \ell_H + w_A \ell_A$, where $w_H$ and $w_A$ follow the same definition as Lemma~\ref{lem:behavioral}, but with the additive term $\ell_{\mathrm{prior}}$ in her decision statistic $\hat{\ell}$. The decision that maximizes accuracy (minimizes 0-1 loss) is $ \hat{Y} := \mathbf{1} \{ \hat{\ell} > 0 \} $.

Unlike with symmetric priors (Lemma~\ref{lem:accuracy}), the expected accuracy sums over two Gaussian CDFs:
\begin{lemma}[Accuracy under arbitrary priors]
	Under Assumption~\ref{ass:worker} and any prior $\pi$, the worker classifying via $ \hat{Y} := \mathbf{1} \{ \hat{\ell} > 0 \} $ attains accuracy
	\begin{align*}
		\mathcal{A}_{HA} \left( \hat{\alpha}, \hat{\beta}; \pi, \tau_H, \tau_A, \lambda \right) 
		&= 
		\pi 
		\Phi \left( \frac {m_{HA} + \ell_{\mathrm{prior}}} {\sigma_{HA}} \right)
		+ \left( 1 - \pi \right) 
		\Phi \left( \frac {m_{HA} - \ell_{\mathrm{prior}}} {\sigma_{HA}} \right),
	\end{align*}
	for any $\hat{\alpha} \in \left[ 0,1 \right], \hat{\beta} > 0$, where
	\begin{align*}
		m_{HA} &= w_H \tau_H + w_A \tau_A,
		&
		\sigma_{HA} &= \sqrt{
			2 \left( w_H^2 \tau_H + w_A^2 \tau_A + 2 w_H w_A \tau_A \lambda \right)
		}.
	\end{align*}
\end{lemma}
Note that $m_{HA} / \sigma_{HA} = r_{HA}$ is the same signal-to-noise ratio as Lemma~\ref{lem:accuracy}.

\subsection{Impairment and Automation Transfer Under All Priors}

\begin{proposition}[Single-agent dominance under symmetric prior transfer to all priors]  \label{prop:single_agent_dominance_transfer}
	Fix $\hat{\alpha} = 0$ and any $\hat{\beta} > 0$. For any admissible $\left( \tau_H, \tau_A, \lambda \right)$ and any prior $\pi$, the expected accuracies as functions of prior $\pi$ satisfy
	\begin{align*}
		\mathcal{A}_{HA} \left( 1/2 \right) 
		&\leq \mathcal{A}_{H} \left( 1/2 \right)
		& 
		&\implies &
		\mathcal{A}_{HA} \left( \pi \right) 
		&\leq \mathcal{A}_{H} \left( \pi \right),
		\qquad
		\forall \pi \in (0,1),
		% \label{eqn:single_agent_dominance_transfer_H}
	\end{align*}
	and
	\begin{align*}
		\mathcal{A}_{HA} \left( 1/2 \right) 
		&\leq \mathcal{A}_{A} \left( 1/2 \right)
		&
		&\implies &
		\mathcal{A}_{HA} \left( \pi \right) 
		&\leq \mathcal{A}_{A} \left( \pi \right),
		\qquad
		\forall \pi \in (0,1).
		% \label{eqn:single_agent_dominance_transfer_A}
	\end{align*}
    This means if the team's accuracy is weakly dominated by the human alone or by the AI alone at prior $1/2$, then the same dominance also holds at $\pi$.
	% Each implication becomes an equivalence if and only if $\pi = 1/2$.
	% For any fixed $\hat{\beta}, \tau_H>0$, they characterize single-agent dominance over all admissible deployments if and only if $\pi = 1/2$.
	% 
	% In both results, equality in the right-hand-side (RHS) condition requires equality in the left-hand-side (LHS) as a necessary condition. If $\priorZ \neq 1/2$, RHS equality in \eqref{eqn:single_agent_dominance_transfer_A} holds iff $\betabelief < 1$, $\ovlpnmlz = \left( 1 - \betabelief \right) / 2$ and $\tau_A = \tau_H / \left( 1 - \betabelief \right)$; whereas the RHS of \eqref{eqn:single_agent_dominance_transfer_H} is strict under $\priorZ \neq 1/2$.
\end{proposition}

\begin{corollary}[Impairment and automation regions under any prior are supersets of symmetric-prior regions]  \label{coro:impairment_automation_any_prior_supersets}
	Fix $\hat{\alpha} = 0$ and any $\hat{\beta}, \tau_H > 0$.
    Let $\tau^*_{\mathrm{aug}} \left( \lambda ; \hat{\beta} \right)$ and $\tau^*_{\mathrm{auto}} \left( \lambda ; \hat{\beta} \right)$ denote the symmetric-prior thresholds in Theorem \ref{thm:phase}, namely
	\begin{align*}
		\tau^*_{\mathrm{aug}} \left( \lambda ; \hat{\beta} \right)
		&:= 
        \textstyle
		\frac{ \hat{\beta} + 2\lambda - 2 }{ \hat{\beta} } \tau_H,
		&
		\tau^*_{\mathrm{auto}} \left( \lambda ; \hat{\beta} \right)
		&:= 
        \textstyle
		\frac
		{ 2 \hat{\beta} - 1 + \sqrt{ \left( 2 \hat{\beta} - 1 \right)^2 + 8 \hat{\beta} \lambda } }
		{ 4 \hat{\beta} \lambda } 
		\tau_H
		\quad
		\text{for } \lambda > 0,
	\end{align*}
	with $\tau^*_{\mathrm{auto}}(0; \hat\beta) = +\infty$ if $\hat\beta \ge 1/2$ and $\tau_H/(1-2\hat\beta)$ otherwise.
	Then, for all priors $\pi \in (0,1)$ and all admissible $\left( \lambda, \tau_A \right)$:
	\begin{itemize}
		\item \emph{Low overlap ($\lambda \leq 1 - \hat{\beta} / 2$, when $\hat{\beta} < 2$):} automation if $\tau_A \geq \tau^*_{\mathrm{auto}} \left( \lambda ; \hat{\beta} \right)$.
		\item \emph{Moderate or high overlap ($1 - \hat{\beta} / 2 < \lambda \leq 1$):} impairment if $\tau_A \leq \tau^*_{\mathrm{aug}} \left( \lambda ; \hat{\beta} \right)$; automation if $\tau_A \geq \tau^*_{\mathrm{auto}} \left( \lambda ; \hat{\beta} \right)$.
	\end{itemize}
	These conditions characterize the full impairment and automation regions if and only if $\pi = 1/2$.
\end{corollary}

% Note that when $\pi \neq 1/2$, Corollary~\ref{coro:impairment_automation_any_prior_supersets} does not fully characterize any of the three regions, nor the augmentation and automation thresholds as functions of $\lambda$. However, this result does not restrict $\pi$.

\subsection{Full Phase Diagram Under Sufficiently Balanced Priors}

\begin{theorem}[Phase diagram under correlation neglect with sufficiently balanced priors]  \label{thm:regimes_horizontal_reasonably_balanced_prior}
	Fix $\hat{\alpha} = 0$, $\pi \in (0,1), \tau_H>0$ and $\hat{\beta} > 0$.
	Assume $\hat{\beta} \neq 1/2$ and 
	$\left| \ell_{\mathrm{prior}} \right| < \bar{\ell} \left( \tau_H, \hat{\beta} \right)$ for a bound $\bar{\ell} \left( \tau_H, \hat{\beta} \right) > 0$ constructed explicitly in the proof. 
	% \footnote{$\lpriorZboundcomp \left( \tau_H, \betabelief \right)$ is a non-trivial upper bound characterized exactly in the proof, and can possibly equal $+\infty$.} 
	Then, there exists 
	$\lambda_{\mathrm{low}} := 1 - \hat{\beta} / 2$, 
	$\lambda_{\mathrm{high}} \in \left( \lambda_{\mathrm{low}}, 1 \right]$,
	and functions $\tau_{\mathrm{aug}}$, $\tau_{\mathrm{auto}}$ such that
	\begin{itemize}
		\item \emph{Low overlap ($\lambda \leq \lambda_{\mathrm{low}}$, when $\hat{\beta} < 2$):} complementarity if $\tau_A < \tau_{\mathrm{auto}} \left( \lambda ; \hat{\beta} \right)$; automation if $\tau_A \geq \tau_{\mathrm{auto}} \left( \lambda ; \hat{\beta} \right)$.
		\footnote{When $\lambda = 0$, $\tau_{\mathrm{auto}} \left( 0 ; \hat{\beta} \right) := +\infty$ if $\hat{\beta} > 1/2$, and is finite if $\hat{\beta} < 1/2$.}
		\item \emph{Moderate overlap ($\lambda_{\mathrm{low}} < \lambda < \lambda_{\mathrm{high}}$):} impairment if $\tau_A < \tau_{\mathrm{aug}} \left( \lambda ; \hat{\beta} \right)$; complementarity if $\tau_{\mathrm{aug}} \left( \lambda ; \hat{\beta} \right) \leq \tau_A \leq \tau_{\mathrm{auto}} \left( \lambda ; \hat{\beta} \right)$; automation if $\tau_A > \tau_{\mathrm{auto}} \left( \lambda ; \hat{\beta} \right)$.
		\item \emph{High overlap ($\lambda \geq \lambda_{\mathrm{high}}$):} impairment if $\tau_A < \tau_H$; automation if $\tau_A \geq \tau_H$.
	\end{itemize}
	Moreover, $\tau_{\mathrm{aug}} \left( \lambda ; \hat{\beta} \right)$ is strictly increasing in $\lambda$; $\tau_{\mathrm{auto}} \left( \lambda ; \hat{\beta} \right)$ is strictly decreasing in $\lambda$ (where finite); and $\tau_{\mathrm{aug}} \left( \lambda ; \hat{\beta} \right) \leq \tau_H \leq \tau_{\mathrm{auto}} \left( \lambda ; \hat{\beta} \right)$, both with equality iff $\lambda = \lambda_{\mathrm{high}}$.
	Also, $\lambda_{\mathrm{high}} = 1$ if and only if $\pi = 1/2$.
\end{theorem}

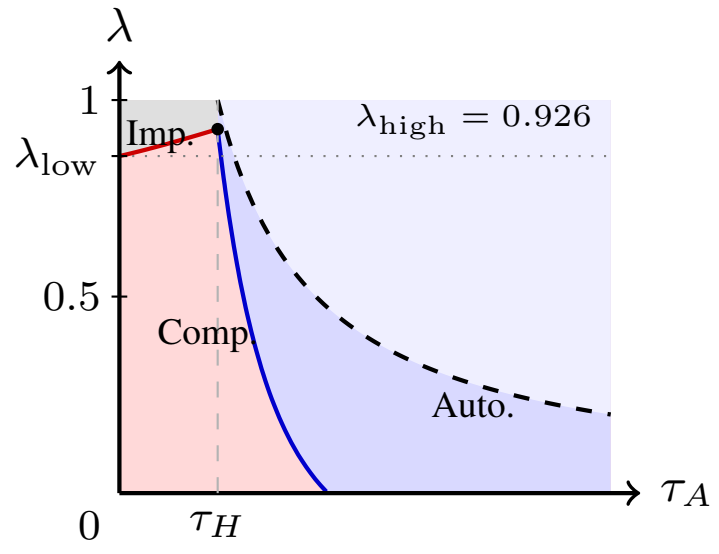
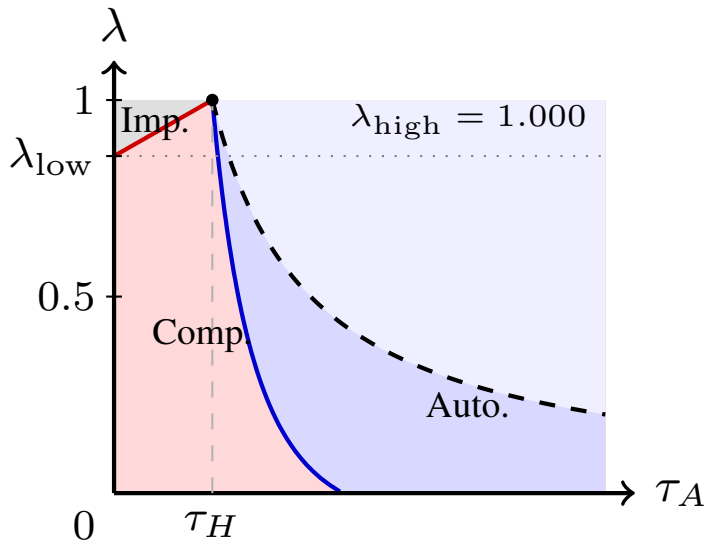
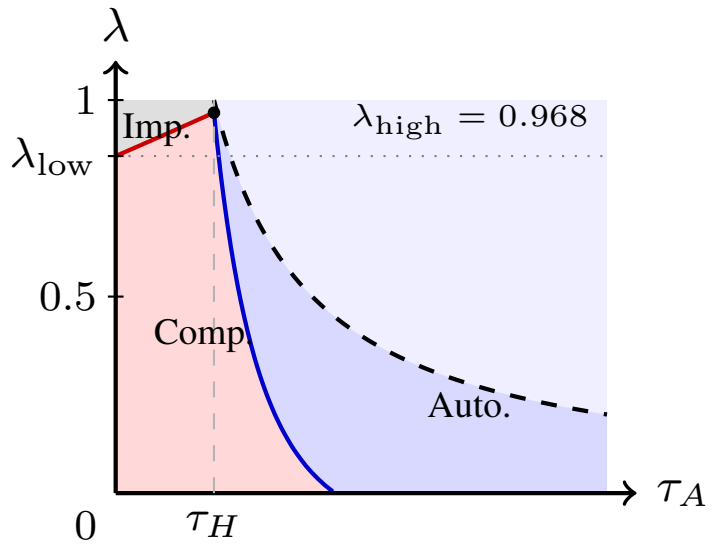
\begin{figure}[t]
\centering
\begin{minipage}{\textwidth}
\centering
\begin{subfigure}{0.31\textwidth}
\centering
\begin{tikzpicture}[scale=0.65]
% pi=0.2500, beta=0.285, lam_low=0.8575, lam_high=0.9260, tau_auto(0)=2.105
\fill[red!15] (0,0) rectangle (5,4);
\fill[blue!15] plot coordinates {(2.105,0.000) (2.105,0.016) (2.047,0.079) (1.994,0.141) (1.946,0.204) (1.901,0.266) (1.859,0.329) (1.819,0.391) (1.783,0.454) (1.748,0.516) (1.716,0.579) (1.685,0.641) (1.656,0.704) (1.628,0.766) (1.602,0.829) (1.577,0.891) (1.552,0.954) (1.529,1.016) (1.507,1.079) (1.486,1.141) (1.466,1.204) (1.446,1.266) (1.428,1.329) (1.410,1.391) (1.392,1.454) (1.375,1.516) (1.359,1.579) (1.343,1.641) (1.328,1.704) (1.313,1.766) (1.298,1.829) (1.284,1.891) (1.271,1.954) (1.258,2.016) (1.245,2.079) (1.232,2.141) (1.220,2.204) (1.209,2.266) (1.197,2.329) (1.186,2.391) (1.175,2.454) (1.164,2.516) (1.154,2.579) (1.144,2.641) (1.134,2.704) (1.124,2.766) (1.114,2.829) (1.105,2.891) (1.096,2.954) (1.087,3.016) (1.078,3.079) (1.070,3.141) (1.061,3.204) (1.053,3.266) (1.045,3.329) (1.037,3.392) (1.030,3.454) (1.022,3.517) (1.014,3.579) (1.007,3.642) (1.000,3.704) (1.000,3.704) (1.000,4.000) (5.000,4.000) (5.000,0.000)} -- cycle;
\fill[gray!25] plot coordinates {(0.000,3.430) (0.000,4.000) (1.000,4.000) (1.000,3.704) (1.000,3.704) (0.977,3.697) (0.954,3.690) (0.930,3.683) (0.907,3.676) (0.883,3.669) (0.860,3.662) (0.836,3.655) (0.813,3.648) (0.789,3.641) (0.765,3.634) (0.741,3.627) (0.717,3.620) (0.693,3.613) (0.668,3.606) (0.644,3.599) (0.619,3.592) (0.595,3.585) (0.570,3.578) (0.545,3.571) (0.520,3.564) (0.495,3.556) (0.470,3.549) (0.444,3.542) (0.418,3.535) (0.392,3.528) (0.366,3.521) (0.340,3.514) (0.313,3.507) (0.287,3.500) (0.260,3.493) (0.232,3.486) (0.205,3.479) (0.177,3.472) (0.148,3.465) (0.120,3.458) (0.091,3.451) (0.061,3.444) (0.031,3.437) (0.000,3.430)} -- cycle;
\fill[white, opacity=0.6] plot[domain=1.0:5.0, samples=40, smooth, variable=\x] ({\x},{4/\x}) -- (5,4) -- (1,4) -- cycle;
\draw[thick, red!80!black] plot coordinates {(0.000,3.430) (0.031,3.437) (0.061,3.444) (0.091,3.451) (0.120,3.458) (0.148,3.465) (0.177,3.472) (0.205,3.479) (0.232,3.486) (0.260,3.493) (0.287,3.500) (0.313,3.507) (0.340,3.514) (0.366,3.521) (0.392,3.528) (0.418,3.535) (0.444,3.542) (0.470,3.549) (0.495,3.556) (0.520,3.564) (0.545,3.571) (0.570,3.578) (0.595,3.585) (0.619,3.592) (0.644,3.599) (0.668,3.606) (0.693,3.613) (0.717,3.620) (0.741,3.627) (0.765,3.634) (0.789,3.641) (0.813,3.648) (0.836,3.655) (0.860,3.662) (0.883,3.669) (0.907,3.676) (0.930,3.683) (0.954,3.690) (0.977,3.697) (1.000,3.704)};
\draw[thick, blue!80!black] plot coordinates {(2.105,0.016) (2.047,0.079) (1.994,0.141) (1.946,0.204) (1.901,0.266) (1.859,0.329) (1.819,0.391) (1.783,0.454) (1.748,0.516) (1.716,0.579) (1.685,0.641) (1.656,0.704) (1.628,0.766) (1.602,0.829) (1.577,0.891) (1.552,0.954) (1.529,1.016) (1.507,1.079) (1.486,1.141) (1.466,1.204) (1.446,1.266) (1.428,1.329) (1.410,1.391) (1.392,1.454) (1.375,1.516) (1.359,1.579) (1.343,1.641) (1.328,1.704) (1.313,1.766) (1.298,1.829) (1.284,1.891) (1.271,1.954) (1.258,2.016) (1.245,2.079) (1.232,2.141) (1.220,2.204) (1.209,2.266) (1.197,2.329) (1.186,2.391) (1.175,2.454) (1.164,2.516) (1.154,2.579) (1.144,2.641) (1.134,2.704) (1.124,2.766) (1.114,2.829) (1.105,2.891) (1.096,2.954) (1.087,3.016) (1.078,3.079) (1.070,3.141) (1.061,3.204) (1.053,3.266) (1.045,3.329) (1.037,3.392) (1.030,3.454) (1.022,3.517) (1.014,3.579) (1.007,3.642) (1.000,3.704)};
\draw[dashed, thick, black] plot[domain=1.0:5.0, samples=40, smooth, variable=\x] ({\x},{4/\x});
\draw[dotted, gray] (0,3.430) -- (5,3.430);
\draw[->, thick] (0,0) -- (5.3,0) node[right, font=\small] {$\tau_A$};
\draw[->, thick] (0,0) -- (0,4.4) node[above, font=\small] {$\lambda$};
\draw[font=\footnotesize] (0,0) node[below left] {$0$};
\draw (-0.08,4) -- (0.08,4); \draw[font=\footnotesize] (0,4) node[left] {$1$};
\draw (-0.08,2) -- (0.08,2); \draw[font=\footnotesize] (0,2) node[left] {$0.5$};
\draw (-0.08,3.430) -- (0.08,3.430); \draw[font=\footnotesize] (0,3.430) node[left] {$\lambda_{\mathrm{low}}$};
\draw[dashed, gray!60] (1,0) -- (1,4); \draw[font=\footnotesize] (1,0) node[below] {$\tau_H$};
\filldraw[black] (1,3.704) circle (1.5pt);
\node[font=\tiny, anchor=east] at (5.0,3.78) {$\lambda_{\mathrm{high}}=0.926$};
\node[font=\scriptsize] at (0.9,1.6) {Comp.};
\node[font=\scriptsize] at (0.42,3.62) {Imp.};
\node[font=\scriptsize] at (3.6,0.9) {Auto.};
\end{tikzpicture}
\caption{$\pi=1/4$}
\end{subfigure}
\hfill
\begin{subfigure}{0.31\textwidth}
\centering
\begin{tikzpicture}[scale=0.65]
% pi=0.5000, beta=0.285, lam_low=0.8575, lam_high=1.0000, tau_auto(0)=2.298
\fill[red!15] (0,0) rectangle (5,4);
\fill[blue!15] plot coordinates {(2.298,0.000) (2.298,0.016) (2.193,0.084) (2.104,0.151) (2.028,0.219) (1.961,0.286) (1.902,0.354) (1.849,0.421) (1.801,0.489) (1.757,0.556) (1.717,0.624) (1.679,0.691) (1.645,0.759) (1.613,0.826) (1.583,0.894) (1.555,0.961) (1.529,1.029) (1.504,1.096) (1.480,1.164) (1.458,1.231) (1.437,1.299) (1.417,1.367) (1.397,1.434) (1.379,1.502) (1.362,1.569) (1.345,1.637) (1.329,1.704) (1.313,1.772) (1.298,1.839) (1.284,1.907) (1.270,1.974) (1.257,2.042) (1.244,2.109) (1.232,2.177) (1.219,2.244) (1.208,2.312) (1.197,2.379) (1.186,2.447) (1.175,2.514) (1.165,2.582) (1.155,2.649) (1.145,2.717) (1.136,2.785) (1.126,2.852) (1.117,2.920) (1.109,2.987) (1.100,3.055) (1.092,3.122) (1.084,3.190) (1.076,3.257) (1.068,3.325) (1.061,3.392) (1.053,3.460) (1.046,3.527) (1.039,3.595) (1.032,3.662) (1.026,3.730) (1.019,3.797) (1.013,3.865) (1.006,3.932) (1.000,4.000) (1.000,4.000) (1.000,4.000) (5.000,4.000) (5.000,0.000)} -- cycle;
\fill[gray!25] plot coordinates {(0.000,3.430) (0.000,4.000) (1.000,4.000) (1.000,4.000) (1.000,4.000) (0.974,3.985) (0.949,3.971) (0.923,3.956) (0.897,3.942) (0.872,3.927) (0.846,3.912) (0.821,3.898) (0.795,3.883) (0.769,3.868) (0.744,3.854) (0.718,3.839) (0.692,3.825) (0.667,3.810) (0.641,3.795) (0.615,3.781) (0.590,3.766) (0.564,3.752) (0.538,3.737) (0.513,3.722) (0.487,3.708) (0.462,3.693) (0.436,3.678) (0.410,3.664) (0.385,3.649) (0.359,3.635) (0.333,3.620) (0.308,3.605) (0.282,3.591) (0.256,3.576) (0.231,3.562) (0.205,3.547) (0.179,3.532) (0.154,3.518) (0.128,3.503) (0.103,3.488) (0.077,3.474) (0.051,3.459) (0.026,3.445) (0.000,3.430)} -- cycle;
\fill[white, opacity=0.6] plot[domain=1.0:5.0, samples=40, smooth, variable=\x] ({\x},{4/\x}) -- (5,4) -- (1,4) -- cycle;
\draw[thick, red!80!black] plot coordinates {(0.000,3.430) (0.026,3.445) (0.051,3.459) (0.077,3.474) (0.103,3.488) (0.128,3.503) (0.154,3.518) (0.179,3.532) (0.205,3.547) (0.231,3.562) (0.256,3.576) (0.282,3.591) (0.308,3.605) (0.333,3.620) (0.359,3.635) (0.385,3.649) (0.410,3.664) (0.436,3.678) (0.462,3.693) (0.487,3.708) (0.513,3.722) (0.538,3.737) (0.564,3.752) (0.590,3.766) (0.615,3.781) (0.641,3.795) (0.667,3.810) (0.692,3.825) (0.718,3.839) (0.744,3.854) (0.769,3.868) (0.795,3.883) (0.821,3.898) (0.846,3.912) (0.872,3.927) (0.897,3.942) (0.923,3.956) (0.949,3.971) (0.974,3.985) (1.000,4.000)};
\draw[thick, blue!80!black] plot coordinates {(2.298,0.016) (2.193,0.084) (2.104,0.151) (2.028,0.219) (1.961,0.286) (1.902,0.354) (1.849,0.421) (1.801,0.489) (1.757,0.556) (1.717,0.624) (1.679,0.691) (1.645,0.759) (1.613,0.826) (1.583,0.894) (1.555,0.961) (1.529,1.029) (1.504,1.096) (1.480,1.164) (1.458,1.231) (1.437,1.299) (1.417,1.367) (1.397,1.434) (1.379,1.502) (1.362,1.569) (1.345,1.637) (1.329,1.704) (1.313,1.772) (1.298,1.839) (1.284,1.907) (1.270,1.974) (1.257,2.042) (1.244,2.109) (1.232,2.177) (1.219,2.244) (1.208,2.312) (1.197,2.379) (1.186,2.447) (1.175,2.514) (1.165,2.582) (1.155,2.649) (1.145,2.717) (1.136,2.785) (1.126,2.852) (1.117,2.920) (1.109,2.987) (1.100,3.055) (1.092,3.122) (1.084,3.190) (1.076,3.257) (1.068,3.325) (1.061,3.392) (1.053,3.460) (1.046,3.527) (1.039,3.595) (1.032,3.662) (1.026,3.730) (1.019,3.797) (1.013,3.865) (1.006,3.932) (1.000,4.000)};
\draw[dashed, thick, black] plot[domain=1.0:5.0, samples=40, smooth, variable=\x] ({\x},{4/\x});
\draw[dotted, gray] (0,3.430) -- (5,3.430);
\draw[->, thick] (0,0) -- (5.3,0) node[right, font=\small] {$\tau_A$};
\draw[->, thick] (0,0) -- (0,4.4) node[above, font=\small] {$\lambda$};
\draw[font=\footnotesize] (0,0) node[below left] {$0$};
\draw (-0.08,4) -- (0.08,4); \draw[font=\footnotesize] (0,4) node[left] {$1$};
\draw (-0.08,2) -- (0.08,2); \draw[font=\footnotesize] (0,2) node[left] {$0.5$};
\draw (-0.08,3.430) -- (0.08,3.430); \draw[font=\footnotesize] (0,3.430) node[left] {$\lambda_{\mathrm{low}}$};
\draw[dashed, gray!60] (1,0) -- (1,4); \draw[font=\footnotesize] (1,0) node[below] {$\tau_H$};
\filldraw[black] (1,4.000) circle (1.5pt);
\node[font=\tiny, anchor=east] at (5.0,3.78) {$\lambda_{\mathrm{high}}=1.000$};
\node[font=\scriptsize] at (0.9,1.6) {Comp.};
\node[font=\scriptsize] at (0.42,3.76) {Imp.};
\node[font=\scriptsize] at (3.6,0.9) {Auto.};
\end{tikzpicture}
\caption{$\pi=1/2$ (symmetric)}
\end{subfigure}
\hfill
\begin{subfigure}{0.31\textwidth}
\centering
\begin{tikzpicture}[scale=0.65]
% pi=0.6667, beta=0.285, lam_low=0.8575, lam_high=0.9679, tau_auto(0)=2.210
\fill[red!15] (0,0) rectangle (5,4);
\fill[blue!15] plot coordinates {(2.210,0.000) (2.210,0.016) (2.128,0.081) (2.057,0.147) (1.993,0.212) (1.936,0.277) (1.885,0.343) (1.838,0.408) (1.794,0.473) (1.755,0.539) (1.718,0.604) (1.683,0.669) (1.651,0.735) (1.620,0.800) (1.592,0.866) (1.565,0.931) (1.539,0.996) (1.515,1.062) (1.492,1.127) (1.470,1.192) (1.449,1.258) (1.429,1.323) (1.410,1.388) (1.392,1.454) (1.374,1.519) (1.358,1.584) (1.341,1.650) (1.326,1.715) (1.311,1.780) (1.296,1.846) (1.282,1.911) (1.268,1.976) (1.255,2.042) (1.242,2.107) (1.230,2.172) (1.218,2.238) (1.206,2.303) (1.195,2.368) (1.184,2.434) (1.173,2.499) (1.163,2.565) (1.153,2.630) (1.143,2.695) (1.133,2.761) (1.124,2.826) (1.115,2.891) (1.106,2.957) (1.097,3.022) (1.089,3.087) (1.080,3.153) (1.072,3.218) (1.064,3.283) (1.057,3.349) (1.049,3.414) (1.042,3.479) (1.034,3.545) (1.027,3.610) (1.020,3.675) (1.013,3.741) (1.007,3.806) (1.000,3.871) (1.000,3.871) (1.000,4.000) (5.000,4.000) (5.000,0.000)} -- cycle;
\fill[gray!25] plot coordinates {(0.000,3.430) (0.000,4.000) (1.000,4.000) (1.000,3.871) (1.000,3.871) (0.975,3.860) (0.949,3.849) (0.924,3.838) (0.899,3.826) (0.873,3.815) (0.848,3.804) (0.822,3.792) (0.797,3.781) (0.772,3.770) (0.746,3.758) (0.721,3.747) (0.695,3.736) (0.670,3.724) (0.644,3.713) (0.619,3.702) (0.593,3.690) (0.568,3.679) (0.542,3.668) (0.517,3.656) (0.491,3.645) (0.465,3.634) (0.440,3.622) (0.414,3.611) (0.388,3.600) (0.363,3.588) (0.337,3.577) (0.311,3.566) (0.285,3.555) (0.260,3.543) (0.234,3.532) (0.208,3.521) (0.182,3.509) (0.156,3.498) (0.130,3.487) (0.104,3.475) (0.078,3.464) (0.052,3.453) (0.026,3.441) (0.000,3.430)} -- cycle;
\fill[white, opacity=0.6] plot[domain=1.0:5.0, samples=40, smooth, variable=\x] ({\x},{4/\x}) -- (5,4) -- (1,4) -- cycle;
\draw[thick, red!80!black] plot coordinates {(0.000,3.430) (0.026,3.441) (0.052,3.453) (0.078,3.464) (0.104,3.475) (0.130,3.487) (0.156,3.498) (0.182,3.509) (0.208,3.521) (0.234,3.532) (0.260,3.543) (0.285,3.555) (0.311,3.566) (0.337,3.577) (0.363,3.588) (0.388,3.600) (0.414,3.611) (0.440,3.622) (0.465,3.634) (0.491,3.645) (0.517,3.656) (0.542,3.668) (0.568,3.679) (0.593,3.690) (0.619,3.702) (0.644,3.713) (0.670,3.724) (0.695,3.736) (0.721,3.747) (0.746,3.758) (0.772,3.770) (0.797,3.781) (0.822,3.792) (0.848,3.804) (0.873,3.815) (0.899,3.826) (0.924,3.838) (0.949,3.849) (0.975,3.860) (1.000,3.871)};
\draw[thick, blue!80!black] plot coordinates {(2.210,0.016) (2.128,0.081) (2.057,0.147) (1.993,0.212) (1.936,0.277) (1.885,0.343) (1.838,0.408) (1.794,0.473) (1.755,0.539) (1.718,0.604) (1.683,0.669) (1.651,0.735) (1.620,0.800) (1.592,0.866) (1.565,0.931) (1.539,0.996) (1.515,1.062) (1.492,1.127) (1.470,1.192) (1.449,1.258) (1.429,1.323) (1.410,1.388) (1.392,1.454) (1.374,1.519) (1.358,1.584) (1.341,1.650) (1.326,1.715) (1.311,1.780) (1.296,1.846) (1.282,1.911) (1.268,1.976) (1.255,2.042) (1.242,2.107) (1.230,2.172) (1.218,2.238) (1.206,2.303) (1.195,2.368) (1.184,2.434) (1.173,2.499) (1.163,2.565) (1.153,2.630) (1.143,2.695) (1.133,2.761) (1.124,2.826) (1.115,2.891) (1.106,2.957) (1.097,3.022) (1.089,3.087) (1.080,3.153) (1.072,3.218) (1.064,3.283) (1.057,3.349) (1.049,3.414) (1.042,3.479) (1.034,3.545) (1.027,3.610) (1.020,3.675) (1.013,3.741) (1.007,3.806) (1.000,3.871)};
\draw[dashed, thick, black] plot[domain=1.0:5.0, samples=40, smooth, variable=\x] ({\x},{4/\x});
\draw[dotted, gray] (0,3.430) -- (5,3.430);
\draw[->, thick] (0,0) -- (5.3,0) node[right, font=\small] {$\tau_A$};
\draw[->, thick] (0,0) -- (0,4.4) node[above, font=\small] {$\lambda$};
\draw[font=\footnotesize] (0,0) node[below left] {$0$};
\draw (-0.08,4) -- (0.08,4); \draw[font=\footnotesize] (0,4) node[left] {$1$};
\draw (-0.08,2) -- (0.08,2); \draw[font=\footnotesize] (0,2) node[left] {$0.5$};
\draw (-0.08,3.430) -- (0.08,3.430); \draw[font=\footnotesize] (0,3.430) node[left] {$\lambda_{\mathrm{low}}$};
\draw[dashed, gray!60] (1,0) -- (1,4); \draw[font=\footnotesize] (1,0) node[below] {$\tau_H$};
\filldraw[black] (1,3.871) circle (1.5pt);
\node[font=\tiny, anchor=east] at (5.0,3.78) {$\lambda_{\mathrm{high}}=0.968$};
\node[font=\scriptsize] at (0.9,1.6) {Comp.};
\node[font=\scriptsize] at (0.42,3.70) {Imp.};
\node[font=\scriptsize] at (3.6,0.9) {Auto.};
\end{tikzpicture}
\caption{$\pi=2/3$}
\end{subfigure}
\end{minipage}
\caption{Phase diagram (regions as in Figure~\ref{fig:phase-diagrams}) under correlation neglect at trust $\hat\beta=0.285$, for three priors. As the prior tilts away from $\pi=1/2$, complementarity pinches off at $\lambda_{\mathrm{high}}<1$ rather than at $\lambda=1$ (Theorem~\ref{thm:regimes_horizontal_reasonably_balanced_prior}).}
\label{fig:phase-prior}
\end{figure}

This is the same phase diagram as Theorem~\ref{thm:phase}, with the additional property that $\lambda_{\mathrm{high}} < 1$ when $\pi \neq 1/2$: the high-overlap regime does not collapse to a single $\lambda=1$. See Figure~\ref{fig:phase-prior} for an illustration.

%Note the following limitations:
% \begin{itemize}
% 	\item $\tau_{\mathrm{aug}}$, $\tau_{\mathrm{auto}}$ and $\lambda_{\mathrm{high}}$ are defined implicitly as unique crossings of accuracies, without closed forms.
% 	\item $\bar{\ell} \left( \tau_H, \hat{\beta} \right) := \min \left\{ \bar{\ell}_{\mathrm{aug}} \left( \tau_H, \hat{\beta} \right), \bar{\ell}_{\mathrm{auto}} \left( \tau_H, \hat{\beta} \right) \right\}$ is a sufficient positive bound fully specified by the proof.
% 	% , uniform over all admissible $\left( \lambda, \tau_A \right)$. 
% 	Its two components are sufficient upper bounds on $\left| \ell_{\mathrm{prior}} \right|$ for establishing the respective thresholds. While $\bar{\ell}_{\mathrm{aug}}$ has an elementary closed form, $\bar{\ell}_{\mathrm{auto}}$ requires infima over parameter ranges. Thus, the full bound $\bar{\ell}$ has no simple closed form nor an overall interpretation.
% 	\footnote{
% 		The closed-form \emph{augmentation-side} prior bound $\bar{\ell}_{\mathrm{aug}}$ does provide some conceptual insights: $\bar{\ell}_{\mathrm{aug}} \left( \tau_H, \hat{\beta} \right)$ becomes less restrictive with higher human capability or lower trust in AI, in which case the ``regular'' single-crossing structure of the augmentation threshold survives under a wider range of priors. Note that it does \emph{not} indicate whether the impairment region itself shrinks or expands.
% 	}
% 	% Note that $\bar{\ell}$ does not rely on a local neighborhood argument or mere existence claims.
% \end{itemize}

\subsection{Proof Sketch}

The expected accuracy for any agent $B \in \left\{ H, A, HA \right\}$ can be written in terms of a shared function $\mathcal{A}_B = \mathcal{A} \left( m_{B}, \sigma_{B}; \pi \right)$, where $\left( m_B, \sigma_B \right)$ and $\ell_{\mathrm{prior}}$ summarize the distribution of $\ell_B$ or $\hat{\ell}$. When the prior is symmetric, $\mathcal{A} \left( m_{B}, \sigma_{B}; 1/2 \right) = \Phi \left( m_B / \sigma_B \right)$, where $r_B = m_B / \sigma_B$ is the signal-to-noise ratio (SNR) in Lemma~\ref{lem:accuracy}.

Now fix an asymmetric prior $\pi \neq 1/2$. In the $\left( m, \sigma \right)$ plane, \footnote{Here we use $\sigma$ as a short hand for $\sigma_B$; note the notational conflict with the $\sigma$ used for defining individual primitive cues.} we define \emph{SNR-rays} along which $r = m / \sigma$ is constant.
We show that each SNR-ray has a unique \emph{calibrated point} where $\sigma^2 = 2m$, where the statistic is a true log-likelihood ratio and its accuracy equals that of a calibrated single agent with signal-to-noise ratio $r$; moreover, accuracy along the SNR-ray is maximized at the calibrated point, which is the unique maximum iff $\pi \neq 1/2$. Additionally, accuracy among all calibrated points is strictly increasing in $r$.
This yields a \emph{two-step SNR-ray projection lemma} that upper bounds $\mathcal{A} \left( m, \sigma \right)$ by a selected single agent's accuracy: if the team is weakly dominated by a single agent at prior $1/2$, its signal-to-noise ratio is at most that agent's; thus, under any prior $\pi$, the team performance $\mathcal{A} \left( m, \sigma \right)$ is still at most the accuracy at the calibrated point $\mathcal{A} \left( 2 m^2/\sigma^2, 2 m/\sigma \right)$ with the same signal-to-noise ratio, which is itself upper bounded by the single agent's accuracy.
This projection gives the single-agent dominance transfer property in Proposition~\ref{prop:single_agent_dominance_transfer}, which immediately leads to Corollary~\ref{coro:impairment_automation_any_prior_supersets}.

Corollary~\ref{coro:impairment_automation_any_prior_supersets} implies that if any single-crossing augmentation and automation thresholds exist, they are lower- and upper-bounded respectively by the symmetric-prior thresholds $\tau^*_{\mathrm{aug}} \left( \lambda ; \hat{\beta} \right)$ and $\tau^*_{\mathrm{auto}} \left( \lambda ; \hat{\beta} \right)$.
It remains to show the existence of such single crossings. For any fixed $\lambda$ and under sufficient conditions, we show that for augmentation, $\mathcal{A}_{HA} \left( \tau_A; \lambda \right)$ is either increasing in $\tau_A$, or has a single minimum and then increases and approaches $1$; and for automation, $\mathcal{A}_{HA} \left( \tau_A; \lambda \right) - \mathcal{A}_{A} \left( \tau_A \right)$ has at most one single crossing to negative before $\tau_A = \tau^*_{\mathrm{auto}} \left( \lambda ; \hat{\beta} \right)$, after which Corollary~\ref{coro:impairment_automation_any_prior_supersets} gives AI-alone dominance. The theorem's upper bound on $\left| \ell_{\mathrm{prior}} \right|$ collects sufficient conditions for the correct signs of such derivatives. These results establish the existence of single-crossing thresholds $\tau_{\mathrm{aug}} \left( \lambda ; \hat{\beta} \right)$ and $\tau_{\mathrm{auto}} \left( \lambda ; \hat{\beta} \right)$ under sufficiently balanced priors. Additionally, the implicit function theorem and derivative signs show that both thresholds are monotone in $\lambda$.
Combining them with the $H$ vs. $A$ threshold ($\tau_A = \tau_H$) proves Theorem~\ref{thm:regimes_horizontal_reasonably_balanced_prior}.

\section{Sample construction}
\label{app:sample}

The ChatBench public release~\citep{chang2025chatbench} contains $5{,}871$ conversation logs (\texttt{conversations.json}), $10{,}450$ user-answer rows (\texttt{user\_answers\_filtered.csv}) covering $643$ workers, and $792$ per-question AI-accuracy records (\texttt{ai\_alone\_answers.csv}) summarizing up to $50$ free-text generations per AI-question pair (mean $49.9$; $7$ pairs have fewer due to sampling failures). We construct the matched answer-first sample as follows.

\paragraph{Filtering and merging.}
Restrict \texttt{conversations.json} to rows with \texttt{condition = answer-first} and \texttt{answer\_type = user-AI} (the answer-first user-AI rows). Restrict \texttt{user\_answers\_filtered.csv} to \texttt{answer\_type = user-alone} with non-missing confidence (the user-alone rows). Inner-join on \texttt{(worker\_id, question\_id, model)} to pair each post-AI answer with its unaided counterpart. Merge \texttt{ai\_alone\_answers.csv} on \texttt{(model, question\_id)} to attach the per-question AI accuracy \texttt{free\_text\_acc}. This produces $2{,}894$ matched pairs.

\paragraph{Ground truth.}
Each MMLU question has a unique correct letter. We recover ground truth from the user-alone sample by taking \texttt{selected\_answer} of any worker who got the question right unaided. For two of the $396$ questions in the matched sample, no user got the question right alone; we drop the resulting $10$ rows ($0.3\%$ of $2{,}894$) so that the AI-correctness indicator is well-defined for every observation. The remaining sample size is $2{,}884$.

\paragraph{Has-Rec partition.}
The conversation logs include a flag \texttt{conv\_has\_answer} indicating whether the AI committed to a definite letter recommendation in its message. Pairs with \texttt{conv\_has\_answer = True} form the \emph{Has-Rec} sample ($2{,}199$ rows; $315$ unique workers; $389$ unique questions); the rest form the No-Rec sample ($685$ rows). All analyses in \S\ref{sec:experiments} use the Has-Rec sample, where the AI's signal is unambiguously a labeled recommendation.

\paragraph{Variable definitions.}
For each Has-Rec observation $i = (w, q, m)$ (worker $w$, question $q$, AI model $m$):
\begin{itemize}
  \item $h\_acc_i \in \{0, 1\}$: whether the worker's unaided answer is correct.
  \item $h\_conf_i \in \{\text{not-, somewhat-, very-confident}\}$: the worker's stated confidence on the unaided answer.
  \item $F_i \in \{0, 1\}$: whether the worker's post-AI answer matches ground truth.
  \item $ai\_correct_i \in \{0, 1\}$: whether the AI's letter recommendation matches ground truth.
  \item $p_A^{m,q} \in [0, 1]$: the AI's empirical accuracy on question $q$ across up to $50$ independent free-text generations (the column \texttt{free\_text\_acc}).
\end{itemize}

\paragraph{Confidence calibration.}
We construct the empirical accuracy-by-confidence map $p_H(c) := \widehat{\Pr}(h\_acc = 1 \mid h\_conf = c)$ from the matched sample:
\[
\textstyle
p_H(\text{not-confident}) = 0.287, \quad p_H(\text{somewhat-confident}) = 0.569, \quad p_H(\text{very-confident}) = 0.787.
\]
The signals used in the analysis are
\[
\textstyle
\ell_H^i = \mathrm{logit}\!\bigl(\mathrm{clip}_\epsilon(p_H(h\_conf_i))\bigr), \qquad \ell_A^i = \mathrm{logit}\!\bigl(\mathrm{clip}_\epsilon(p_A^{m_i, q_i})\bigr),
\]
where $\mathrm{clip}_\epsilon(p) := \max(\epsilon, \min(1 - \epsilon, p))$ with $\epsilon = 0.005$ to handle $p_A \in \{0, 1\}$. About $62\%$ of Has-Rec observations have $p_A \in \{0, 1\}$ (the AI gets every generation right or every generation wrong); $\ell_A$ is therefore clipped to $\pm 5.293$ for the majority of rows, and $|\ell_A|$ should be read as a coarse three-level indicator of AI question-difficulty. Sensitivity to the choice of $\epsilon$ is examined in \S\ref{app:identification}.

\paragraph{Per-question $H_q$.}
For Approach~2 of \S\ref{app:identification}, we use a finer human-side signal: $H_q := \frac{1}{n_q} \sum_{i: q_i = q} h\_acc_i$, the empirical unaided accuracy on question $q$ in the user-alone sample, and $\ell_H^{H_q,i} := \mathrm{logit}(\mathrm{clip}_\epsilon(H_{q_i}))$. This gives $94$ unique values of $\ell_H^{H_q}$ versus only $3$ unique values of $\ell_H$ from confidence.

\paragraph{Sample-size summary.}
\begin{center}
\small
\begin{tabular}{lr}
\toprule
Quantity & Value \\
\midrule
Workers in raw study & $643$ \\
Matched answer-first pairs (post truth-recovery) & $2{,}884$ \\
Has-Rec pairs (used in \S\ref{sec:experiments}) & $2{,}199$ \\
\quad of which GPT-4o & $1{,}160$ \\
\quad of which Llama-3.1-8b & $1{,}039$ \\
Unique workers in Has-Rec & $315$ \\
Unique questions in Has-Rec & $389$ \\
\bottomrule
\end{tabular}
\end{center}

\section{Identification of the worker's parameters and regime tests}
\label{app:identification}

\subsection{Theoretically principled method}
\label{sec:identification}
To use Theorem~\ref{thm:phase} for empirical analysis, we need to recover $(\hat\alpha, \hat\beta)$ from data on the worker's behavior. The two parameters are separately identified if the same worker population is observed paired with two AI systems that differ in their overlap with the human's evidence. We refer to a (worker population, AI system, task) configuration with parameters $(\tau_H, \tau_A, \lambda)$ as a \emph{deployment}, and assume the analyst observes data from two deployments $d$ and $d'$ (e.g., the same workers paired with GPT-4o on $d$ and Llama on $d'$, on the same task) that share the worker population but differ in overlap: $\lambda^d \neq \lambda^{d'}$. The distinguishing property of BR vs.\ CN (\S\ref{sec:benchmarks}) provides the identification lever: at BR, the worker's AI weight $w_A$ depends on $\lambda$ (she corrects for overlap); at CN, $w_A$ does not (she treats the AI signal as independent). Across two deployments with $\lambda^d \neq \lambda^{d'}$, the difference in $w_A$ is informative about $\hat\alpha$.

\begin{definition}[Cross-deployment AI-weight ratio]  \label{def:cross_deploy_weight_ratio}
    $R_A \left( \hat{\alpha}, \hat{\beta}; d, d' \right) := w_A^d \left( \hat{\alpha}, \hat{\beta} \right) / w_A^{d'} \left( \hat{\alpha}, \hat{\beta} \right)$, with $w_A^d, w_A^{d'}$ from Lemma~\ref{lem:behavioral}. The empirical analogue is $\hat{R}_A := \hat{w}_A^d / \hat{w}_A^{d'}$, where $\hat{w}_A^d, \hat{w}_A^{d'}$ are estimated within each deployment $d:=(\tau_H, \tau_A^d, \lambda^d), d':=(\tau_H, \tau_A^{d'}, \lambda^{d'})$.
\end{definition}

\begin{proposition}[Cross-deployment signature]  \label{prop:cross_deploy_signature_with_ratio}
    Under Assumption~\ref{ass:worker}, 
    \begin{itemize}
        \item \emph{Correlation neglect ($\hat{\alpha}=0$, any $\hat{\beta}>0$):} $R_A^{\mathrm{CN}} = 1$.
        \item \emph{Best response ($\hat{\alpha}=1, \hat{\beta}=1$):} When $\lambda^d \neq \lambda^{d'}$, $R_A^{\mathrm{BR}} \neq 1$ generically. 
        \item \emph{Interior ($\hat{\alpha} \in \left( 0,1 \right)$)}: $R_A$ is continuous in $\hat{\alpha}$, and $R_A \neq 1$ generically. Moreover, when $\lambda^d \neq \lambda^{d'}$, $R_A \left( \hat{\alpha} \right)$ is locally invertible near $\hat{\alpha}=0$ with partial derivative $\lambda^{d'} - \lambda^{d} \neq 0$.
    \end{itemize}
\end{proposition}

The empirical $\hat{R}_A$ thus separates correlation-neglect workers from best-response workers, and locally identifies an interior $\hat{\alpha}$ near correlation neglect.

\paragraph{Two-step identification.} 
\emph{Step 1 (recover $\hat{\alpha}$ from $\hat{R}_A$).}
Compute the BR-predicted ratio from deployment parameters $(\tau_H, \tau_A^d, \lambda^d)$, $(\tau_H, \tau_A^{d'}, \lambda^{d'})$,
\begin{align*}
  \textstyle
    R_A^{\mathrm{BR}} \left( d, d' \right) 
    = \frac{
        \left.
        \left( 1 - \lambda^{d} \right)
        \middle/
        \left( \tau_H  - \tau_A^{d} \left( \lambda^{d} \right)^2 \right)
        \right.
    }{
        \left.
        \left( 1 - \lambda^{d'} \right)
        \middle/
        \left( \tau_H  - \tau_A^{d'} \left( \lambda^{d'} \right)^2 \right)
        \right.
    },
\end{align*}
and compare to $\hat{R}_A$: $\hat{R}_A \approx 1$ supports $\hat{\alpha} = 0$, $\hat{R}_A \approx R_A^{\mathrm{BR}}$ supports $\hat{\alpha} = 1$, and intermediate values can pin down $\hat{\alpha} \in \left( 0,1 \right)$ by solving
\begin{align*}
  \textstyle
    \hat{R}_A
    = \frac{
        \left.
        \left( 1 - \hat{\alpha} \lambda^{d} \right)
        \middle/
        \left( \tau_H  - \tau_A^{d} \hat{\alpha}^2 \left( \lambda^{d} \right)^2 \right)
        \right.
    }{
        \left.
        \left( 1 - \hat{\alpha} \lambda^{d'} \right)
        \middle/
        \left( \tau_H  - \tau_A^{d'} \hat{\alpha}^2 \left( \lambda^{d'} \right)^2 \right)
        \right.
    }.
\end{align*}
\emph{Step 2 (recover $\hat{\beta}$ from $\hat{w}_A$ given $\hat{\alpha}$).}
Inverting Lemma~\ref{lem:behavioral},
\begin{align*}
    \hat{\beta} = \hat{w}_A^d \frac
        { \tau_H  - \tau_A^d \hat{\alpha}^2 \left( \lambda^d \right)^2 }
        { \tau_H \left( 1 - \hat{\alpha} \lambda^d \right) }.
\end{align*}
At $\hat{\alpha} = 0$, this reduces to $\hat{\beta} = \hat{w}_A^d$: the empirical AI weight estimates trust directly.
\begin{proof}[Proof of Proposition~\ref{prop:cross_deploy_signature_with_ratio}]
    By Lemma~\ref{lem:behavioral}, the worker's AI weight at deployment $d$ is
    \begin{align*}
        w_A^d \left( \hat{\alpha}, \hat{\beta} \right)
        =
        \hat{\beta} \cdot \frac
        { \tau_H \left( 1 - \hat{\alpha} \lambda^d \right) }
        { \tau_H  - \tau_A^d \hat{\alpha}^2 \left( \lambda^d \right)^2 }.
    \end{align*}
    The structural cross-deployment ratio is
    \begin{align*}
        R_A \left( \hat{\alpha}, \hat{\beta}; d, d' \right)
        = \frac{w_A^d} {w_A^{d'}}
        % = \frac{
        %     \frac
        %     { \tau_H \left( 1 - \hat{\alpha} \lambda^{d} \right) }
        %     { \tau_H  - \tau_A^{d} \hat{\alpha}^2 \left( \lambda^{d} \right)^2 }
        % }{
        %     \frac
        %     { \tau_H \left( 1 - \hat{\alpha} \lambda^{d'} \right) }
        %     { \tau_H  - \tau_A^{d'} \hat{\alpha}^2 \left( \lambda^{d'} \right)^2 }
        % }.
        = \frac{
            \left.
            \left( 1 - \hat{\alpha} \lambda^{d} \right)
            \middle/
            \left( \tau_H  - \tau_A^{d} \hat{\alpha}^2 \left( \lambda^{d} \right)^2 \right)
            \right.
        }{
            \left.
            \left( 1 - \hat{\alpha} \lambda^{d'} \right)
            \middle/
            \left( \tau_H  - \tau_A^{d'} \hat{\alpha}^2 \left( \lambda^{d'} \right)^2 \right)
            \right.
        }.
    \end{align*}
    This is a rational function of $\hat{\alpha}$ and deployment parameters.
    In particular, $R_A$ does not depend on $\hat{\beta}$, which allows us to potentially identify $\hat{\alpha}$ from $\left( d, d' \right)$ and the observed weights $w_A^d$ and $w_A^{d'}$. We thus drop $\hat{\beta}$ from the notation of $R_A$.

    Note that the denominator after simplification is $\left( 1 - \hat{\alpha} \lambda^{d'} \right) \left( \tau_H  - \tau_A^{d} \hat{\alpha}^2 \left( \lambda^{d} \right)^2 \right)$. On $\hat{\alpha} \in \left[ 0, 1 \right]$ (no over-correlation), the second term is bounded away from zero under the model's assumed parameter restriction $\tau_A^{d} \left( \lambda^{d} \right)^2 < \tau_H$, and the first term is bounded away from zero iff $\hat{\alpha} \lambda^{d'} \neq 1$. Hence $R_A$ is continuous in $\hat{\alpha} \in \left[ 0, 1 \right]$, except at $\hat{\alpha} = 1$ if $\lambda^{d'} = 1$.
    
    % \emph{Best response benchmark.} At $\hat{\alpha}=1, \hat{\beta}=1$,
    % \begin{align*}
    %     R_A^{\mathrm{BR}}
    %     = \frac{
    %         \left.
    %         \tau_H \left( 1 - \lambda^{d} \right)
    %         \middle/
    %         \left( \tau_H  - \tau_A^{d} \left( \lambda^{d} \right)^2 \right)
    %         \right.
    %     }{
    %         \left.
    %         \tau_H \left( 1 - \lambda^{d'} \right)
    %         \middle/
    %         \left( \tau_H  - \tau_A^{d'} \left( \lambda^{d'} \right)^2 \right)
    %         \right.
    %     }.
    % \end{align*}
    % \redtext{Deferring this case in order to attempt a more general claim about $R_A=1$.}

    \emph{Special case: Correlation neglect.} At $\hat{\alpha}=0$, $R_A^{\mathrm{CN}} = \frac{ 1 / \tau_H } { 1 / \tau_H } = 1$ for every $\hat{\beta}$.

    \emph{Ratio is generically not equal to $1$ outside of correlation neglect.} 
    $R_A \left( \hat{\alpha}; d, d' \right) = 1$ under the following necessary and sufficient condition:
    \begin{align}
        % \frac
        %     { \tau_H \left( 1 - \hat{\alpha} \lambda^{d} \right) }
        %     { \tau_H  - \tau_A^{d} \hat{\alpha}^2 \left( \lambda^{d} \right)^2 }
        % &=
        % \frac
        %     { \tau_H \left( 1 - \hat{\alpha} \lambda^{d'} \right) }
        %     { \tau_H  - \tau_A^{d'} \hat{\alpha}^2 \left( \lambda^{d'} \right)^2 },
        % \nonumber \\
        % \left( 1 - \hat{\alpha} \lambda^{d} \right)
        % \left( \tau_H  - \tau_A^{d'} \hat{\alpha}^2 \left( \lambda^{d'} \right)^2 \right)
        % &=
        % \left( 1 - \hat{\alpha} \lambda^{d'} \right)
        % \left( \tau_H  - \tau_A^{d} \hat{\alpha}^2 \left( \lambda^{d} \right)^2 \right),
        % \nonumber \\
        \tau_H \hat{\alpha}  \left( \lambda^{d'} - \lambda^{d} \right)
        &= 
        \left( 1 - \hat{\alpha} \lambda^{d} \right)
        \tau_A^{d'} \hat{\alpha}^2 \left( \lambda^{d'} \right)^2
        -
        \left( 1 - \hat{\alpha} \lambda^{d'} \right)
        \tau_A^{d} \hat{\alpha}^2 \left( \lambda^{d} \right)^2 .
        \label{eqn:weight_ratio_1_iff}
    \end{align}
    Observe that~\eqref{eqn:weight_ratio_1_iff} describes a specific knife-edge case that depends on all deployment parameters in $d$ and $d'$, not just $\lambda$. To see this, note that it is equivalent to a unique requirement for $\tau_A^{d'}$, the second deployment's AI capability:
    \begin{align*}
        \tau_A^{d'} = \frac
        { \tau_H \hat{\alpha}  \left( \lambda^{d'} - \lambda^{d} \right) 
        + 
        \left( 1 - \hat{\alpha} \lambda^{d'} \right)
        \tau_A^{d} \hat{\alpha}^2 \left( \lambda^{d} \right)^2 }
        { \left( 1 - \hat{\alpha} \lambda^{d} \right)
        \hat{\alpha}^2 \left( \lambda^{d'} \right)^2 }
    \end{align*}
    whenever the denominator is nonzero.
    Also note that the proposition's claim about best response (BR, $\hat{\alpha}=1, \hat{\beta}=1$) is a direct consequence of \eqref{eqn:weight_ratio_1_iff}.

    \emph{Derivatives (for solving $\hat{\alpha}$): general case.} The global log-derivative in $\hat{\alpha}$ is
    \begin{align}
        \frac{\partial}{\partial \hat{\alpha}} \log R_A \left( \hat{\alpha}; d, d' \right)
        % &= - \frac{ \lambda^{d} } { 1 - \hat{\alpha} \lambda^{d} }
        % + \frac{ \lambda^{d'} } { 1 - \hat{\alpha} \lambda^{d'} }
        % + \frac
        % { 2 \hat{\alpha} \tau_A^{d} \left( \lambda^{d} \right)^2 }
        % { \tau_H  - \tau_A^{d} \hat{\alpha}^2 \left( \lambda^{d} \right)^2 }
        % - \frac
        % { 2 \hat{\alpha} \tau_A^{d'} \left( \lambda^{d'} \right)^2 }
        % { \tau_H  - \tau_A^{d'} \hat{\alpha}^2 \left( \lambda^{d'} \right)^2 }
        % \\
        &= \frac
        { \lambda^{d'} - \lambda^{d} }
        { \left( 1 - \hat{\alpha} \lambda^{d} \right) \left( 1 - \hat{\alpha} \lambda^{d'} \right) }
        + \frac
        { 2 \hat{\alpha} \tau_H
          \left[ \tau_A^{d} \left( \lambda^{d} \right)^2 - \tau_A^{d'} \left( \lambda^{d'} \right)^2 \right] }
        { \left( \tau_H  - \tau_A^{d} \hat{\alpha}^2 \left( \lambda^{d} \right)^2 \right)
          \left( \tau_H  - \tau_A^{d'} \hat{\alpha}^2 \left( \lambda^{d'} \right)^2 \right) }.
        \label{eqn:weight_ratio_log_deriv}
    \end{align}
    Specifically, the local derivative near $\hat{\alpha} = 0$ is
    $
        \left. \frac{\partial}{\partial \hat{\alpha}} \log R_A 
        % \left( \hat{\alpha}; d, d' \right) 
        \right|_{\hat{\alpha} = 0}
        = \lambda^{d'} - \lambda^{d},
    $
    so $R_A$ is locally invertible near $\hat{\alpha} = 0$ iff $\lambda^{d'} - \lambda^{d} \neq 0$.

  \end{proof}
In practice, $\left( \hat{\alpha}, \hat{\beta} \right)$ may vary across topical domains, so identification applies within a domain (math, physics, etc.) rather than pooled.

\subsection{Implementation details}
\S\ref{sec:identification} presents the two-step identification result theoretically. In the empirical work of \S\ref{sec:experiments} we implement two complementary procedures, both inversions of Lemma~\ref{lem:behavioral}, differing in how the human-side signal $\ell_H$ and the overlap $\lambda$ are constructed.

\paragraph{Approach 1: joint inversion using confidence-based $\ell_H$.}
For each deployment $d \in \{\text{GPT-4o}, \text{Llama-3.1-8b}\}$, fit a logistic regression of $F$ on $(\ell_H, \ell_A)$ with intercept (no regularization, BFGS) on the Has-Rec rows for that deployment, recovering empirical weights $(\hat w_H^d, \hat w_A^d)$. Estimate the conditional-mean primitives by their plug-in moments,
\[
\textstyle
\hat\tau_H = \tfrac{1}{2}\!\left(\overline{\ell_H \mid h\_acc = 1} - \overline{\ell_H \mid h\_acc = 0}\right), \qquad
\hat\tau_A^d = \tfrac{1}{2}\!\left(\overline{\ell_A \mid ai\_correct = 1, d} - \overline{\ell_A \mid ai\_correct = 0, d}\right),
\]
treating $\hat\tau_H$ as deployment-invariant (the empirical per-deployment values $0.374$ and $0.411$ agree to within $0.04$). Then solve the four-equation system from Lemma~\ref{lem:behavioral},
\[
\textstyle
\hat w_H^d = \frac{\hat\tau_H - \hat\tau_A^d \,\hat\alpha\, \hat\lambda^d}{\hat\tau_H - \hat\tau_A^d \,\hat\alpha^2 (\hat\lambda^d)^2}, \qquad
\hat w_A^d = \hat\beta \cdot \frac{\hat\tau_H (1 - \hat\alpha \,\hat\lambda^d)}{\hat\tau_H - \hat\tau_A^d \,\hat\alpha^2 (\hat\lambda^d)^2}, \qquad d \in \{\text{GPT}, \text{Llama}\},
\]
for the four unknowns $(\hat\alpha, \hat\beta, \hat\lambda^{\text{GPT}}, \hat\lambda^{\text{Llama}})$ by trust-region-reflective nonlinear least squares with bounds $\hat\alpha \in [0, 1]$, $\hat\beta \in (0, 5]$, $\hat\lambda^d \in [0, 0.99]$, and $100$ random restarts. The global minimizer is
\[
\textstyle
\hat\alpha = 0.315, \quad \hat\beta = 0.285, \quad \hat\lambda^{\text{GPT}} = 0.478, \quad \hat\lambda^{\text{Llama}} = 0.365, \quad \mathrm{SSR} = 5.5 \times 10^{-3}.
\]
A $200$-replication non-parametric bootstrap gives $\hat\beta \in [0.246, 0.322]$ at the $95\%$ level.

\paragraph{Approach 2: $\hat\lambda$ from cross-question variation.}
This procedure replaces $\ell_H$ with the per-question signal $\ell_H^{H_q}$ from \S\ref{app:sample} and estimates $\lambda$ directly as the within-$Y$ pooled covariance,
\[
\textstyle
\hat\lambda^d := \frac{\sum_{y \in \{0,1\}} n_{d,y}\, \widehat{\mathrm{Cov}}(\ell_H^{H_q}, \ell_A \mid Y = y, d)}{\sum_{y \in \{0,1\}} n_{d,y}\, \widehat{\mathrm{Var}}(\ell_A \mid Y = y, d)},
\]
mirroring the population identity $\lambda = \mathrm{Cov}(\ell_H, \ell_A \mid Y)/\mathrm{Var}(\ell_A \mid Y)$ implied by Lemma~\ref{lem:joint}. We obtain
\[
\textstyle
\hat\lambda^{\text{GPT}} = 0.071 \in [0.041, 0.100]_{95\%}, \qquad \hat\lambda^{\text{Llama}} = 0.127 \in [0.100, 0.154]_{95\%},
\]
with both bootstrap CIs strictly above zero. We then re-run the per-deployment logistic regressions of $F$ on $(\ell_H^{H_q}, \ell_A)$ to obtain weights under this finer signal, and solve the analogous system for $(\hat\alpha, \hat\beta)$ with $\hat\lambda^d$ now treated as plug-in. The resulting fit gives $\hat\beta = 0.254$, in close agreement with Approach~1.

\paragraph{On the discrepancy in $\hat\alpha$.}
The two approaches agree on $\hat\beta$ (the headline parameter) but not on $\hat\alpha$: Approach~1 gives an interior $\hat\alpha = 0.315$, Approach~2 returns $\hat\alpha = 1$ at the upper bound. We interpret this as evidence that $\hat\alpha$ is weakly identified by these moments — small differences in how $\ell_H$ is constructed propagate through the nonlinear inversion to large differences in the $\hat\alpha$ point estimate. The qualitative claim that $\hat\alpha > 0$ (strict correlation neglect rejected) is robust across approaches and across three direct tests:
\begin{enumerate}
  \item[(i)] Approach~1's residual sum of squares jumps $141$-fold when $\hat\alpha = 0$ is forced.
  \item[(ii)] A likelihood-ratio test of whether $|\ell_A|$ adds information beyond 4-cell $(h\_acc, ai\_correct)$ fixed effects gives $\chi^2(1) = 3.93$, $p = 0.048$.
  \item[(iii)] Approach~2's direct $\hat\lambda^d$ are strictly positive in $100\%$ of bootstrap samples for both deployments.
\end{enumerate}

\paragraph{On the discrepancy in $\hat\lambda$.}
Both approaches yield positive estimates of overlap, but differ in their relative ordering across the two AIs. We adopt Approach 2 as our primary measure, as it estimates overlap directly from the conditional covariance structure of human and AI signals, rather than indirectly through a nonlinear inversion of worker behavior.

\paragraph{Robustness of $\hat\beta$.}
Varying the clipping constant $\epsilon \in \{0.001, 0.005, 0.01, 0.02\}$ (\S\ref{app:sample}) shifts $\hat\beta$ across $[0.22, 0.37]$ but leaves it well below $1$ at every value, so the rejection of $\hat\beta = 1$ is robust. A worker-cluster bootstrap that resamples the $315$ workers (rather than $2{,}199$ rows i.i.d.) gives $\hat\beta \in [0.241, 0.332]$ at the $95\%$ level, slightly wider than the row-bootstrap CI but with the same qualitative conclusion across all three headline quantities ($\mathrm{GAP}$, $\hat\beta$, $\hat\lambda^{\text{Llama}} - \hat\lambda^{\text{GPT}}$).

\paragraph{Test for $\mathcal{A}_{HA}$ versus $\mathcal{A}_{AI}$.}
Regime classification in Table~\ref{tab:emp-summary} compares team accuracy $\mathcal{A}_{HA}$ to AI accuracy $\mathcal{A}_{AI}$ on the same observations. Define the discordant pair counts
\[
\textstyle
n_{10}^d := |\{i \in d : F_i = 1, \, ai\_correct_i = 0\}|, \qquad n_{01}^d := |\{i \in d : F_i = 0, \, ai\_correct_i = 1\}|.
\]
Under $H_0: \mathcal{A}_{HA} = \mathcal{A}_{AI}$, the conditional distribution of $n_{10}^d$ given $n_{10}^d + n_{01}^d$ is $\mathrm{Binomial}(n_{10}^d + n_{01}^d, \tfrac{1}{2})$. The reported $p$-values are exact two-sided binomial $p$-values from this conditional distribution (the exact form of the McNemar test). The relevant counts are $(n_{10}^{\text{GPT}}, n_{01}^{\text{GPT}}) = (51, 89)$ and $(n_{10}^{\text{Llama}}, n_{01}^{\text{Llama}}) = (74, 66)$. We classify a deployment as automation if the test rejects in favor of $n_{01} > n_{10}$, complementarity if it rejects in favor of $n_{10} > n_{01}$, and boundary otherwise.

\section{Capability-versus-behavior decomposition}
\label{app:capability}

The cross-deployment GAP gap of $+0.081$ admits two non-exclusive interpretations: (i) AI-capability composition — different fractions of pairs in each $(h\_acc, ai\_correct)$ cell across deployments — or (ii) workers responding differently to the two AIs at the per-cell level. We separate them by holding worker behavior at the pooled per-cell rate $\widehat{\Pr}(F = 1 \mid h\_acc, ai\_correct)$ across both deployments and letting only the cell composition vary by AI.

\paragraph{Counterfactual GAP gap.}
Let $\bar p_{(h, a)} := \widehat{\Pr}(F = 1 \mid h\_acc = h, ai\_correct = a)$ on the pooled Has-Rec sample. For each deployment $d$, define the counterfactual team accuracy
\[
\textstyle
\mathcal{A}_{HA}^{d, \mathrm{cf}} := \sum_{h, a \in \{0,1\}} \widehat{\Pr}_d(h\_acc = h,\, ai\_correct = a) \cdot \bar p_{(h, a)},
\]
which assigns the pooled per-cell behavior to deployment $d$'s observed cell composition. The capability share is the ratio of the counterfactual GAP gap (driven by composition alone) to the observed GAP gap:
\[
\textstyle
\mathrm{CapShare} := \frac{(\mathcal{A}_{HA}^{\text{GPT}, \mathrm{cf}} - \mathcal{A}_H^{\text{GPT}}) - (\mathcal{A}_{HA}^{\text{Llama}, \mathrm{cf}} - \mathcal{A}_H^{\text{Llama}})}{\mathrm{GAP}^{\text{GPT}} - \mathrm{GAP}^{\text{Llama}}}.
\]
The point estimate is $\mathrm{CapShare} = 92.6\%$. A non-parametric row bootstrap with $1{,}000$ replications gives a median of $93.2\%$ and a 95\% confidence interval of $[68\%, 148\%]$. The wide upper end reflects the fact that the denominator $\mathrm{GAP}^{\text{GPT}} - \mathrm{GAP}^{\text{Llama}}$ is itself a small estimated quantity ($+0.081$); the lower bound stays well above $50\%$ across resamples.

\paragraph{Direct test of behavioral homogeneity.}
A complementary test compares two logistic regressions of $F$ on $(h\_acc, ai\_correct)$ cell fixed effects: a reduced model with no AI interaction versus a full model interacting cell with deployment. The likelihood-ratio statistic is $\chi^2(4) = 0.96$ ($p = 0.92$); we cannot reject the null that workers respond identically to the two AIs at the per-cell level. The capability-share decomposition and the LR test thus point in the same direction: the cross-deployment welfare difference is driven primarily by the composition of $(h\_acc, ai\_correct)$ pairs each AI generates, not by deployment-specific worker behavior.

\section{Overlap heterogeneity across AIs and subjects}
\label{app:lambda}

The direct $\hat\lambda^d$ estimator from Approach~2 of \S\ref{app:identification} can be applied at finer-than-deployment granularity, by restricting the within-$Y$ covariance computation to subsets of the Has-Rec sample. We report two cuts.

\paragraph{Across AIs.}
Pooled across all subjects, $\hat\lambda^{\text{GPT}} = 0.071$ and $\hat\lambda^{\text{Llama}} = 0.127$, a difference of $+0.056$. A row bootstrap on the difference gives a 95\% CI of $[+0.015, +0.097]$ (two-sided $p \approx 0.009$), with $99.7\%$ of $1{,}000$ resamples having $\hat\lambda^{\text{Llama}} > \hat\lambda^{\text{GPT}}$. The smaller AI's evidence overlaps more with what humans access on the same questions; this is consistent with the cue-pool reading of \S\ref{sec:model}, in which a smaller model relies more heavily on cues that humans also access while a larger model contributes more genuinely novel evidence.

\paragraph{Across subjects.}
The Has-Rec sample is split as math ($n = 1{,}388$), physics ($n = 497$), and moral reasoning ($n = 314$). Direct $\hat\lambda$ varies dramatically across subjects, and the same ordering holds within each AI separately:

\begin{center}
\small
\begin{tabular}{lcccc}
\toprule
Subject  & $n$ (Has-Rec) & GPT-4o $\hat\lambda$ & Llama $\hat\lambda$ & Pooled $\hat\lambda$ \\
\midrule
Math     & $1{,}388$ & $+0.19$ & $+0.25$ & $+0.20$ \\
Physics  & $497$     & $+0.04$ & $+0.07$ & $+0.04$ \\
Moral    & $314$     & $\phantom{+}0.00$ & $\phantom{+}0.00$ & $\phantom{+}0.00$ \\
\midrule
Pooled   & $2{,}199$ & $+0.07$ & $+0.13$ & $+0.08$ \\
\bottomrule
\end{tabular}
\end{center}

Domains where humans and AI rely on shared procedural cues (math) exhibit substantial overlap; domains where reasoning paths are more idiosyncratic (moral reasoning) exhibit essentially none. This provides empirical content for the cue-pool framing: under the model, $\lambda$ is a property of the human-domain pair (Proposition~\ref{prop:overlap-domain}); the data shows the predicted variation across domains and additionally a cross-AI effect that the uniform-sampling model does not predict.

\paragraph{Limitations of these estimates.}
Both findings rely on the per-question $\ell_H^{H_q}$ proxy, which is a population-level signal rather than an individual worker's signal. Replacing $\ell_H^{H_q}$ with the confidence-based $\ell_H$ (Approach~1) does not give a directly comparable $\hat\lambda$ because the latter has only three values per worker. We report the cross-AI and cross-subject patterns as evidence that $\lambda$ varies meaningfully and in interpretable directions, not as precise structural estimates.

\section{Empirical welfare evaluation}
\label{app:welfare}

We evaluate the welfare consequences of trust calibration by simulating a trust-calibrated decision rule. For each observation $i$, recall $h_i := h\_acc_i$ indicate whether the human's unaided answer is correct, $a_i := ai\_correct_i$ whether the AI's actual recommendation is correct, and $F_i$ whether the worker's final answer is correct. We simulate a reliance policy that chooses between the human's and AI's answers, and its accuracy is evaluated against the ground truth of the question.

\paragraph{Constructing the simulated policy.}
The policy uses $X_i = (\ell_H^i, \ell_A^i)$, where $\ell_H^i$ is constructed from the worker's stated confidence and $\ell_A^i$ from the AI's question-level accuracy, as described in Appendix~\ref{app:sample}. These predictors summarize human confidence and AI performance on the question; they do not encode the content of the AI's specific response.

Within the disagreement sample, define
\[
p_H^d(x) := \Pr(h_i = 1 \mid X_i = x,\ i \in \mathcal{D}_d),
\qquad
p_A^d(x) := \Pr(a_i = 1 \mid X_i = x,\ i \in \mathcal{D}_d).
\]
We estimate these conditional correctness probabilities using logistic regression on $(\ell_H, \ell_A)$ with five-fold cross-validation. Each observation receives predictions from models fitted on the other four folds.

For each held-out observation, define the reliance score and decision
\[
s_i := \hat p_A^d(X_i) - \hat p_H^d(X_i),
\qquad
\pi_i := \mathbf{1}\{s_i > 0\}.
\]
The simulated policy submits the AI's answer when $\pi_i = 1$ and retains the human's answer otherwise. We fix the threshold at zero rather than selecting it to maximize accuracy on the evaluation observations. With the true conditional probabilities, this rule is optimal among policies choosing between the two answers using $X_i$. The fitted logistic rule provides an empirical approximation to this benchmark.

\begin{table}[h]
\centering
\caption{Accuracy gains from the logistic answer-selection policy. All entries are percentage points. The full-sample contribution weights the disagreement gain by the disagreement share.}
\label{tab:welfare-decomp}
\small
\begin{tabular}{lcc}
\toprule
Deployment & Gain on disagreements & Full-sample contribution \\
\midrule
GPT-4o       & $+5.9$ & $+2.6$ \\
Llama-3.1-8b & $-3.4$ & $-1.4$ \\
Pooled       & $+1.4$ & $+0.6$ \\
\bottomrule
\end{tabular}
\end{table}

\paragraph{Capability-dependent benefits of trust calibration.}
For GPT-4o, the simulated policy raises accuracy from approximately $0.851$ to $0.877$, bringing the team close to the AI-alone accuracy of $0.884$. It follows the AI on approximately $92\%$ of disagreements, compared with the workers' observed reliance rate of $70\%$. The policy's greater reliance and higher accuracy are consistent with a welfare cost from under-trust on the stronger AI.

For Llama-3.1-8b, the observed workers outperform the fitted policy. Within this comparison, replacing their decisions with the coarse-signal rule offers no improvement. The benefits of greater trust therefore depend on AI capability and cannot be assumed to transfer uniformly across deployments.

\paragraph{Information and the limits of the benchmark.}
In reality, the worker observes more than the two predictors available to the fitted rule. In particular, she reads the AI's actual answer and can assess its reasoning and relevance to the question. The question-level accuracy measure averages across responses and omits these response-specific information. Workers' superior performance relative to the fitted rule in the Llama deployment is consistent with their using this additional information.

The broader implication is that improving human--AI performance may require attention to information as well as behavior. Better-calibrated trust offers a measurable opportunity in the stronger-AI deployment, but a rule based only on coarse confidence and accuracy measures does not capture the worker's full decision problem. Richer response-specific information would allow a more informative benchmark and a clearer assessment of the remaining scope for behavioral improvement.

\section{Departures from theory and what they mean}
\label{app:departures}

The framework of \S\S\ref{sec:model}--\ref{sec:analysis} is a binary-classification model with a Gaussian generative process, symmetric prior, and linear-rule decision-maker. The empirical setting departs from each of these in ways the analysis must accommodate.

\paragraph{Binary reduction of a four-way task.}
ChatBench is a four-way multiple-choice task; the model is binary. We define the outcome $Y_i$ as the indicator that the worker's chosen letter matches the unique correct letter on item $i$, collapsing the three wrong letters into $Y_i = 0$. Two consequences follow. \emph{(i)} The model's closed-form AI accuracy $\Phi(\sqrt{\hat\tau_A^d/2})$ does not match empirical accuracy: at $\hat\tau_A^{\text{Llama}} = 1.64$ the formula gives $0.82$ versus empirical $\mathcal{A}_{AI}^{\text{Llama}} = 0.787$, and at $\hat\tau_A^{\text{GPT}} = 2.02$ it gives $0.84$ versus empirical $0.884$. We do not use the closed-form accuracy formula anywhere in the empirical section; every accuracy reported in \S\ref{sec:experiments} and Tables~\ref{tab:emp-summary} and~\ref{tab:welfare-decomp} is the empirical sample mean. \emph{(ii)} The empirical base rate $\widehat{\Pr}(h\_acc = 1) = 0.612$ is not $0.5$, so the symmetric-prior assumption underlying $\Phi(\sqrt{\hat\tau_H/2}) = \Pr(h\_acc = 1)$ does not literally hold. We use prior-free conditional-mean estimators for $\hat\tau_H$ and $\hat\tau_A^d$ (Appendix~\ref{app:identification}), which are invariant to the base rate.

\paragraph{$\ell_A$ saturation and cross-deployment scale.}
About $62\%$ of Has-Rec rows have $p_A^{m,q} \in \{0, 1\}$ (Appendix~\ref{app:sample}), saturating $\ell_A$ at $\pm 5.293$. Saturation is more severe for GPT-4o, where $p_A^{m,q} = 1$ on the majority of items, than for Llama-3.1-8b. As a consequence, per-deployment regression coefficients $\hat w_A^d$ are not on the same scale across deployments: $\hat w_A^{\text{GPT}}$ effectively measures the response to a near-binary "AI is right" indicator, while $\hat w_A^{\text{Llama}}$ uses more variation. Approach~1 (\S\ref{app:identification}) absorbs this scale difference through the structural inversion, which divides through $\hat\tau_A^d$ when recovering $\hat\beta$. Approach~2 uses cross-question covariance and is invariant to monotone rescaling of $\ell_A$. Direct comparison of $\hat w_A^d$ across deployments would not be valid; the structural recovery of $\hat\beta$ is.

\paragraph{Cue-pool generative story.}
The cue-pool model of \S\ref{sec:model} motivates $\lambda$ as the fraction of the AI's cues also accessed by the human. Language-model decisions are not literally generated by sampling discrete cues from a finite pool. We do not commit to the cue-pool generative process. The structural objects we estimate — the joint distribution of $(\ell_H, \ell_A, Y)$ and its summary statistics $(\tau_H, \tau_A, \lambda)$ — are well-defined regardless of the underlying generative process. The cue-pool story provides intuition for the cross-domain $\lambda$ heterogeneity in Appendix~\ref{app:lambda}, but it is not required for the identification or welfare results.

%\paragraph{Linear decision rule.}
%The framework assumes the worker uses a linear-weighting decision rule (Lemma~\ref{lem:behavioral}). Empirically, workers may use non-linear rules. The per-deployment logistic regressions of \S\ref{app:identification} fit the best linear approximation to actual worker behavior; the inversion identifies the model parameters that would produce these linear weights. The framework is informative about workers to the extent that linear weights describe their behavior. The small empirical $\mathrm{TWL}_{\mathrm{lin}}$ ($+0.014$) suggests workers in this sample are within $1.4$ percentage points of the best linear classifier, so the linear-rule restriction is not where most of the worker--Bayes-rational gap lives. The further $\sim 7$pp to $\mathcal{A}^{\mathrm{oracle}}$ (Appendix~\ref{app:welfare}) is the welfare reachable only by non-linear use of the joint signal, which the framework cannot capture.

\paragraph{Between-subjects design.}
ChatBench randomizes each worker to exactly one AI, so the GPT-4o and Llama-3.1-8b samples are disjoint worker pools. Cross-deployment comparisons therefore rely on random assignment rather than within-worker variation. Random assignment makes large population-level imbalances unlikely in expectation but does not rule out them in any specific sample. The behavioral-homogeneity test of Appendix~\ref{app:capability} ($\chi^2(4) = 0.96$, $p = 0.92$) provides a direct check: per-cell response rates are statistically indistinguishable across the two AI populations, supporting the cross-deployment comparability of $\hat\beta$ and other behavioral parameters.

\section{Reproducibility}
\label{app:repro}

All empirical claims in Section~\ref{sec:experiments} and Appendices~\ref{app:sample}--\ref{app:welfare} are computed by a single self-contained Python script that takes the three raw ChatBench files as input. It reproduces sample construction, signal calibration, the headline findings with bootstrap CIs ($1{,}000$ direct, $200$ structural, $500$ welfare), all regime, correlation-neglect, capability-share, and $\lambda$-heterogeneity tests, the $\hat\beta = 1$ counterfactual, and three robustness checks (worker-cluster bootstrap on all four headline quantities, $\epsilon$ sensitivity, $5$-fold cross-validated linear ceiling). All seeds are fixed; reference values appear in the script header; total runtime is approximately $4$ minutes on a single CPU core. The script will be provided with the camera-ready version.

%A Jupyter notebook \texttt{verify\_chatbench.ipynb} accompanies this paper and reproduces every numerical claim in \S\ref{sec:experiments} and Appendices~\ref{app:sample}--\ref{app:welfare} from the three raw ChatBench files. The notebook (i) loads the raw data and verifies sample sizes, (ii) constructs the matched answer-first Has-Rec sample including the truth-recovery drop, (iii) calibrates $\ell_H$ (both proxies) and $\ell_A$, (iv) computes the eight headline findings of the empirical section with bootstrap CIs ($1{,}000$ replications for direct quantities, $200$ for the structural inversion, $500$ for welfare benchmarks), (v) executes the regime tests, the three correlation-neglect tests, the capability-share decomposition, the $\lambda$-heterogeneity bootstraps, and the $\hat\beta = 1$ counterfactual, and (vi) runs three robustness checks (worker-cluster bootstrap, $\epsilon$ sensitivity, $5$-fold cross-validated linear ceiling). All random seeds are fixed; reference values are listed in the notebook header for verification. Total runtime is approximately $4$ minutes on a single CPU core.

%%%%%%%%%%%%%%%%%%%%%%%%%%%%%%%%%%%%%%%%%%%%%%%%%%%%%%%%%%%%

\newpage
\section*{NeurIPS Paper Checklist}

\begin{enumerate}

\item {\bf Claims}
    \item[] Question: Do the main claims made in the abstract and introduction accurately reflect the paper's contributions and scope?
    \item[] Answer: \answerYes{}
    \item[] Justification: The abstract and introduction state three contributions: a phase diagram in $(\tau_A, \lambda)$ partitioning impairment, complementarity, and automation regions; identifiability of $(\hat\alpha, \hat\beta)$ from cross-deployment variation in overlap; and a welfare decomposition into correlation-neglect and trust-miscalibration costs --- which are delivered respectively in Theorem~\ref{thm:phase}, Proposition~\ref{prop:cross_deploy_signature_with_ratio}, and Proposition~\ref{prop:welfare}, and instantiated empirically on ChatBench in \S\ref{sec:experiments}.

\item {\bf Limitations}
    \item[] Question: Does the paper discuss the limitations of the work performed by the authors?
    \item[] Answer: \answerYes{}
    \item[] Justification: Appendix~\ref{app:departures} enumerates the main limitations of the empirical study as departures from the theoretical model.

\item {\bf Theory assumptions and proofs}
    \item[] Question: For each theoretical result, does the paper provide the full set of assumptions and a complete (and correct) proof?
    \item[] Answer: \answerYes{}
    \item[] Justification: All theoretical results are stated with full assumptions. Complete proofs are given in Appendices~\ref{app:behavioral} and the proof sections referenced from \S\ref{sec:model} and \S\ref{sec:analysis}.

    \item {\bf Experimental result reproducibility}
    \item[] Question: Does the paper fully disclose all the information needed to reproduce the main experimental results of the paper to the extent that it affects the main claims and/or conclusions of the paper (regardless of whether the code and data are provided or not)?
    \item[] Answer: \answerYes{}
    \item[] Justification: Appendix~\ref{app:sample} fully describes sample construction; Appendix~\ref{app:identification} specifies the estimation procedures; Appendix~\ref{app:repro} describes reproducibility details.

\item {\bf Open access to data and code}
    \item[] Question: Does the paper provide open access to the data and code, with sufficient instructions to faithfully reproduce the main experimental results, as described in supplemental material?
    \item[] Answer: \answerYes{}
    \item[] Justification: All experiments use the publicly released ChatBench dataset~\citep{chang2025chatbench}. Reproducibility details (sample construction, estimators, bootstrap settings, runtime) are in Appendix~\ref{app:repro}; the code will be released publicly with the camera-ready version.

\item {\bf Experimental setting/details}
    \item[] Question: Does the paper specify all the training and test details (e.g., data splits, hyperparameters, how they were chosen, type of optimizer) necessary to understand the results?
    \item[] Answer: \answerYes{}
    \item[] Justification: \S\ref{sec:experiments} and Appendices~\ref{app:sample}--\ref{app:identification} specifies necessary details.

\item {\bf Experiment statistical significance}
    \item[] Question: Does the paper report error bars suitably and correctly defined or other appropriate information about the statistical significance of the experiments?
    \item[] Answer: \answerYes{}
    \item[] Justification: Headline empirical claims in \S\ref{sec:experiments} (e.g., $\mathrm{GAP} = +0.213$, $\hat\beta = 0.285$, $\hat\lambda^{\text{Llama}} - \hat\lambda^{\text{GPT}} = +0.056$) are reported with $95\%$ non-parametric bootstrap confidence intervals; and a worker-cluster bootstrap is reported in Appendix~\ref{app:identification} as a robustness check.

\item {\bf Experiments compute resources}
    \item[] Question: For each experiment, does the paper provide sufficient information on the computer resources (type of compute workers, memory, time of execution) needed to reproduce the experiments?
    \item[] Answer: \answerYes{}
    \item[] Justification: Appendix~\ref{app:repro} reports total runtime of approximately $4$ minutes on a single CPU core for the entire pipeline.
    
\item {\bf Code of ethics}
    \item[] Question: Does the research conducted in the paper conform, in every respect, with the NeurIPS Code of Ethics \url{https://neurips.cc/public/EthicsGuidelines}?
    \item[] Answer: \answerYes{}
    \item[] Justification:

\item {\bf Broader impacts}
    \item[] Question: Does the paper discuss both potential positive societal impacts and negative societal impacts of the work performed?
    \item[] Answer: \answerYes{}
    \item[] Justification: The framework's positive design implications are stated in \S\ref{sec:experiments} and \S\ref{sec:analysis}. On the negative side, the analysis quantifies regimes where adding an AI assistant degrades worker accuracy, which cautions against blanket deployment. The paper introduces no new model or dataset with foreseeable misuse pathways.
    
\item {\bf Safeguards}
    \item[] Question: Does the paper describe safeguards that have been put in place for responsible release of data or models that have a high risk for misuse (e.g., pre-trained language models, image generators, or scraped datasets)?
    \item[] Answer: \answerNA{}
    \item[] Justification: The paper releases no pretrained model, image generator, or scraped dataset.

\item {\bf Licenses for existing assets}
    \item[] Question: Are the creators or original owners of assets (e.g., code, data, models), used in the paper, properly credited and are the license and terms of use explicitly mentioned and properly respected?
    \item[] Answer: \answerYes{}
    \item[] Justification: All external assets are cited at first use: ChatBench~\citep{chang2025chatbench} for the human-AI interaction data and MMLU~\citep{hendrycks2020measuring} for the underlying questions.

\item {\bf New assets}
    \item[] Question: Are new assets introduced in the paper well documented and is the documentation provided alongside the assets?
    \item[] Answer: \answerNA{}
    \item[] Justification:

\item {\bf Crowdsourcing and research with human subjects}
    \item[] Question: For crowdsourcing experiments and research with human subjects, does the paper include the full text of instructions given to participants and screenshots, if applicable, as well as details about compensation (if any)?
    \item[] Answer: \answerNA{}
    \item[] Justification: We did not run any new crowdsourcing or human-subjects experiment.

\item {\bf Institutional review board (IRB) approvals or equivalent for research with human subjects}
    \item[] Question: Does the paper describe potential risks incurred by study participants, whether such risks were disclosed to the subjects, and whether Institutional Review Board (IRB) approvals (or an equivalent approval/review based on the requirements of your country or institution) were obtained?
    \item[] Answer: \answerNA{}
    \item[] Justification: We did not run any new human-subjects research; all data come from the publicly released ChatBench study, which was conducted by its original authors.

\item {\bf Declaration of LLM usage}
    \item[] Question: Does the paper describe the usage of LLMs if it is an important, original, or non-standard component of the core methods in this research? Note that if the LLM is used only for writing, editing, or formatting purposes and does \emph{not} impact the core methodology, scientific rigor, or originality of the research, declaration is not required.
    %this research?
    \item[] Answer: \answerNA{}
    \item[] Justification:  LLMs are studied as objects in this paper (GPT-4o and Llama-3.1-8b in the ChatBench data) but were not used as components of the methodology.

\end{enumerate}

\end{document}